\documentclass[a4paper,11pt]{article}
\usepackage{jheppub} 
\usepackage{lineno}
\usepackage{amsfonts,amssymb,epsfig,amsmath,mathtools}
\usepackage{color}
\usepackage{comment}
\usepackage{url}
\usepackage{mathrsfs}
\usepackage{graphicx}
\usepackage{subcaption}
\usepackage{hyperref}

\def\Tr{{\rm Tr}}

\def\det{{\rm det}}

\newcommand{\be}{\begin{eqnarray}}
\newcommand{\ee}{\end{eqnarray}}

\newcommand{\bn}{\begin{enumerate}}
\newcommand{\en}{\end{enumerate}}
\newcommand{\mL}{\mathcal{L}}
\newcommand{\bL}{{\bar L}}

\preprint{TIFR/TH/23-8}
\title{\boldmath 
\vspace*{-0.3cm}
`Grey Galaxies' as an endpoint of the Kerr-AdS superradiant instability 
\vspace{-0.2cm}
}

\author[a,1]{Seok Kim,\note{seokkimseok@gmail.com}}
\author[b,2]{Suman Kundu,\note{suman.kundu@weizmann.ac.il}}
\author[a,3]{Eunwoo Lee,\note{eunwoo42@snu.ac.kr}}
\author[c,4]{Jaeha Lee,\note{jaeha@caltech.edu}}
\author[d,5]{Shiraz Minwalla,\note{minwalla@theory.tifr.res.in}}
\author[d, 6]{Chintan Patel.\note{chintan.patel@tifr.res.in}}
\affiliation[a]{Department of Physics and Astronomy \& Center for Theoretical Physics,\\
Seoul National University, Seoul 08826, Korea}
\affiliation[b]{Department of Particle Physics and Astrophysics,\\
The Weizmann Institute of Science, Rehovot 76100, Israel}
\affiliation[c]{Walter Burke Institute for Theoretical Physics,\\
California Institute of Technology, Pasadena, CA 91125, U.S.A.}
\affiliation[d]{Department of Theoretical Physics, \\ Tata Institute
of Fundamental Research, Homi Bhabha Rd, Mumbai 400005, India \vspace{-0.3cm}} 

\abstract{
Kerr-AdS$_{d+1}$ black holes for $d\geq 3$ suffer from classical superradiant instabilities over a range of masses above extremality. We conjecture that these instabilities settle down into Grey Galaxies (GGs) - a new class of coarse-grained solutions to Einstein's equations which we construct in $d=3$. Grey Galaxies are made up of a black hole with critical angular 
velocity $\omega=1$  in the `centre' of $AdS$, surrounded by a large flat disk of thermal bulk gas that revolves around the centre of $AdS$ at the speed of light. The gas carries a finite fraction of the total energy, as its parametrically low energy density and large radius are inversely related. GGs exist at masses that extend all the way down to the unitarity bound. Their thermodynamics is that of a weakly interacting mix of Kerr-AdS black holes and the bulk gas. Their boundary stress tensor is the sum of a smooth `black hole' contribution and a peaked gas contribution that is delta function localized around the equator of the boundary sphere
in the large $N$ limit. We also construct another class of solutions with the same charges;  `Revolving Black Holes (RBHs)'. RBHs are macroscopically charged  $SO(d,2)$ descendants of 
AdS-Kerr solutions, and consist of $\omega=1$ black holes revolving around the centre of $AdS$ at a fixed radial location but in a quantum wave function in the angular directions. RBH solutions are marginally entropically subdominant to GG solutions and do not constitute the endpoint of the superradiant instability. Nonetheless, we argue that supersymmetric versions of these solutions have interesting implications for the spectrum of supersymmetric states in, e.g.  ${\cal N}=4$ 
Yang-Mills theory.}

\begin{document}
\maketitle
\flushbottom

\section{Introduction} \label{intro}

It was observed almost 20 years ago \cite{Cardoso:2004hs} that Kerr-AdS$_{d+1}$ black holes in $d\geq 3$ suffer from classical super-radiant instabilities \cite{Zel'Dovich71} over a range of energies above the extremality bound. In this paper, we present a proposal for the endpoint of this instability. 

The rest of this introductory section is structured as follows. In subsection \ref{quest} we first elaborate on the question addressed in this paper and emphasize its importance for understanding the spectrum of operators in CFTs with a bulk gravity dual description. We then proceed, in subsections \ref{me-intro}-\ref{esri}, to outline our proposal. In the interests of concreteness, we focus, through this paper, mainly on the special case $d=3$, though we expect the generalization of our analysis to arbitrary $d >3$ to be relatively straightforward.

\subsection{The question} \label{quest}

The basis of local operators in any $CFT_3$ may be chosen to have definite values of the scaling dimension $\Delta$  and $z$ component of angular momentum, $J_z$. 
$n(\Delta, J_z)$, the number of operators with dimension $\Delta$ and angular momentum $J_z$, is one of the most fundamental CFT observables. When $\Delta$ and $J_z$ are large, it is also natural to define an `entropy of operators', $S(\Delta, J_z)$ via 
\begin{equation}\label{entdef} 
	n(\Delta, J_z)= e^{S(\Delta, J_z)}\ .
\end{equation} 	

The operator spectrum of the CFT may equivalently be characterized by the partition function $Z$ defined by  
\begin{equation}\label{pfops}
	\begin{split} 	
		&Z=\sum_{\Delta, J_z} n(\Delta, J_z) e^{-\beta(\Delta -\omega J_z)}\ . \\
	\end{split} 
\end{equation}
In an appropriate thermodynamical limit (of either the large $N$ or high-temperature variety), the partition function and the entropy function are Legendre transforms of each other.   

The state operator map allows us to interpret $Z$ and $S$ in thermodynamical terms. $Z$ is the thermal partition function over the Hilbert Space of the $CFT_3$ on $S^2$
\begin{equation}\label{pfopss}
	\begin{split} 	
		&Z= {\rm Tr} e^{-\beta(E -\omega J_z)}\ ,  \\
	\end{split} 
\end{equation}
while  $S(E, J_z)$ is the thermodynamical entropy of this system.\footnote{Through 
	this paper we use the term `angular velocity' for $\omega$,  the chemical potential dual to  
	angular momentum $J_z$ in \eqref{pfopss}.}

The partition function $Z$ and entropy $S$ are effectively computable in large $N$ $CFT$s that admit a  two-derivative AdS bulk dual. A standard entry in the AdS/CFT dictionary asserts that $S(E, J)$  equals the entropy of the dominant bulk black hole that carries the same charges.\footnote{This rule applies at leading order in $N$ and at energies and angular momenta of order $N^\alpha\sim G^{-1}$ with a suitable 
	constant $\alpha>0$, 
	where $G$ is the Newton constant of the AdS$_{d+1}$. 
	(For instance, $\alpha=2$ for $d=4$ maximal super-Yang-Mills theory and $\alpha=\frac{3}{2}$ for $d=3$ ABJM theory \cite{Aharony:2008ug}.)
	If more than one black holes exist at any given charges then $S(E, J)$ is given by the largest of the black hole entropies. } One might thus guess that the entropy function, $S_{BH}(E, J)$, of the explicitly known Kerr-AdS black holes \cite{Carter:1973rla} computes the entropy of the dual field theory. While this guess is believed to be correct at large values of $E$, it cannot hold at all energies, as Kerr-AdS black holes are unstable at low energies, as we now review.

\begin{figure}[!t]
	\centering
	\includegraphics[width=0.7\textwidth]{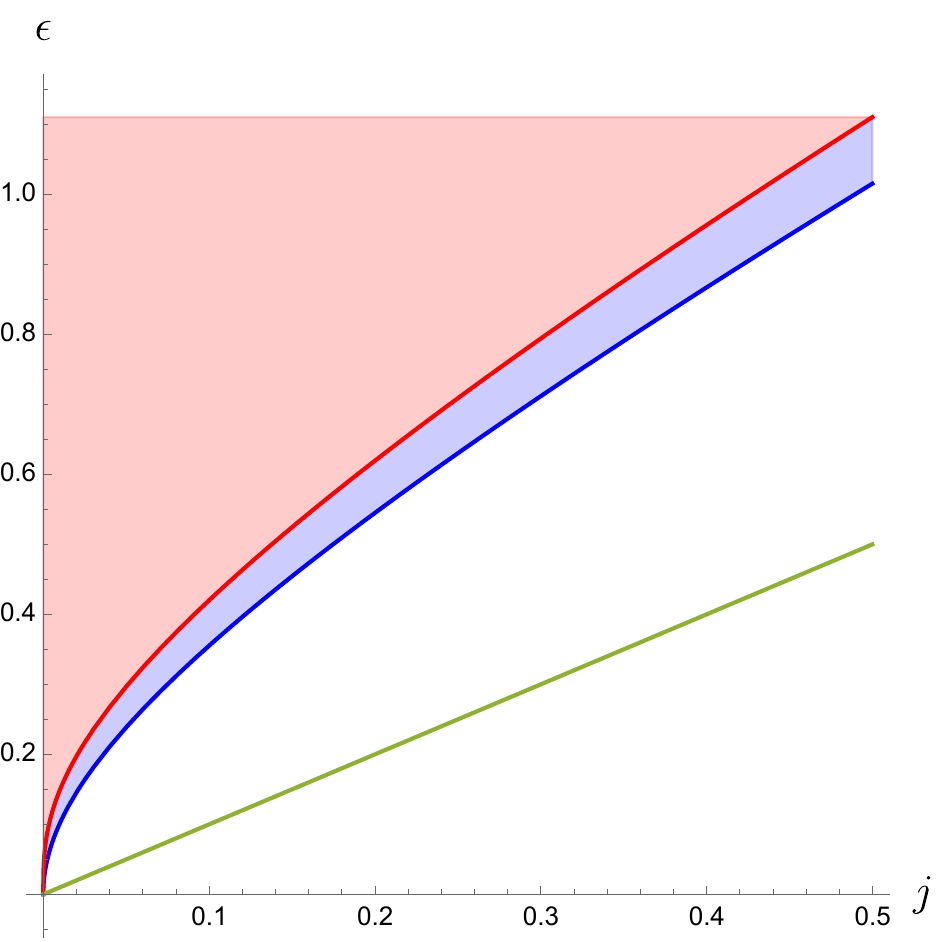}
	\caption{The `phase diagram' of Kerr-AdS black holes in $E$-$J$ space($\epsilon=G E$ and $j=G J$ are the rescaled energy and angular momentum). Black holes are stable in the red region that is bounded from below by the bold red line, $E=E_{\rm SR}(J)$.
		Black holes are unstable in the blue region bounded from above by the bold red line and from below by the bold blue line: the latter is the curve $E=E_{\rm ext}(J)$. Kerr-AdS black holes do not exist in the unitarity allowed region between the dark blue curve and the green line, $E=J$.}
	\label{kerr-ads4}
\end{figure}

At any given value of $J$, Kerr-AdS black holes exist only for energies $E \geq E_{\rm ext}(J)$ \footnote{Solutions of the negative 
	cosmological constant Einstein equations formally exist even at $E< E_{\rm ext}(J)$. However these solutions have naked singularities, and so are, presumably, unphysical.}
where $E_{\rm ext}(J)$ \footnote{Although Kerr-AdS black holes do not 
	exist at energies smaller than $E_{\rm ext}(J)$, previous attempts to 
	determine the endpoint of the superradiant instability have already 
	led to the construction of new - although still unstable - solutions 
	of general relativity called black resonators \cite{Dias:2015rxy, Ishii:2018oms, Chesler:2018txn, Ishii:2020muv, Chesler:2021ehz}. Black resonators are the angular momentum analogs of the hairy black hole
	solutions of \cite{Basu:2010uz, Bhattacharyya:2010yg, Dias:2011tj}, 
	and exist down to energies smaller than $E_{\rm ext}(J)$, though not all the way to the unitarity bound.}
is the energy of an extremal Kerr-AdS black hole with angular momentum $J$. $E_{\rm ext}(J)$ is a complicated known function of $J$ (see subsection \ref{tcext}). It turns out that  $E_{\rm ext}(J) \geq  J$ at all values of  $J$,
\footnote{The equality holds only at $J=0$.} as expected on general grounds.\footnote{All states in a CFT obey the inequality $E \geq J$.  This inequality is saturated only by a vacuum state. If we work with a normalization in which derivative operator $\partial_z$ carries $E=1$ and $J_z=1$, then primaries with $J_z=0$ and $J_z=\frac{1}{2}$ (and hence also their descendants) obey $E\geq J_z+\frac{1}{2}$. Primaries at all higher $J_z$ (and hence also their descendants)  obey the still tighter inequality $E \geq J_z+1$. At the parametrically large values of $E$ and $J$ of interest to this paper, however, all these fine distinctions are washed away, and the unitarity bound is effectively 
	simply $E \geq J$.}

It is easily verified that the angular velocity, $\omega$,  of Kerr-AdS black holes is a monotonically decreasing function of energy at fixed $J$. It is also easily verified that the angular velocity of {\it extremal} Kerr Black holes exceeds unity at every value of $J$. It follows from these facts that there exists a unique energy $E_{\rm SR}(J)$, at every value of $J$, at which the black hole angular velocity, $\omega$, equals unity. Clearly 
\begin{equation}\label{ineqaul}
	E_{\rm SR}(J) \geq E_{\rm ext}(J) > J\ .
\end{equation}
$E_{\rm SR}(J)$ is a known but complicated function of $J$. Kerr-AdS black holes with energy larger than $E_{\rm SR}(J)$ have $\omega<1$, while black holes with energies in the range 
\begin{equation}\label{bhrange}
	E_{\rm ext}(J)\leq E <	E_{\rm SR}(J)
\end{equation} 
have $\omega>1$. 

$E_{\rm SR}(J)$ marks an important dividing line in the phase space of Kerr-AdS black holes, because black holes in $AdS_D$ space with $\omega>1$ are always unstable \cite{Cardoso:2004hs, Green:2015kur} (see also  \cite{Kunduri:2006qa,Murata:2008xr,Cardoso:2013pza, Niehoff:2015oga, Dias:2015rxy, Chesler:2018txn,Ishii:2018oms, Ishii:2020muv, Chesler:2021ehz, Kodama:2009rq}). This instability is superradiant in nature. See Appendix \ref{superradiance} for an intuitive explanation of this fact, and a comparison with the superradiant phenomenon in charged black holes.

The existence of Kerr-AdS black holes in the energy range \eqref{bhrange} tells us that the dual CFT possesses 
a large entropy (order $\frac{1}{G}$) of states at these energies. Since these black holes are unstable they cannot represent the dual of CFT thermal equilibrium. There must, thus,  exist a new bulk black hole solution (one with a larger horizon area - hence larger entropy - as compared to the Kerr-AdS solution with the same charges) that describes the true thermal equilibrium of the CFT at these charges. The nature of this new solution is the topic of this paper.

\subsection{New solutions involving the quantum gas} \label{me-intro} 

In the steady state, a black hole in $AdS$ is in equilibrium 
with a thermal gas (made up of its own Hawking radiation). As the black hole and the gas are in equilibrium with each other
they have the same value of the temperature and $\omega$. \footnote{Through this subsection we work in the micro-canonical ensemble. The temperature and angular velocities should be thought of as derived quantities defined by $$\frac{1}{T_{BH}}= \partial_{E_{\rm BH}} S_{BH},~~  \frac{1}{T_{gas}}= \partial_{E_{\rm gas}} S_{gas},~~ \frac{\omega_{BH}}{T_{BH}}= -\partial_{J_{BH}} S_{BH},~~ \frac{\omega_{gas}}{T_{gas}}= -\partial_{J_{gas}} S_{gas}.$$ 
	From this viewpoint, the conditions 
	$$T_{BH}=T_{gas}, ~~~~\omega_{BH}=\omega_{gas} $$
	are derived from the maximization of total system entropy.  The same principle also implies that occupation numbers of the gas follow the Boltzmann distribution, (see e.g. Appendix I of \cite{Minwalla:2022sef}).} Generically,  the order unity energy and angular momentum of the gas are negligible compared to their order $\frac{1}{G}$ black hole counterparts. As $\omega$ approaches unity, however, the gas energy can easily be shown to diverge (see subsection \ref{tce}).  This divergence has its origin in gas modes that of large angular momentum. These modes live at very large radial locations (see Appendix \ref{fnll}),  and so effectively in global $AdS$ space.  Now the Hilbert Space of a gas made out of a bulk field in $AdS_4$ is the Fock Space over a single particle Hilbert space, whose states are in one-to-one correspondence with dual operators. The $\omega\rightarrow 1$ divergence arises from the contribution of infinite sequences of operators of
increasing angular momentum, such as 
$$ \partial_z^l O, ~~~~~~~~~~~~~l=1 \ldots \infty$$
where $O$ is the operator dual to the scalar field. The contribution of this 
sequence to the gas partition function diverges because the ratio of Boltzmann suppression factor for successive terms, $e^{-\beta(1-\omega) l}$, 
becomes unity when $\omega=1$. At $\omega = 1$ the partition function diverges because all terms in the sequence contribute equally to it. It follows that all thermodynamical formulae blow up as $\omega \to 1$ from below, and we find 
(see \eqref{sej})
\begin{equation}\label{sejint} 
	E=J\propto \frac{1}{(1-\omega)^2}, ~~~
	S\propto \frac{1}{1-\omega}\ . 
\end{equation}
\footnote{The energy and angular momentum of the gas are equal because the dominant contribution is from very high angular momenta, i.e. the gas is effectively chiral. The entropy of this gas is
	$\propto \sqrt{E}$ because the divergent contribution to the gas is effectively $1+1$ dimensional. See \eqref{sej} for details.}
where $E$ $J$ and $S$ are the angular momentum and entropy of the gas. 

It follows from \eqref{sejint} that when $1-\omega \sim {\cal O}(\sqrt{G})$, the energy carried in the gas is of order $1/G$. For such black holes - i.e. those that lie an order $G$ distance to the left of the solid red curve in Fig. \ref{kerr-ads4} - 
the classical formulae of black hole thermodynamics must be corrected, even at leading order, to account for the contribution of the gas.

\begin{figure}[!t]
	\begin{subfigure} [b]{\textwidth}
		\centering
		\includegraphics[width=0.7\textwidth]{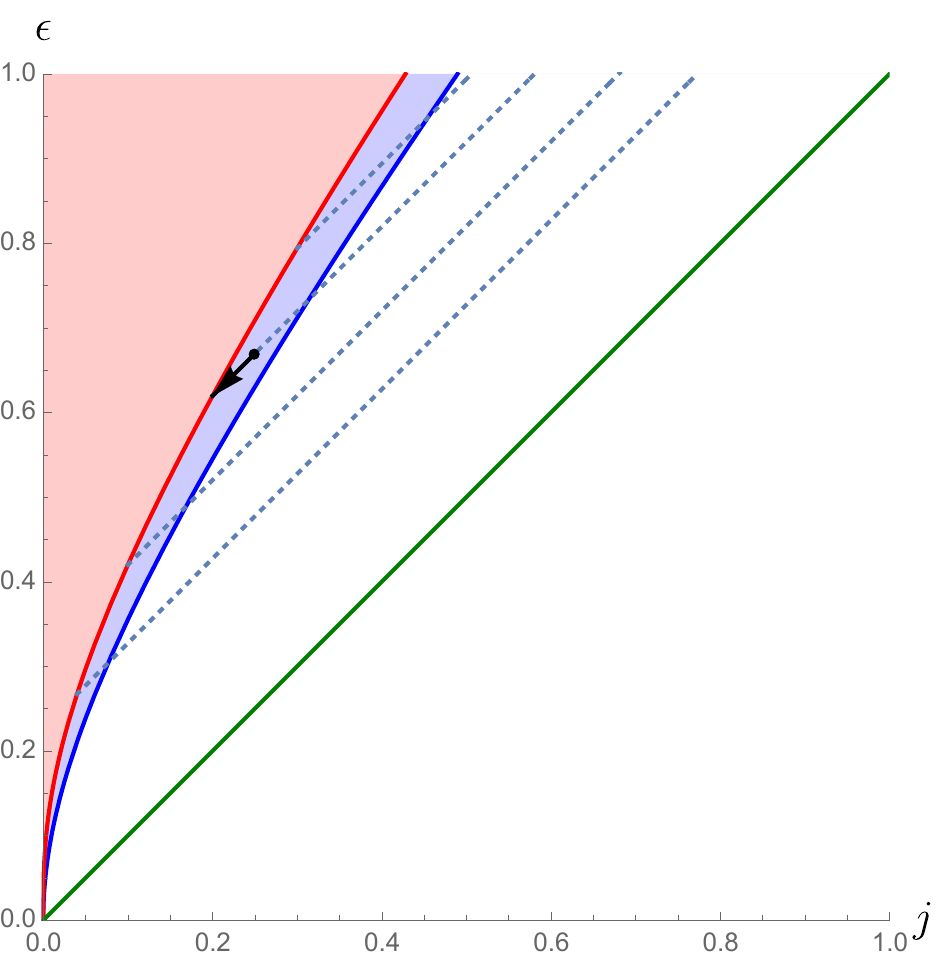}
		
	\end{subfigure}
	\caption{The Phase diagram in the microcanonical ensemble. The $x$ and $y$ axes are the scaled angular momentum and energy defined in \eqref{scaledcharges}. The green,  blue, and red curves respectively, denote the unitarity bound, extremality, and the edge of the superradiant instability. Kerr-AdS black holes are the dominant saddle above the red curve.  Grey Galaxy saddles dominate between the red and green curves. Their entropy equals that of the $\omega=1$ Kerr-AdS black hole located at the intersection of the dotted line that passes through the point of interest and the dark red curve. }
	\label{kerr-ads4-boost}
\end{figure}
The corrected thermodynamical formulae are
\footnote{As $1-\omega \propto \sqrt{G}$ is parametrically small, and as thermodynamical relations of classical black holes are analytic in the neighborhood of the red curve in Fig. \ref{kerr-ads4},  this small deviation of $\omega$ from unity may be ignored, to leading order, when evaluating the classical thermodynamics of these black holes.}
\begin{equation}\label{noin}
	\begin{split} 	
		&E= E_{\rm BH}+ C'\\
		&J= J_{\rm BH}+ C'\\
		&S= S_{\rm BH}\\
	\end{split} 
\end{equation} 
where $E_{\rm BH}$, $J_{\rm BH}$ and $S_{\rm BH}$
represent the classical energy, angular momentum, and entropy of $\omega=1$ Kerr-AdS black holes and 
the positive number $C^\prime$ is the energy and angular momentum of the Hawking gas. 

It follows from \eqref{noin} that the values of $(E, J)$ of our new gas-corrected black holes (we will later call these Grey Galaxies) are obtained as follows. We start with any point 
on the red curve of Fig. \ref{kerr-ads4}, and then move to the right along a 45-degree line for an arbitrary distance. The two parameters for this new family of solutions can be taken to be 
\begin{itemize} 
	\item Where on the red curve we start. This parameter completely determines the entropy (and also the temperature) of the new saddle.
	\item How far along the 45-degree line we go. This parameter determines the shift in energy and angular momentum, $C'$, of the new saddles.
\end{itemize} 
This discussion is illustrated in Fig. \ref{kerr-ads4-boost}.

It may be shown that the slope of the red curve in Fig. \ref{kerr-ads4} is everywhere greater than unity (see e.g. subsection \ref{omegone}). It follows that our new saddles always lie below the red curve in Fig. \ref{kerr-ads4-boost}. Note also that our new saddles exist at every point in Fig. \ref{kerr-ads4-boost} that lies below the red curve but also lies above the unitarity bound, $E=J$. 

As the entropy of the gas is parametrically sub-leading compared to that of the black hole (see \eqref{sejint}) the entropy of any of our new saddles may be obtained as follows. Given any point under the red line of the $E, J$ plane, we follow one of the 45-degree lines in Fig.\ref{kerr-ads4-boost} backward, until we hit a point on the red curve. To leading order in $G$, the entropy of our new saddle simply equals the classical entropy of this $\omega=1$ Kerr-AdS black hole. 

Our new solutions may be thought of as a very weakly interacting mix of the black hole (at the centre) and the gas (which lives far towards the boundary of $AdS$).  The thermodynamical charges of this mix are simply a sum of the different components 
because they interact so weakly. 
Energy (and angular momentum) can be exchanged between the two components: the final equilibrium 
configuration is the one that maximizes entropy at fixed $E$ and $J$. In section \ref{me} we demonstrate that this 
maximum is attained precisely when the black hole part of the mix has $\omega=1$. The discussion, in this regard, is very similar to that presented in  \cite{Basu:2010uz} for small charged black holes. While the noninteracting mix picture of \cite{Basu:2010uz} is precise only for a very small black hole,  however,  the weakly interacting model is exact for rotating black holes,  even when they are large.

We have just seen that one of our new solutions exists at every point between the red and unitarity curves in   Fig. \ref{kerr-ads4}. As Kerr-AdS black holes are stable at every point above the red curve in Fig. \ref{kerr-ads4},  It follows that every point above the unitarity bound in the $E, J$ plane hosts exactly one of either a stable classical black hole (above the red curve ) or one of the new saddle points (below the red curve but above the unitarity bound), leading to the micro-canonical phase diagram depicted in Fig. \ref{phas}. As we know how to compute the entropy of the new solutions, we now have a formula for $S(E, J)$ for every value of $E$ and $J$ that satisfies the unitarity inequality $E \geq J$. 
We conjecture that this formula correctly captures the entropy (at ${\cal O}(\frac{1}{G})$) of states (and operators) of every large $N$ $CFT_3$ that admits a two-derivative gravity dual description, at energies and angular momenta of 
order $\frac{1}{G}$.

One obtains several additional insights into the nature of the new gas-corrected black holes described above by approaching their thermodynamics from a canonical viewpoint, explored in subsection \ref{ans}.

\subsection{Grey Galaxies}\label{bs}

Our discussion of the gas-corrected black holes has proceeded by accounting for the energy and angular momentum of the gas but ignoring its backreaction on the metric, and all other interactions. The skeptical reader may wonder how this is consistent, given that the gas energy is of order $\frac{1}{G}$. The answer to this question lies in the fact that  
our gas is spread over a parametrically large spatial region in $AdS$, and so carries a parametrically small energy density. The low energy density of this gas ensures that its backreaction is parametrically small in an appropriate sense.  We now explain how this works in some detail. 

Consider a mode of angular momentum $l$, with $l \gg 1$, propagating in the unit radius global $AdS_4$
\begin{equation}\label{ads4met} 
	ds^2= \frac{dr^2}{1+r^2} - (1+r^2) dt^2 + r^2 \left( d \theta^2 + \sin^2 \theta d\phi^2 \right)
\end{equation} 
As a consequence of the centrifugal force, the wave function for such a mode is concentrated at values of  $r$ of order $\sqrt{l}$ (see Appendix \ref{fnll}). If the mode also carries $J_z=l$, its wave function is also sharply localized about the equator, $\theta = \frac{\pi}{2}$, with an angular spread $\delta \theta$ of order $\frac{1}{\sqrt{l}}$. The proper thickness of this mode in the angular direction is of order $r\delta \theta\sim\sqrt{l}\times \frac{1}{\sqrt{l}}$ and so is of order unity. It turns out that the range of  $l$ that contributes dominantly to our bulk gas ensemble is $l=1 \ldots  \sim \frac{1}{\beta(1-\omega)}  \sim 1\ldots \frac{1}{\sqrt{G}}$ (see under \eqref{sejint}).  Our bulk gas thus lives in a region that may be visualized as a large but uniformly flat pancake that lives at $\theta=\frac{\pi}{2}$. The parametrically large radius of this pancake is of order  $\sim \frac{1}{G^{\frac{1}{4}}}$.

The large radius of our bulk gas pancake ensures that its energy density of the gas is parametrically low. It is this fact that allows for the backreaction of the gas to be computed in perturbation theory.

In sections \ref{tbs} and \ref{bst} we have performed a detailed study of the backreaction of the $AdS$ sourced by the gas modes of a bulk scalar field dual to a scalar operator of dimension $\Delta$. 
Our final bulk solution is given by patching together two metrics, which are individually valid in distinct but overlapping spatial regions. 
\begin{itemize} 
	\item Over radial distances of order unity (i.e. distances that are not parametrically scaled in 
	power of $\frac{1}{G}$), the leading order metric for our new saddle point is simply that of the  
	$\omega=1$ black hole that lives at its core. 
	\footnote{Deviations away from this solution, caused by the gas, are parametrically suppressed in $G$ because of the smallness of the energy density of the gas. }
	\item Over radial distances of order $ r \sim \frac{1}{(1-\omega)^{\frac{1}{2}}} \sim \frac{1}{G^{\frac{1}{4}}}$, i.e. the size of the gas pancake, we work with the following scaled coordinates
	\begin{equation}\label{scaledcoord} 	
		x=r (1-\omega)^{\frac{1}{2}}, ~~~\zeta = \frac{\theta-\frac{\pi}{2}}{ (1-\omega)^\frac{1}{2}}, ~~~t' =\frac{t}{(1-\omega)^\frac{1}{2}}, ~~~
		\phi' =\frac{\phi}{(1-\omega)^\frac{1}{2}}
	\end{equation}
	The scaling of $r$ and $\theta$ are chosen to ensure that the bulk gas varies over a scale of order unity in the new coordinates. The scaling of $t$ and $\phi$ is chosen to ensure that the background $AdS$ metric, in the new coordinates, is independent of   $G$. It turns out that the black hole tail - the difference between the black hole and the AdS metric in the scaled coordinates - is of order $G^\frac{3}{4}$ - and so is very small - in these coordinates.
	Moreover, the magnitude of Newton's constant times the bulk stress tensor of the gas is of order $G^\frac{1}{2}$, and so is also parametrically small in this new scaling. To leading order, as a consequence, the deviation of our spacetime metric away from that of the background $AdS$ metric is simply the sum of the black hole tail and the linearized gravitational response to the stress tensor of the gas (contributions that are individually of order $G^\frac{3}{4}$ and $G^\frac{1}{2}$). The gas contribution is computed in full detail in section \ref{bst}. 
	\item  The final boundary stress tensor of our solution is the sum of two terms:  the boundary stress tensor of the original $\omega=1$ black hole (reviewed in subsection \ref{bstbh}) and the boundary stress tensor resulting from the metric response to the bulk gas (see in subsection \ref{bdrstr}). The black hole contribution to the boundary stress tensor is smooth on the sphere, and is, of course, of the classical order $\frac{1}{G}$. On the other hand, the gas contribution to the boundary stress tensor turns out to be of order $\frac{1}{G^{\frac{5}{4}}}$ (and so larger than classical). However, this contribution is peaked around the equator over a boundary angular distance of order $G^{\frac{1}{4}}$, and so its integral over the sphere is of the expected classical order $\frac{1}{G}$. In other words, the gas contribution to the boundary stress tensor localized to an angular width of order $G^{\frac{1}{4}}$ at the equator of the boundary sphere. In the classical small $G$ limit this distribution becomes a $\delta$-function (of classical strength). This new contribution to the boundary stress tensor is rightmoving, and so is similar to the contribution of a chiral $1+1$ dimensional gas localized at the equator.  
\end{itemize} 

Our new bulk solution resembles a galaxy, with a big central black hole surrounded by a large flat disk of gas rotating at the speed of light. For this reason, we call these solutions `Grey Galaxies.' We use the word grey rather than black because part of the bulk energy in our solution is `black' (shielded behind an event horizon) while the rest of it is `white' (visible in the form of a gas).

We emphasize that the bulk gas in our `Grey Galaxy' solutions is an ensemble over field configurations rather than a particular field configuration. In particular, we compute the bulk stress tensor of this gas using thermodynamics. How reliable is this approximation? One way of addressing this question is to estimate the fluctuations in the ensemble average. 
Our bulk gas is made up of a collection of bulk fluctuation modes, each of which is thermally occupied. The occupation number of any given mode is of order unity (see e.g. \eqref{sumover}), and so fluctuations in individual mode occupation numbers are also of order unity. However typical individual modes only carry energies only of $\frac{1}{G^\frac{1}{2}}$. We obtain classical energies only by summing over $\frac{1}{G^\frac{1}{2}}$ such modes. This is precisely the number of modes that contributes to the stress tensor of an interval of order unity in the scaled coordinate $x$. This fact explains why we have a new classical solution only in scaled coordinates. The law of large numbers thus tells us that the fractional fluctuations value of the (scaled coordinate) bulk stress tensor is of order ${G^{\frac{1}{4}}}$. As fluctuations are parametrically suppressed in comparison to the mean, fluctuations in the bulk stress tensor (and hence our final bulk metric) are negligible when we work in scaled coordinates. It follows that the classical (scaled coordinate) metric presented in this paper - and in particular the boundary stress tensor it is dual to - is parametrically reliable.

As we have mentioned above, in section \ref{tbs} we have only derived detailed mathematical formulae for the bulk stress tensor for the gas made up of a scalar field. Any given bulk theory of interest will, in general, have several other bulk fields (including, in particular, the graviton). The full bulk gas stress tensor will be the sum of the stress tensors associated with each bulk field. Although we have not computed the bulk stress tensor for higher spin fields in this paper, it seems clear that the result of this computation will be very similar to that of a scalar with a similar dimension (see \eqref{zfin} for a hint of the final answer). This stress tensor, once available, can be plugged into 
the general formulae of subsection \ref{bst}, to 
obtain the final formula of the backreacted 
metric. Structural aspects of the analysis of 
sections \ref{tbs} and \ref{bst} make it clear that the gas contribution to the final boundary stress tensor will be a delta function localized 
around the equator of the boundary sphere in the large $N$ limit. The coefficient of this delta function is given by the total energy of the gas, 
which we have computed, for fields of every spin, in subsection \ref{tce}. While it is, thus, certainly of interest to fill the computational gap of this paper, by obtaining explicit expressions for the bulk stress tensor of the gas 
coming from higher spin fields, and in particular fermions \footnote{For charged hairy black holes in AdS, one finds 
	condensation of one bosonic mode with a macroscopic occupation number \cite{Basu:2010uz,Bhattacharyya:2010yg,Dias:2011tj}. 
	On the other hand, fermions can equally contribute to the hair in our solution 
	since each mode has $O(1)$ average occupation number. 
	We thank Kimyeong Lee for pointing this out.}, we believe that the analysis of this paper already uncovers all qualitative aspects of the solutions that these 
gases will source. 

We note that super-radiant instabilities of Kerr-AdS black holes from scalar fields are studied, for instance, in \cite{Dias:2011at,Ishii:2021xmn}.

Although not addressed in this paper, it may prove technically possible to study subleading corrections to the classical Grey Galaxy solutions constructed in this paper. In order to do this 
we would have to account for the backreaction of 
the black hole metric to the matter. For instance, the back-reacted stress-energy tensor may be computed from the Euclidean 2-point function in the black hole background,  instead of the thermal AdS as we did in section \ref{eucb}. Although the effect of this stress tensor on the black hole metric is parametrically, small, its effects will have to be accounted for in computing the corrections to our solution in a power series expansion in $G$. This should prove technically possible, precisely because the corrections are small. We leave further discussion of this fascinating possibility to future work.

\subsection{Revolving  Black Holes}

The Grey Galaxy solution
consists of an $\omega=1$ black hole in the centre of $AdS$, in equilibrium with a chiral gas
whose `equation of state' is $E \approx J$.

In Appendix \ref{rbs} we note that every black hole has a fluctuation mode with $E=J$ (the equality is precise). This is the mode that sets the black hole revolving around the centre of $AdS$. From the point of view of the dual CFT, populating this mode corresponds to taking $\partial_z$ descendants of the primaries that make up the classical black hole at the centre of $AdS$. 

In Fig. \ref{kerr-ads4-boost} we have explained that an $\omega>1$ black hole increases its entropy by transferring some of its $E=J$ energy into the chiral gas that makes up Grey Galaxies. Similarly, an $\omega>1$ black hole can also raise its entropy by transferring a significant fraction of its energy and angular momentum into the descendent mode described in the previous paragraph. The physical interpretation of such a state is explored in Appendix \ref{rbs}, where we explain that the state obtained by the macroscopic occupation of descendents is a quantum wave function for a spinning black hole that is also revolving around the centre of $AdS$ (the wave function describes the orbital motion of the black hole).  
We call these new configurations Revolving Black Holes (RBH)s (see Appendix \ref{rbs}). 

As we explain at the end of subsection \ref{me}, 
RBH solutions are marginally entropically subdominant compared to Grey Galaxy Solutions. 
For this (and other) reasons we do not expect these solutions to represent the endpoint of the super-radiant instability. Nonetheless, these solutions are of interest for several reasons. First, they are both elegant and precise. As we explain in Appendix \ref{fpa}, these solutions are constructed entirely out of the action of the symmetry group $SO(3,2)$. Second, had we not known about the existence of Grey Galaxy solutions, we could anyway have used the (easily constructed) 
RBH solutions to put a lower bound on the entropy function $S(E, J)$. \footnote{As the leading order construction of $S(E, J)$ (see Fig. \ref{kerr-ads4-boost}) is identical for RBHs and GGs, this would actually have given us the correct (i.e. GG) answer at leading order.}

The last point may prove practically useful in situations where the entropy function is not yet clearly known. In section \ref{susy} we argue that supersymmetric versions of RBH solutions allow us to place lower bounds on the five charge entropy of supersymmetric states in ${\cal N}=4$ Yang-Mills theory. 

To end this subsection we reiterate that an RBH is a quantum wave function over classical geometries. In this respect, it differs qualitatively from classical Kerr-AdS black holes and also from
Grey Galaxies (which are also described by classical metrics, albeit in 
a coarse grained sense). In contrast, the  RBH is a quantum state that is time independent over classical time scales. We expect, of course, that RBHs eventually decay into Grey Galaxies; see  section \ref{discussion} for some discussion on this point.

\subsection{End point of the superradiant instability} \label{esri}

Consider a Kerr-AdS black hole with $\omega>1$. As we have reviewed
above, such black holes are unstable; when perturbed they evolve to new solutions. What is the endpoint of this instability? The results of this paper suggest the following scenario. 

Any particular perturbation will seed a time-dependent solution of General Relativity. At the level of differential equations, we expect that this solution will continue to evolve without ever reaching a terminal endpoint. In the purely classical theory, we expect this evolution to drive the solution to ever smaller angular scales (i.e. ever larger values of $l$) and for this process to continue without stopping (this scenario was first suggested in \cite{Dias:2011at}).\footnote{This is a manifestation of the ultraviolet catastrophe that classical field theories always suffer from, and that led to the discovery of quantum mechanics.} 

Once quantum effects are accounted for, on the other hand, 
we expect the cascade to ever smaller angular scales to stop at $l \sim \frac{1}{\beta(1-\omega)}$. The main quantum effect that is relevant here is the quantization of modes: the fact that field configurations 
at a given frequency $\nu$ admit excitations only in packets of energy  $h\nu$. It is this quantum effect that effectively cuts off the summation over $l$ in \eqref{sumover} at $l \sim \frac{1}{\beta(1-\omega)}$. 

Even though the (now quantized) configuration will
stop evolving to smaller angular scales,  we do not expect it to settle down to any particular microscopic state. We expect that the 
quantum state will continue to evolve in time within the phase space of the bulk `gas' that we have described in this paper. However coarse-grained observables - like the leading order term in the metric in scaled variables $x$ and $\zeta$ -  should settle down into the Grey Galaxy metric presented in this paper, with fluctuations that are parametrically suppressed (see above). The scenario we have 
sketched above is similar, in many respects, to the picture of 
\cite{Niehoff:2015oga}.

In the sense described in the paragraphs above, we conjecture that the Grey Galaxy solutions are the end-point of the superradiant instability of an $\omega>1$ Kerr-AdS black hole. 

In more detail, we expect the evolution of the superradiant instability to proceed as follows.  As the emission time of a quasinormal mode with angular momentum $l$ scales like $e^{b l}$ at large $l$,  
\footnote{This nonperturbatively long time scale has its origin in the fact that modes at large $l$ need to `tunnel' through the centrifugal barrier before making it out to infinity. Note that at any fixed  $r$ of order unity, the modes presented in \eqref{fnl} decay with $l$ like $e^{-l \left( \frac{1}{2} \ln \frac{1+r^2}{r^2} \right)  }$. A very rough estimate of the order of the constant $b$ is thus 
	$b \sim \ln \frac{1+r_0^2}{r_0^2}$, where $r_0$ 
	is the value of $r$ beyond which the black hole metric starts approximating global AdS.} 
the early emission will be almost entirely into modes with small $l$. 
For a while the solutions might look a little bit like the black resonator solutions of \cite{Dias:2015rxy} at the given small values of $l$.
\footnote{Recall that a black resonator is an $\omega>1$ black hole 
	in equilibrium with a Bose condensate of a mode at a given particular value of $l$.}  As time passes, the largest accessible value of $l$, $l_{m}$, increases. As long as $l_m$ is not too large -  specifically, when $l_{m} \ll \frac{1}{\sqrt{G}}$ - (see \eqref{rc} for a more precise formula) it seems plausible to us that the time-dependent configuration will continue to resemble a black resonator with the condensate in the mode with $l \approx l_m$. 
At still later times, when $l_m$ first becomes comparable to $\frac{1}{\sqrt{G}}$ (see \eqref{rc} for a more accurate condition) we expect the nature of our configuration to undergo 
a transition. While a significant fraction of the energy outside the black hole will continue to lie in the Bose condensate at $l\sim l_m$,  an increasingly large fraction of the energy will be spread out among a much larger number of large $l$ modes, i.e. into the bulk gas described extensively above. At even later times, when $l_m$ is large in units of 
$\frac{1}{\beta(1-\omega)}$, the cut-off is irrelevant, and we expect the configuration to begin to closely resemble the Grey Galaxy constructed in this paper. At this point, almost all of the energy 
of the solution is carried by the gas, and Bose condensates play no role.

Note that the time scale for the formation of 
a Grey Galaxy solution is extremely long, likely of order $e^{\frac{1}{\sqrt{G}}}$ (recall $\frac{1}{\sqrt{G}}$ is the order of the largest angular momentum that is significantly occupied in Grey Galaxy solutions). Note this is much larger than the time scale for thermal equilibration of our bulk gas \footnote{This thermalization time scales like an inverse power of $G$, perhaps like $\frac{1}{G^2}$.}. This explains why our treatment of the bulk gas as thermalized is consistent over the relevant time scales, even though the gas is parametrically weakly coupled.

The discussion presented here appears to us to be qualitatively consistent with the results of the numerical simulations  \cite{Chesler:2021ehz}, which we discuss in more detail in section \ref{chesler}, as well as the fact that the entropy of Grey Galaxy solutions is always larger than that of Black Resonators (see section \ref{chesler}).

\section{Kerr-AdS$_4$ black holes}\label{kerr} 

In this section, we review the Kerr-AdS$_4$ solution and its thermodynamics. 

Consider Einstein's equation with a negative cosmological 
constant,
\begin{equation}\label{eeqcc}  
	R_{\mu\nu}=-3g_{\mu\nu}
\end{equation} 	
We have chosen the value of the cosmological constant so that the `unit radius' $AdS_4$ space 
\eqref{ads4met}, is a solution to these equations. 

\subsection{Kerr-AdS black hole solutions} 

Another set of exact solutions to these equations are 
the Kerr-AdS black holes given by 
\cite{Carter:1973rla}
\begin{equation}\label{bsol} 
	ds^2=-\frac{\Delta}{\rho^2}\left(dt-\frac{a}{1-a^2}\sin^2\theta d\phi\right)^2
	+\frac{\Delta_\theta\sin^2\theta}{\rho^2}
	\left(\frac{r^2+a^2}{1-a^2}d\phi-adt\right)^2
	+\frac{\rho^2}{\Delta}dr^2+\frac{\rho^2}{\Delta_\theta}d\theta^2
\end{equation}
The functions $\Delta$, $\Delta_\theta$ and $\rho$ which appear in \eqref{bsol} are given by 
\begin{eqnarray}\label{funcdef} 
	\Delta&=&(r^2+a^2)\left(1+ {r^2}\right)-2mr\ ,\ \
	\Delta_\theta=1-{a^2}\cos^2\theta\nonumber\\
	\rho^2&=&r^2+a^2\cos^2\theta\ .
\end{eqnarray}
The numbers $a$ and $m$, that occur in \eqref{bsol} and \eqref{funcdef} are constants. These two constant parameters determine the mass and angular momentum (as well as the temperature and angular velocity) of the black hole, according to the formulae we will report below. 

\subsection{Parametric ranges of variation} 

The parameter $a$ lies in the range
\footnote{This can be seen from the fact that $\Delta_\theta$ changes sign as a function of 
	$\theta$ when $a$ lies outside this range, causing - for instance - the coordinate $\theta$ to
	switch signature, resulting in a singular metric (presumably the curvature also blows up at the
	value of $\theta$ where $\Delta_\theta$ switches sign).} 
\begin{equation}\label{rangeofa}
	a \in [-1, 1]\ .
\end{equation}

We will now determine the range of variation of the 
parameter $m$. 

The outer horizon of the black hole \eqref{bsol}  is located at $r=r_+$, where $r_+$ is the  largest root of 
the equation $\Delta(r_+)=0$, i.e. the largest root of the equation 
\begin{equation}\label{outerhor} 
	2m=\frac{(r_+^2+a^2)\left(1+ r_+^2 \right)}{r_+}\ .
\end{equation}
Black hole solutions only exist for $m\geq m_{\rm ext}$, where 
$m_{\rm ext}$ is the value of $m$ at which \eqref{outerhor} has a double root, i.e. the 
value of $m$ at which $r_+^2$ solves the equation
\begin{equation}\label{xext} 
	1+a^2+{3(r_+)^2_{\rm ext}}-\frac{a^2}{(r_+)^2_{\rm ext}}=0
\end{equation} 
(\eqref{xext} is obtained by differentiating \eqref{outerhor}, and setting the result to zero, as is appropriate for a double root). The black hole with $m=m_{\rm ext}$ is extremal. Upon solving \eqref{xext} we find 
\begin{equation} \label{xstar}
	(r_+)^2_{\rm ext}=-\frac{1}{6}\left(1+ a^2\right)
	+\sqrt{\frac{1}{36}\left(1+ a^2\right)^2
		+\frac{a^2}{3}}
\end{equation}
Plugging \eqref{xstar} into \eqref{outerhor} and squaring we find 
and
\begin{equation}\label{extremal}
	4m_{ext}^2=\frac{2}{27}
	\left[{\textstyle
		-1+33 a^2+33 a^4
		-a^6+\left(1+14 a^2
		+a^4\right)^{\frac{3}{2}}
	}\right]\ .
\end{equation}
It follows that, for all allowed black holes, the parameter $m$ obeys the inequality 
\begin{equation}\label{bigextremal}
	4m^2\geq \frac{2}{27}
	\left[{\textstyle
		-1+33 a^2+33 a^4
		-a^6+\left(1+14 a^2
		+a^4\right)^{\frac{3}{2}}
	}\right]\ .
\end{equation}

\subsection{The function $r_+(m, a)$} 

As we have explained above, the outer radius of the black hole \eqref{bsol} is given by $r_+$, the largest root of 
\eqref{outerhor}. As \eqref{outerhor} determines $r_+$ only implicitly, in this subsection we pause to review some important qualitative properties of $r_+(m, a)$. 

For values of $m$ that obey \eqref{bigextremal}, $r_+$  is an increasing function of $m$ at fixed $a$. This plot looks as follows. The minimum allowed value of $m$ (at fixed $a$) is given 
by \eqref{extremal}. At this minimum value, $r_+$ takes the value \eqref{xstar}. As we further increase $m$, 
$r_+$ increases, asymptoting to $r_+= (2 m)^{\frac{1}{3}}$ at large $m$. 

We can also plot $r_+$ as a function 
of $a$ at fixed $m$; we find that $r_+$ is a decreasing function of $a$, reaching the smallest allowed value 
at $a=1$ (at this value $r_+$ is such that it obeys the equation $(1+r_+^2)^2=2 mr_+$). Note, in particular, that $r_+$ does not increase without bound in the large mass limit $a \to 1$. 

\subsection{Thermodynamical charges as  functions of $m$ and $a$} 

The energy $E$, angular momentum $J$ and the entropy $S$ are given by \cite{Gibbons:2004ai}
\begin{equation} \label{chargeshh0}
	E=\frac{1}{G}\frac{m}{(1-a^2)^2}\ ,\ \ J=\frac{1}{G}\frac{ma}{(1-a^2)^2}\ ,\ \
	S= 
	\frac{\pi(r_+^2+a^2)}{G (1-a^2)}\ .
\end{equation} 
The inverse temperature $\beta$ and angular velocity 
$\omega$ are given by 
\begin{equation}\label{potentials}
	\beta=\frac{4\pi(r_+^2+ a^2)}{r_+\left(1+a^2+ 3r_+^2-a^2r_+^{-2}\right)}, ~~~\omega=\frac{a\left(1+{r_+^2} \right)}{r_+^2+a^2}\ ,
\end{equation}
where $\omega$ is the angular velocity of the event horizon with respect 
to the non-rotating observer at infinity. (See section \ref{nonrotating} for the definition 
of this observer.)

Note that the extremality condition \eqref{xext} is simply the condition that $T=0$ (i.e. that $\beta$ diverges). 
Notice $E$ and $J$ both diverge in the limit $a\rightarrow 1$. \footnote{Note that the inequality 
	\eqref{bigextremal} makes it impossible to take $m \to 0$ while simultaneously scaling 
	$a\rightarrow 1$.}, 
and so $a \to 1$ (at fixed $m$) is the large mass and large angular momentum limit. 

It is convenient to define the scaled energy, angular 
momentum and entropy,  $\epsilon$, $j$ and $s$ by the expressions 
\begin{equation}\label{scaledcharges}
	\epsilon= G E,~~~ j= G J,~~~s= 
	\frac{G S}{\pi}\ .
\end{equation} 
The expressions for the scaled charges are given by 
\begin{equation} \label{chargeshh}
	\epsilon=\frac{m}{(1-a^2)^2}\ ,\ \ j=\frac{ma}{(1-a^2)^2}\ ,\ \
	s=   \frac{r_+^2+ a^2}{1-a^2}
\end{equation}

\subsection{Thermodynamical charges at extremality} 
\label{tcext}

The thermodynamical charges for extremal black holes can be obtained by plugging \eqref{extremal} into 
\eqref{chargeshh}. The resultant expressions are messy in general, but simplify when $a$ is small and when $a \to 1$. 

\subsubsection{Small $a$ } 

At leading order the small $a$ expansion \eqref{xstar} simplifies to  
\begin{equation}\label{xstarsmalla}
	(r_+)_{\rm ext}= a.
\end{equation} 
\eqref{extremal} reduces to 
\begin{equation}\label{extremalsmalla}
	m^2 = a^2 \ .
\end{equation} 
\footnote{Comparing with \eqref{rsrcond} and \eqref{msrcond} we see that these black holes are superradiant unstable.}  

The thermodynamical charges of this small extremal black hole are given by   
\begin{equation} \label{chargessmalla}
	\epsilon= a \ ,\ \ j= a^2\ ,\ \ s= 2a^2 , ~~~
	\omega=\frac{1}{2a}, ~~~\beta=\frac{2 \pi}{a}
\end{equation}
Note, in particular, that 
\begin{equation}\label{jeext} 
	j=\epsilon^2 
\end{equation}

\subsubsection{$a \rightarrow 1$} 

As we have mentioned above, $E$ and $J$ become large when $a$ is taken to unity (at a value of $m$ that is large enough to obey the extremal bound, but is otherwise arbitrary). Let $a=1-\alpha$ and then take the limit $\alpha \to 0$. To first order we find 
\begin{equation}\label{xstaraone}
	(r_+)_{\rm ext}^2= \frac{1-\alpha}{3} .
\end{equation}
\eqref{extremal} reduces to 
\begin{equation}\label{extremalaone}
	m^2 = \frac{64}{27}- \frac{64 \alpha}{9}
\end{equation} 

\eqref{chargeshh} becomes
\begin{equation} \label{chargeslargea}
	\epsilon= \frac{2}{3 \sqrt{3} \alpha^2}\ ,\ \ j= \frac{2}{3 \sqrt{3} \alpha^2} \ ,\ \
	\frac{j}{\epsilon} = 1-\alpha \, \ \ {\rm so}~~~
	\epsilon-j=\frac{2}{3 \sqrt{3} \alpha},  ~~
	\omega=1 + \frac{\alpha}{2} 
\end{equation}

Note that the deviation from extremality,  $\epsilon-j$, scales like $\sqrt{j}$ in this limit. So while the deviation grows in absolute terms, it decreases as a fraction of $j$.

\subsection{$a$ and $m$ as functions of $E$ and $J$} 

We can invert the expressions for $E$ and $J$ to express $m,a$ as functions of 
$E,J$. We find 
\begin{equation}\label{a-m-from-E-J}
	a=\frac{J}{E}\ ,\ \ m=G E\left(1-\frac{J^2}{E^2}\right)^2\ .
\end{equation}
\eqref{a-m-from-E-J} gives us a new perspective on the bound $|a|\leq 1$; this is simply a restatement of 
the unitarity bound $|E| \geq |J|$. 

\subsection{The superradiant instability curve  in the $m$, $a$ plane}

As we have reviewed in the introduction, the black hole suffers from superradiant instabilities
when $\omega>1$. Using \eqref{potentials}, this condition can be rewritten as  
\begin{equation} \label{rsrcond}
	r_+^2< a 
\end{equation}
When $\omega=1$, $r_+^2=a$ \footnote{Since $a<1$, it follows that on this curve $r_+<1$. In the limit that 
	where $a\to 1$, black holes become big, in other words, 
	$r_+ \to 1$.} 
and it follows from  \eqref{outerhor} that 
\begin{equation}\label{eqsr} 
	2m= \sqrt{a}\left(1+ a\right)^2.
\end{equation}  
Using the fact that $m$ is an increasing function of $r_+^2$ at fixed $a$, it follows that the instability 
condition can equivalently be written as 
\begin{equation} \label{msrcond} 
	2m<\sqrt{a}\left(1+ a\right)^2\ .
\end{equation}
In the limit $a \rightarrow 1$, $a=1-\alpha$ with 
$\alpha$ small, \eqref{msrcond} reduces to 
\begin{equation} \label{msrcondaone} 
	m<2-3\alpha\ .
\end{equation}
Note that the condition \eqref{msrcondaone} is met by extremal black holes.

\begin{figure}[h]
	\begin{subfigure} [b]{\textwidth}
		\centering
		\includegraphics[width=0.7\textwidth]{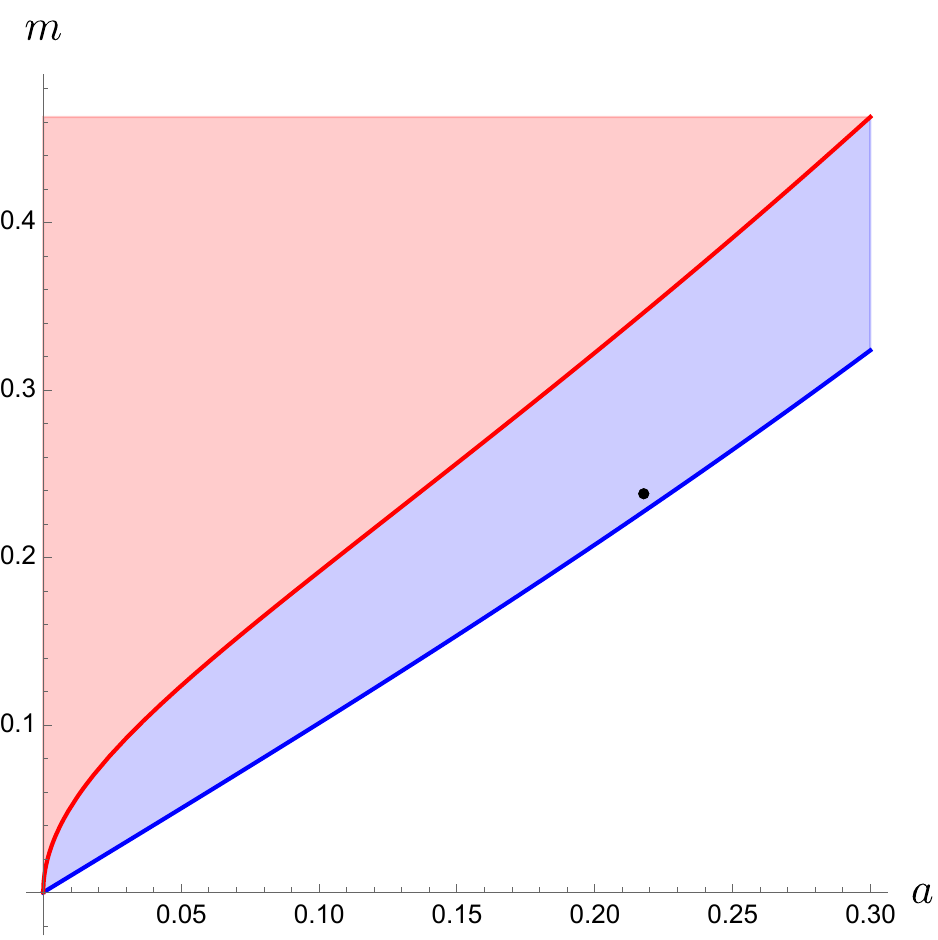}
	\end{subfigure}
	\caption{Plots of $m$ vs $a$ for the extremal black hole (in blue) and the black hole at $\omega=1$ (in red). Black holes in the shaded blue region are unstable. Black holes in the shaded pink region are stable. Black holes do not exist below the blue curve or at values of $a>1$.}
	\label{extvsunstab}
\end{figure}

Plotting the values of $m$, \eqref{msrcond} and \eqref{extremal} simultaneously as a function of $a$, 
one obtains the curve plotted in Fig. \ref{extvsunstab} 

Notice that extremal black holes lie under the Super-radiant instability curve, hence are always unstable as we saw above.

\subsection{Thermodynamical charges as a function of $a$ on the super-radiant instability curve}\label{omegone} 

As we have reviewed above, $\omega=1$ along the curve 
\eqref{eqsr}. Along this curve, the scaled thermodynamical charges of the black hole are given by 
\begin{equation} \label{chargessr}
	\epsilon=\frac{\sqrt{a}}{2(1-a)^2}\ ,\ \ j=\frac{a^{\frac{3}{2}}}{2(1-a)^2}\ ,\ \  s= \frac{a}{1-a} , ~~
\end{equation} 
The inverse temperature and angular velocity of these black holes are given by 
\begin{equation}\label{tempangvel}  		
	\beta=\frac{4\pi \sqrt{a}}{1+a}, ~~~\omega=1
\end{equation}
As $a$ varies from $0$ to $1$, $\beta$ varies from $0$ to $2 \pi$. It follows that the temperature of 
$\omega=1$ black holes is always greater than or equal to $\frac{1}{2 \pi}$. In other words, the $\omega=1$ 
black holes never come close enough to their extremal counterparts to go to zero temperature. 

The energy versus angular momentum curve is easily obtained at small and large values of $a$. At leading order in the small $a$ expansion, one obtains 
\begin{equation}\label{smallaje}
	\epsilon = 2^{\frac{-2}{3}} j^{\frac{1}{3}}\ .
\end{equation} 
In the limit $a\to 1$, on the other hand, we set 
$a=1-\alpha$ and obtain 
\begin{equation}\label{enang}
	\epsilon= \frac{1-\frac{\alpha}{2}}{2 \alpha^2}, ~~~
	j= \frac{1-\frac{3\alpha}{2}}{2 \alpha^2}, ~~~
	\epsilon-j= \sqrt{\frac{j}{2}} 
\end{equation}

The free energy on the $\omega=1$ line takes the following form:
\begin{equation}\label{fene}
	\begin{split}
		G&=E-TS-\omega J\\
		&= \frac{1}{G}\left(\epsilon-j-\pi T\right)\\
		&=\frac{\sqrt{a}}{4G}.
	\end{split}
\end{equation}
Note that the (Grand) Free energy of these black holes is positive for all values of $a$. It follows that these solutions are always subdominant compared to the thermal gas \footnote{As we will see later in this paper, in units of $1/G$, the thermal gas has zero grand free energy 
	as $\omega\to 1$, even though it's energy and angular momentum are order $1/G$, as the energy in this gas-phase equals the angular momentum to leading order, see \eqref{sej} and \eqref{zfin}.}.

It is easy to check that, on the  $\omega=1$, the energy of Kerr-AdS black holes is given in terms of their temperature by  
\begin{equation}
	G E=\frac{\sqrt{4 \pi ^2 T^2-1}-2 \pi  T}{2 \left(4 \pi  T \left(\sqrt{4 \pi ^2 T^2-1}-2 \pi  T\right)+2\right)^2}.
\end{equation}
As $T\to \infty$ (when the $\omega=1$ black hole is small) 
\begin{equation}
	E\approx\frac{1}{8\pi G T}
\end{equation}
the same expression as that for Schwarzschild black holes in flat space.

In the other limit, when $T=\frac{1}{2\pi}$, the energy diverges as 
\begin{equation}
	E\approx \frac{1}{32 \pi G \left(T-\frac{1}{2 \pi }\right)}
\end{equation}

We can now compute the derivative of $E$ with respect to $T$ (holding $\omega$ fixed at unity) to obtain a specific heat. We get:
\begin{equation}\label{sph}
	C=-\frac{\pi  \sqrt{4 \pi  T \left(2 \pi  T-\sqrt{4 \pi ^2 T^2-1}\right)-1} \left(8 \pi ^2 \left(\sqrt{4 \pi ^2 T^2-1}+2 \pi  T\right) T^2+\sqrt{4 \pi ^2 T^2-1}\right)}{4 \left(1-4 \pi ^2 T^2\right)^2}
\end{equation}
Note that the specific heat defined in \eqref{sph} is everywhere negative. In the limit $T\to \frac{1}{2\pi}$, the specific heat diverges as
\begin{equation}
	C\approx -\frac{1}{32 G \pi  \left(T-\frac{1}{2 \pi }\right)^2}.
\end{equation}
On the other hand, when $T$ is very large, the specific heat is very small (but still negative) 
\begin{equation}
	C=-\frac{1}{8 G \pi  T^2}
\end{equation}

The fact that their specific heat is always negative, suggests that Kerr-AdS black holes at $\omega=1$ are locally unstable saddles of the grand canonical ensemble. This fact is also suggested by the fact that $\omega=1$ black holes with a given temperature are unique, and have grand free energy larger than that of the corresponding thermal $AdS$ solution, and so should presumably be thought of as local maxima that flow directly into a local minimum in 
a Landau Ginzburg-type diagram of the sort constructed in 
\cite{Aharony:2003sx, Aharony:2004ig, Aharony:2005bq, Aharony:2005ew}.

\subsection{Large $r$ behaviour of the solution in a non rotating frame} \label{nonrotating}

The metric of pure $AdS$ space takes the form  
\begin{equation}
	ds_{AdS}^2=-\left(1+y^2\right)dT^2+\frac{dy^2}{1+y^2}+y^2 \left({d\Theta }^2+\sin ^2\Theta {d\Phi}^2  \right)\ ,
\end{equation}
\footnote{Note, in section \ref{omegone} we used $T$ for temperature but here in section \ref{nonrotating} we use the same notation $T$ but for Lorentzian time.}
which transforms to \eqref{bsol} with $m=0$ using the coordinate transform
\begin{equation}
	\begin{split}
		T & = t\\
		\Phi & = \phi + at\\
		y \cos\Theta & = r \cos \theta\\
		y^2 &= \frac{r^2\Delta_\theta+a^2 \sin^2\theta}{1-a^2}\ .
	\end{split}
\end{equation}
In these coordinates, the black hole metric has the following form at large $y$,
\begin{equation} \label{bhtail} 
	ds^2 = ds_{AdS}^2 + \frac{2 m}{y} \frac{\left(\text{dT}-a \, \sin ^2\Theta \, d\Phi \right)^2}{\Delta _{\theta }^{5/2}} + \mathcal{O}\left(\frac{1}{y^2}\right)\ .
\end{equation}
At large $y$, the metric takes the following form:
\begin{equation}\begin{split}
		ds^2=\frac{dy^2}{y^2}+ y^2{g_{mn}dh^m dh^n}
	\end{split}
\end{equation}
where $h^m$ are the boundary coordinates ($T,\Theta,\Phi$). 
Note that the first correction to $g_{mn}$ relative to pure $AdS_4$ metric is of the order $\frac{1}{y^3}$, and so is a normalizable 
deformation of the metric. The coefficient of this deformation encodes the stress tensor.

\subsection{Boundary stress tensor}\label{bstbh}

As shown in Appendix C of \cite{Bhattacharyya:2007vs}, the  components of boundary stress tensor dual to the general Kerr-AdS black hole are given by:
\begin{equation}\begin{split} \label{sten} 
		&8\pi G T^{tt}=\frac{m \left( a^2 \sin ^2{\theta }+2\right)}{\left(1-a^2 \sin^2 \theta \right)^{5/2}}\\
		&8\pi G T^{\phi\phi}=\frac{ m \left(\csc ^2{\theta}+2 a^2\right)}{\left( 1-a^2 \sin^2 \theta \right)^{5/2}}\\
		&8\pi G T^{t\phi}=\frac{3 m a }{\left(1-a^2 \sin ^2{\theta}\right)^{5/2}}\\
		&8\pi G T^{\theta\theta}=\frac{ m}{\left(1-a^2 \sin ^2{\theta}\right)^{3/2}}\ .
	\end{split}	
\end{equation}  
The stress tensor of $\omega=1$ black holes may then be obtained as a function of $a$ by plugging in $m=\frac{\sqrt{a}(1+a)^2}{2}$ (see \eqref{msrcond}). If desired, the stress tensor may then be obtained as a function of $j$ or $\epsilon$ by using \eqref{chargessr} to solve for $a$ in terms of $j$ or $\epsilon$.

For generic values of $a$ the stress tensor above is 
smoothly distributed all over the sphere. As $a \rightarrow 1$ (i.e. in the large black hole limit), on the other hand, the stress tensor becomes increasingly peaked about the equator of the sphere. For instance, the total boundary energy is given by
\begin{equation}\label{entot} 
	E=\int d^2 \Omega T^0_0= \frac{m}{4G}  \int_0^\pi d\theta ~ \frac{\sin \theta \left( a^2 \sin ^2{\theta }+2\right)}{\left(1-a^2 \sin^2 \theta \right)^{5/2}} =\frac{\sqrt{a}}{2G(1-a)^2} \equiv E(a)
\end{equation} 
as expected from the first of \eqref{chargessr}. The point of interest here is the shape of the energy density function. Let us define $\rho(\theta)$ to be the differential energy density, 
$\frac{dE}{d \theta}$, normalized so that 
$$\int d\theta \rho(\theta)=1.$$
Clearly 
\begin{equation}\label{rhothet}
	\rho(\theta)= \frac{1}{E(a)} \times \left( \frac{m}{4G} \frac{\sin \theta \left( a^2 \sin ^2{\theta }+2\right)}{\left(1-a^2 \sin^2 \theta \right)^{5/2}} \right)  =(1-a^2)^2\frac{\sin \theta \left( a^2 \sin ^2{\theta }+2\right)}{4\left(1-a^2 \sin^2 \theta \right)^{5/2}}\ .
\end{equation} 
Note that $m$ drops out of the expression for $\rho(\theta)$. 
\begin{figure}[h]
	\begin{subfigure} [b]{\textwidth}
		\centering
		\includegraphics[width=0.6\textwidth]{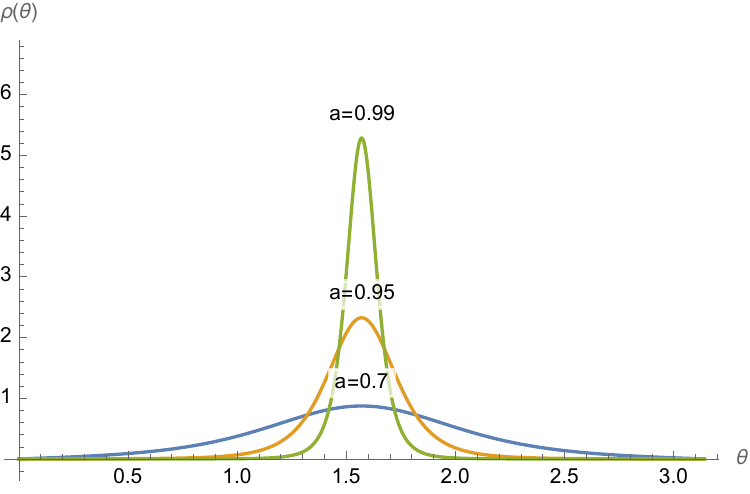}
	\end{subfigure}
	\caption{Plots of the normalized boundary differential energy density, $\rho(\theta)$, as a function of $\theta$ for three different values of $a$. }
	\label{rhotheta}
\end{figure}
In Fig. \ref{rhotheta} we have plotted $\rho(\theta)$ versus $\theta$ at three different values of $a$.  As we see from the figure, at $a=0.7$, $\rho(\theta)$ is smoothly spread all over the sphere, even though its maximum value occurs at $\theta= \frac{\pi}{2}$. At 
$a=0.9$, $\rho(\theta)$ is more distinctly peaked at 
$\theta=\frac{\pi}{2}$, and this peak is even more prominent at $a=0.99$. 

In the limit $a \rightarrow 1$, $\rho(\theta)$ tends to delta function localized about $\theta= \frac{\pi}{2}$. Setting $a=1-\varepsilon$, and $\theta= \frac{\pi}{2} - \delta \theta$,  and expanding for small $\varepsilon$ and small $\delta \theta$, one obtains 
\begin{equation}\label{roht}
	\rho_\varepsilon (\theta) = \frac{3\varepsilon^2}{(2 \varepsilon + (\delta \theta)^2)^{5/2}}\ .
\end{equation} 
It is easily directly verified that the integral of $\rho_\varepsilon(\theta)$ over the real line, is unity. In the limit $\varepsilon \to 0$ $\rho_\varepsilon(\theta)$ vanishes away from $\frac{\pi}{2}$ so it follows that 
$$ \lim_{\varepsilon \to 0 } \rho_\varepsilon (\theta) = \delta(\delta \theta)\;.$$

Continuing to work in the small $\varepsilon$ limit and the neighborhood of the equator, and retaining only terms at leading order, the various components of the boundary stress tensor simplify to 
\begin{equation}\begin{split} \label{steneq} 
		&T^{tt}=T^{\phi\phi}=T^{t\phi}=\frac{3m}{8\pi G(2 \varepsilon + (\delta \theta)^2)^{5/2}} \approx \frac{1}{4\pi G\varepsilon^2} ~\delta(\theta -\frac{\pi}{2}) \\
		&T^{\theta\theta}=0\ .
	\end{split}	
\end{equation} 
Moving to the left and right moving coordinates $\sigma^+=t+\phi$ and $\sigma^-=t-\phi$, we see that \eqref{steneq} can be rewritten as 
\begin{equation}\begin{split} \label{steneq-light} 
		&T^{++}=T_{--}\approx \frac{1}{\pi G\varepsilon^2}~ \delta(\theta - \frac{\pi}{2} ) \\
		&T^{--}=T^{+-}=T^{-+}=T^{\theta\theta}=0\ .\\
	\end{split}	
\end{equation} 

Let us summarize. At generic values of $a$, the boundary stress tensor to the rotating black hole is 
spread all over the sphere. In the large mass limit, $a \to 1$, on the other hand, the stress tensor 
becomes highly peaked on the equator, and reduces, in fact, to the stress tensor of a two-dimensional 
left moving chiral gas localized at the equator. 

\subsection{Comparison with the stress tensor of a free field theory in the limit $\omega \to 1$}

In the previous section, we have studied the boundary stress tensor dual to Kerr-AdS black holes. Recall that this stress tensor is generically smoothly distributed on the $S^2$, but gets increasingly peaked around the equator in the limit that the energy and the angular momentum of the black hole go to infinity, with their ratio tending to unity.

In order to have a point of comparison for this result, in Appendix 
\ref{tstfb}, we have presented the computation of the stress tensor of a CFT composed of $N^2$ free bosonic fields on $S^2$, at inverse temperature $\beta$ and 
angular velocity (chemical potential dual to angular momentum) $\omega$. We have performed this computation using two different methods: first by taking the coincident limit of an (easily evaluated)  Euclidean thermal two-point function of the free scalars, and second by explicitly constructing the thermal density matrix and evaluating the expectation value of the bulk stress tensor in this field theory state. \footnote{Apart from providing the point of comparison described above, these two computations also turn out to be a useful warm-up for a similar computation we perform in the bulk of $AdS_4$ in section \ref{tbs} below. } Unsurprisingly, the result for the field theory boundary stress 
tensor, in this case, is qualitatively similar to that of the black hole. 
\footnote{While the spatial distribution of the energy density in the boundary stress tensor dual to classical Kerr-AdS black holes is qualitatively similar to the distribution of energy density in free theories with similar energies and angular momenta, the entropy as a function of energy and angular momenta differ qualitatively in these two cases. As a consequence, the angular velocity (as a function of energy and angular momentum) is such that the classical black hole stress tensor reaches 
	$\omega=1$ at finite energy and angular momentum (see Fig. \ref{kerr-ads4}) while the free CFT only attains $\omega=1$ at asymptotically large values of the energy and angular momentum. One of the main points of this paper is that this difference is less stark when considering quantum corrections around the black hole background, which diverge in the limit $\omega \to 1$.} 
The stress tensor is smooth on the sphere when $E/N^2$ and $L/N^2$ are 
of unit order, but becomes increasingly localized near the equator of 
the $S^2$ when $E/N^2$ and $L/N^2$ become large.

We have spent so much discussing the distribution 
of the energy density on $S^2$, both for black holes and free boundary theories, because this distribution will turn out to be qualitatively different in the boundary stress tensor dual to the new Grey Galaxy saddles we study in this paper.
We turn to a study of these new solutions in the next section.

\section{Thermodynamics} \label{tggs} 

In the previous section, we have 
presented a detailed review of the thermodynamics of the classical black hole. In this section, we first work out the thermodynamics of the gas in the canonical ensemble (see subsection \ref{tce} below).  In subsection 
\ref{me} we then work 
out the thermodynamics of Grey Galaxy solutions (and briefly compare the same to Revolving Black Hole solutions) in the microcanonical ensemble. 
Finally, in subsection \ref{ans}, we also work out the thermodynamics of Grey Galaxy solutions in the `canonical ensemble'. 

\subsection{Thermodynamics of the gas in the canonical  ensemble} \label{tce}

The space of finite energy solutions of any bulk field in $AdS_4$  transforms in an irreducible representation of $SO(3,2)$. Consider a spin $s$ field whose mass is such that its solutions correspond 
to the representation $(\Delta, s)$ (where we label representations by the dimension and spin of the primary). 
Via the state operator map, the (single particle) state space of this representation is in one-to-one correspondence with the set of operators we can make by acting on the spin $s$ primary with derivatives.

\subsubsection{State Content of long representations} 

Let us first assume that $\Delta >s+1$, so that the representation is generic (long) (see e.g. \cite{Minwalla:1997ka}). In this case, there are no null states (or conservation equations) so the derivatives act on the primary in an unconstrained manner. Consider acting on the primary with $C^{\mu_1 \ldots \mu_j} \partial_{\mu_1} \ldots \partial_{\mu_j}$, where $C^{\mu_1 \ldots \mu_j}$ is any traceless symmetric tensor. Note that the linear space of $j$ index traceless symmetric tensors 
transform in the spin $j$ representation of $SO(3)$. As a consequence, the action of these derivatives yields the representations with scaling dimension $\Delta +j$, and angular momenta $l$ that lie in the Clebsh Gordon decomposition of the tensor product of $j$ and $s$ irreps, i.e. angular momenta  
$$ \left( |j-s|, |j-s|+1, \ldots j+s \right).$$ 
In other words, this set of operators carries angular momenta $l$ with 
$l$ given by 
\begin{equation}\label{sumang} 
	\sum_{\alpha=|j-s| -j}^s (j+ \alpha)\ .
\end{equation} 
This decomposition holds for all values of $j$. In addition, we can dress each of the operators listed above with $(\nabla^2)^n$. It follows that the full operator content of this representation is given by 
\begin{equation}\label{sumang2} 
	\sum_{n=0}^\infty\sum_{j=0}^\infty \sum_{\alpha=|j-s| -j}^s 
	\left( \Delta + j + 2n , j+ \alpha \right) 
\end{equation} 
where we have denoted $SO(3)$ representations with spin $l$ and scaling dimension $\Delta$ as $(\delta, l)$. We denote the $J_z$ quantum number by $l-a$, where $a$ runs from $0$ to $2a$. 

In the specially simple case of a scalar of dimension $\Delta$, the operators in the representation take the form 
$$ (\nabla^2)^n C^{\mu_1 \ldots \mu_l} \partial_{\mu_1} \ldots \partial_{\mu_l} O$$
for all $n$ and $l$. These multiplets carry dimension $= \Delta +2n +l$ and angular momentum $l$ and so constitute the sum of representations  
\begin{equation}\label{sumangscal} 
	\sum_{n=0}^\infty\sum_{l=0}^\infty  
	\left( \Delta + l + 2n , l \right) 
\end{equation} 
which is a special special case of \eqref{sumang2} with $s=0$. 

\subsubsection{Partition Function over Fock Space}

It follows from \eqref{sumang2} that the multi-particle partition function over the single-particle Hilbert space \eqref{sumang2} is given by 
\begin{equation}\label{mutparp} 
	\ln Z= \sum_{n=0}^\infty \sum_{l=0}^\infty \sum_{\alpha=-s}^{l-|l-s|}\sum_{a=0}^{2l}   - \ln \left(1-e^{-\beta(\Delta + 2n-\alpha +\omega a )- \beta l(1-\omega) }\right)\ .
\end{equation} 
In the limit $\omega\to 1$, the sum over $l$ in \eqref{mutparp} is effectively cut off  at the parametrically large value $l= \frac{1}{\beta(1-\omega)}$. As the typical value of $l$ that occurs in the summation is of order $\frac{1}{\beta(1-\omega)}$, at leading order in $1-\omega$ we can replace the summation with an integral \footnote{More precisely, the contribution to the sum of terms from $l=0$ to $M$, where $M$ is any finite number, is subleading in the limit $\omega \rightarrow 1$.} and obtain 
\begin{equation}\label{zfini}
	\ln Z \approx  -\frac{1}{\beta(1-\omega)}\sum_{\alpha=-s}^s \sum_{n=0}^\infty  \sum_{a=0}^{\infty} \int_0^\infty  dx
	\ln\left( 1-e^{-\beta( \Delta -\alpha+2n+a)-x} \right)   
\end{equation} 

In the special case of a scalar of dimension $\Delta$, the Fock Space partition function is given by 
\begin{equation}\label{sumover} \begin{split}
		\ln Z&= \sum_{n, l=0}^\infty 	\sum_{a=0}^{2l} - \ln \left( 1- e^{-\beta(\Delta 
			+ 2n + \omega a) -\beta(1-\omega) l} \right) \\
		&\approx -\frac{1}{\beta(1-\omega)}\sum_{n, a=0}^\infty  \int_0^\infty  dx  \ln \left( 1- e^{-\beta(\Delta + 2n +a) -x} \right) \\
		&\equiv -\frac{C_\Delta(\beta)}{\beta(1-\omega)},
	\end{split}
\end{equation} 
where we have defined 
\begin{equation}\label{senab}
	C_{\Delta} (\beta)= \sum_{n, a=0}^\infty  \int dx  \ln \left( 1- e^{-\beta(\Delta + 2n +a) -x} \right) 
\end{equation}

\eqref{zfini} can be rewritten in terms of $C_{\Delta}$ as 
\begin{equation}\label{zfin}
	\ln Z \approx -\frac{1}{\beta(1-\omega)} \left( \sum_{\alpha=-s}^s C_{\Delta-\alpha}(\beta) \right) 
\end{equation} 
In the limit $\omega \to 1$, in other words, the contribution of a single long spin $s$ representation to the partition function is the same as the contribution of $2s+1$ scalars, whose dimensions are 
$\Delta-s, \Delta -s+1 \ldots \Delta +s$. 

\subsubsection{Contribution of short representations}

A short spin $s$ representation has $\Delta =s+1$. The state content of this representation is that of 
the long representation $(s+1, s)$ minus the state content of the null state representation, i.e. the 
long representation with quantum numbers $(s+2, s-1)$. Using \eqref{zfin} and taking the difference, we see that the partition function of the short spin $s$ representation is given by 
\begin{equation}\label{zshort}
	\ln Z =  -\frac{1}{\beta(1-\omega)} \left( C_{1}(\beta) +C_{2}(\beta)\right) 
\end{equation}
The fact that we receive contributions from only two series of modes (independent of $s$) reflects the fact that 
particles of all spin have only two polarizations in the four-dimensional bulk. 

\subsubsection{The full partition function of the gas} 

In the limit, $\omega \to 1$, the logarithm of the full partition function of the determinant (the gas) equals the sum of contributions from each bulk field, i.e. the sum of contributions of the form \eqref{zfin} and 
\eqref{zshort}. It follows that the full partition function is given by 
\begin{equation}\label{zfull}
	\ln Z =  -\frac{C(\beta) }{\beta(1-\omega)}  
\end{equation}
where $C(\beta)$ is a summation over $C$ factors for each bulk field. Recall long spin $s$ field of dimension $\Delta$ contributes a $C$ factor equal to $\left( \sum_{\alpha=-s}^s C_{\Delta-\alpha}(\beta) \right)$, while each 
short spin $s$ field contributes a factor of $C_1(\beta)+ C_2(\beta)$. 

The thermodynamical charges that follow from \eqref{zfull} are 
\begin{equation}\begin{split} 
		\label{sej} 
		J&= -\frac{C(\beta)}{\beta^2 (1-\omega)^2} \\
		E&=-\frac{C(\beta)}{\beta^2 (1-\omega)^2} + \frac{C'(\beta)}{\beta(1-\omega)}\\
		S&
		=\frac{C'(\beta)}{(1-\omega)}-\frac{2C(\beta)}{\beta(1-\omega)}
	\end{split} 
\end{equation} 
Note that $C(\beta)<0$ and $C'(\beta)>0$,  
so the entropy is a positive number, and the energy is always larger than the angular momentum. At the leading order, we can drop the second term in the RHS of the expression for energy in \eqref{sej}, so $E=J$. Note also that the entropy is proportional to $\frac{1}{(1-\omega)}$ and so is of order $\frac{1}{\sqrt{G}}$ when the energy and angular momentum are of order $\frac{1}{G}$. It follows that the entropy of the gas is parametrically smaller than that of the black hole, and so can be ignored for thermodynamical purposes. 

The formulae of this section compute the thermodynamics (in the microcanonical ensemble) 
of the thermal gas phase in $AdS_4$ in the limit 
$\omega \to 1$. \footnote{
	When the energy and angular momentum are of order $\frac{1}{G}$ (and so of leading order) in this phase, the grand free energy and entropy are both of order $\frac{1}{\sqrt{G}}$ (and so are subleading). } As explained in the introduction, however, they also determine the thermodynamics of the gas component of Grey Galaxy solutions. Indeed, for thermodynamical purposes, Grey Galaxies can be thought of as a non-interacting mix of the Kerr-Ads black hole and the thermal 
AdS phases.

\subsection{Microcanonical ensemble} \label{me} 

As we have explained in the introduction, our new solution describes a weakly interacting mix of the black hole and the gas. In this subsection, we analyze this mix in the microcanonical ensemble. In this ensemble, the energy and angular momentum of the full system is fixed. However, the system is free to exchange energy and angular momentum between the gas and the black hole. Recall that (always working at leading order in $1/G$) the entropy of the gas is negligible compared to the entropy of a black hole with similar energy and angular momentum as that of the gas.

Within the microcanonical ensemble, we maximize the entropy of our system. 
The thermodynamical relation 
\begin{equation}\label{entropy-change}
	T\Delta S=\Delta E-\omega\Delta J
\end{equation} 
helps us understand the nature of this maximum. Let us first imagine that all of the system's energy and angular momentum is in the black hole, and none of it is in the gas.  Now let this black hole emit some energy and angular momentum to the gas. Let the change in the black hole energy and angular momentum, respectively, be denoted by $\Delta E$ and $\Delta J$. Note that $\Delta E$ and $\Delta J$ are both negative. ($|\Delta E|$ and $|\Delta J|>0$ are the energy and angular momentum gained by the gas). Applying the unitarity bound to the gas, we conclude that  
\begin{equation}\label{guni}
	|\Delta E|\geq |\Delta J|.
\end{equation}
If the core black hole has $\omega<1$, then \eqref{entropy-change} implies 
\begin{equation} \label{entinc}
	T\Delta S\leq \Delta J(1-\omega)<0\ .
\end{equation}
(the first inequality uses \eqref{guni} and the last inequality follows because $\Delta J$ is negative; recall that the entropy of the gas is negligible compared to that of the black hole).
So the emission of energy and angular momentum (to the gas) from an $\omega <1$ black hole  can only decrease its entropy. In this situation, the black hole we started with is the 
maximum entropy configuration, and so is the dominant 
saddle point. This saddle point lies above the red curve in Fig. \ref{kerr-ads4}.  

Let us now suppose the charges are such that the starting black hole (which contains all of the angular momentum and energy of our system) has $\omega>1$. It is now possible for the black hole to increase its entropy by losing 
energy and charge. From \eqref{entinc} and \eqref{guni}, it is clear that the entropy gain for the black hole is largest when it loses energy and angular momentum to modes of gas with $|\Delta E|=|\Delta J|$. It is precisely modes of this nature that make up the gas 
studied in the previous subsection. In this situation 
\begin{equation} \label{entinco}
	T\Delta S= \Delta J(1-\omega)>0\ .
\end{equation}
(where we have used $\omega>1$ and the fact that $\Delta J$
is negative). So the black hole increases its entropy by 
emitting modes with $|\Delta E|=|\Delta J|$ to the gas. 

Since the black hole under study has $\omega>1$, it lies in the region depicted in blue in Fig. \ref{kerr-ads4}.  As the black hole emits the gas, the charges carried by the black hole itself 
move along the $ 45$-degree line toward the bottom left. The black hole continues to emit gas as long as this increases its entropy, and so its charges keep moving to the bottom left till we reach the $\omega=1$ curve.  Once we reach this curve, any further emission results in a black hole with $\omega<1$, and emission into the gas reduces (rather than increases) the entropy of the black hole.  It follows that the endpoint of the superradiant emission should involve a black hole whose charge, $E_{BH}$ and $J_{BH}$, are such that $\omega=1$ (i.e. are at the intersection of the 45-degree line and the red curve).

Hence in equilibrium, we have a non-interacting mix of a black hole with entropy corresponding to the entropy of a black hole with charges $E_{BH}$ and $J_{BH}$. The entropy of these black holes on $\omega=1$ line is listed in section \ref{omegone}.
\footnote{Let us say this in equations. Using the fact that the energy and the angular momentum of the 
	gas are equal to each other, the full entropy of our system equals 
	\begin{equation}\label{fent}
		S(E, J)= S_{\rm BH}(E-x, J-x)
	\end{equation} 
	where $x$ is the energy and angular momentum of the gas, and $E-x$ and $J-x$, respectively, are the energy 
	and angular momentum of the black hole. $x$ in \eqref{fent} must be chosen to ensure that 
	\begin{equation}\label{extremizing}
		\partial_x \left( S_{\rm BH}(E-x, J-x) \right) = -\partial_E S(E-x, J-x) -\partial_J S(E-x, J-x)=0
	\end{equation} 
	\eqref{entropy-change}  allows us to recast \eqref{extremizing} as 
	$-\beta \left( 1-\omega \right)=0$. \eqref{extremizing} is satisfied - hence our entropy has an extremum - when $\omega =1$. }

From the viewpoint of the microcanonical ensemble, it is the entropy in the gas (see the last line of \eqref{sej}) that explains why the Grey Galaxy solutions dominate over the Revolving Black Holes (RBH)s (see Appendices \ref{fpa} and \ref{rbs}). The comparison goes as follows. Consider the Grey Galaxy and RBH solutions 
at the same value of $E$ and $J$. The entropy of the Grey Galaxy solution is that of its central black hole plus that of its gas, while the entropy of the RBH solution is just that of the black hole (the condensate of derivatives in the RBH solution carries no entropy). The two black holes 
are not identical. The central black hole of the Grey Galaxy solution lies along the dotted line, to the left of the red curve in Fig. \ref{kerr-ads4-boost} by an amount $\delta j = {\cal O} \left( {\sqrt{G}} \right)$. The RBH solution, on the other hand, also lies along the same dotted line, to the left of the red curve by an amount $\delta j = {\cal O} \left({G}\right)$. Now recall that $\omega$ equals unity on the solid red curve. It follows that the derivative of the 
entropy (for motion along the dotted line) vanishes on the solid red curve 
(see \eqref{entropy-change}). Consequently, the entropy of the black hole 
at the centre of the Grey Galaxy/RBH  is smaller than the entropy of the black hole at the intersection of the solid red curve and dotted lines by order 
$\frac{1}{G} \times (\delta j)^2$ \footnote{We have expanded to second order in the Taylor expansion and used the fact that the entropy function is $1/G$ times a function of $\epsilon$ and $j$.}. This quantity is of order unity for Grey Galaxies but of order $G$ for RBHs. In other words, the entropy of the black hole in the RBH solution is larger than the entropy of the black hole in the centre of the Grey Galaxy by a number of order unity. On the other hand, the entropy of the gas in the Grey Galaxy solution is of order $\frac{1}{\sqrt{G}}$, and so overwhelms the relatively small difference in black hole entropies, explaining the entropic dominance of Grey Galaxies over RBH solutions. 

\subsection{Thermodynamics of Grey Galaxies and Revolving Black Holes in the `canonical ensemble'} \label{ans} 

Our interest, in this paper, is in solutions at fixed conserved energy and angular momentum, i.e. solutions in the micro-canonical ensemble. 
\footnote{While the Grey Galaxy solutions we construct in this paper will (we conjecture) dominate the microcanonical ensemble of the theory over a range of angular momenta and energies, they will never dominate the canonical ensemble.}
It is often useful, however, as a technical device to first work in the grand canonical ensemble, 
and then reconstruct the entropy as a function of angular momenta by performing the relevant inverse Laplace transforms. In this subsection, we will employ this strategy to study Grey Galaxy solutions. As we will see below, the use of this strategy is complicated by the fact that Grey Galaxies correspond to unstable saddles in the grand canonical ensemble. In this subsection, we 
first propose a strategy to deal with this complication, and then proceed with the analysis proper. 

While the discussion below will deepen our intuitive appreciation for Grey Galaxy saddles, it is not needed for the logical development of this paper. The conservative reader who finds herself uncomfortable with adventurous manipulations involving unstable saddle points can safely skip this subsection.  

\subsubsection{Black Hole Saddles and Determinants} 

Consider the canonical partition function
\begin{equation}\label{partfun}
	Z={\rm Tr} e^{- \beta H + \beta \omega J_z}\ .
\end{equation}
In the bulk description, the partition function \eqref{partfun} is computed by evaluating the (appropriately twisted) Euclidean gravitational path integral in the saddle point approximation. The relevant saddle points are 
\begin{itemize}
	\item Thermal AdS with temperature $1/\beta$, and twisted boundary conditions determined by $\omega$. 
	\item The Euclidean continuation of  Kerr-AdS black holes that have temperature $\beta$ and 
	horizon angular velocity $\omega$. 
\end{itemize} 	
The contribution of each of these saddle points equals 
$e^{-S}$ (where $S$ is the Euclidean action of the saddle point), multiplied by a one-loop determinant and other subleading quantum corrections.

In general, the saddle points that contribute to \eqref{partfun} include both those that are locally stable and those that are locally 
unstable. An example of a locally unstable saddle
\cite{Gross:1982cv}\footnote{The instability of this saddle can be intuitively understood from the viewpoint of the boundary theory as the small black hole is expected to correspond to a maximum rather than a minimum in the `Landau Ginzburg effective action for holonomies \cite{Aharony:2003sx, Aharony:2004ig,  Aharony:2005bq,Aharony:2005ew}}
is the small Schwarzschild Black hole in $AdS_4$. As we are interested in the canonical ensemble to eventually perform the inverse Laplace transform back to the microcanonical ensemble, it is clear that our analysis should include unstable saddles such as those described above, as the small Schwarschild black hole 
dominates the microcanonical ensemble in the appropriate energy 
range. The practical question we are faced with is the following: how do 
we meaningfully compute quantum corrections around an unstable saddle point \footnote{We thank J. Santos and T. Ishii for highlighting this question.}? This  question, which is of relevance to us (as $\omega=1$
Kerr-AdS black holes all have negative specific heat) has generated some recent discussion (see e.g. \cite{Marolf:2022jra}). We proceed as follows. 

As large black holes have positive specific heat, their determinants are well-defined. In a paper written some time ago, Denef, Hartnoll, and Sachdev (DHS) \cite{Denef:2009kn,Denef:2009yy} derived a beautiful formula for the Euclidean determinant around any such black hole in asymptotically $AdS$ space. The DHS formula for the black hole determinant is presented as an infinite product over a set of factors. Each factor in this product is associated with a quasinormal mode in the black hole background and is determined by the corresponding quasi-normal mode frequency.  The final one-loop contribution to the microcanonical ensemble, for such black holes, is thus given by the inverse Laplace Transform of the product of the saddle point partition function and the  DHS formula for the determinant. 

In the micro-canonical ensemble, we expect the (loop-corrected) entropy of black holes to be an analytic function of their mass and angular momentum. \footnote{ Recall that specific heat does not determine the stability of small black holes in a micro-canonical ensemble, and so we expect the entropy to be everywhere analytic.}  It follows that the one-loop contribution to the micro-canonical entropy at small masses can be obtained from the analytic continuation of the large mass result.  The DHS formula gives us an easy way to perform the required analytic continuation. It seems reasonable to expect  that black hole quasinormal modes frequencies (which, after all, are micro-canonical data)  are themselves analytic functions of the black hole mass and angular momentum. It follows that the analytic continuation
of the one-loop contribution to the entropy can be obtained by simply applying the DHS formula even for black holes of small mass, and then proceeding to take the inverse Laplace transform. As the resultant expression is presumably analytic in the mass and angular momentum and gives the correct answer at large masses, it should  yield the correct one-loop contribution to the microcanonical entropy at every value of the black hole mass and angular momentum.
\footnote{We thank J. Santos and T. Ishii for extremely useful probing questions on this point. We also thank F. Denef and S. Hartnoll for discussions on this point.}

The net upshot of this discussion is simple. 
The DHS formula effectively provides a definition 
of the determinant about negative specific heat black holes, one that seems guaranteed to correctly reproduce micro-canonical physics. \footnote{ While the arguments presented above seem relatively convincing to us, they do not constitute a proof.  We need, for instance, to show the absence of obstructions to analytic continuation. If the continuation involves branch ambiguities, we would also need to specify a choice of branch. It would certainly be useful to justify our prescription more carefully. In this paper we proceed assuming the correctness of our prescription, leaving more careful justification to further work.}

Armed with this prescription, we return to the sum over saddles. At large $N$ (i.e. small $G$), and at generic values of $\omega$ and $\beta$, the order unity determinant is negligible compared to the classical contribution and can be ignored. We will now explain that the neglect of quantum corrections fails as $\omega \rightarrow 1$, as the formally subleading determinant actually diverges in this limit. 

As we have explained above, the logarithm of the determinant around a Euclidean Kerr-AdS black hole is a sum of an infinite number of terms, one associated with each of the black hole quasi-normal modes. 
There are two ways in which 
this determinant might diverge in the limit $\omega \rightarrow 1$. First, the contribution of one or more of the quasi-normal modes may individually diverge in this limit. Second, the sum over the contributions of an infinite family of modes may diverge, even though the contribution of any member of this family stays finite. We examine each of these possibilities in turn. 

\subsubsection{Divergent contributions from `centre of mass' motion} \label{divzero}

The unitarity bound assures us that it is impossible for the contribution of any single mode to diverge about the thermal AdS saddle. Consider a fluctuation mode of frequency $f$ and angular momentum $j$. The Boltzmann suppression factor for this mode is $e^{-\beta(f-\omega j)}$. At $\omega=1$ this factor simplifies to $e^{-\beta(f-j)}$. It follows that the contribution of this fluctuation to the determinant diverges if - and only if - $j\geq f$, a condition that is inconsistent with the unitarity inequality that all fluctuation modes around $AdS$ are constrained to satisfy, namely $f \geq j+\frac{1}{2}$.  Consequently, the contribution of any single mode to the determinant around the thermal AdS saddle is always finite at $\omega=1$. 

The result of the previous paragraph, however,  does not apply to the black hole saddle. There exists at least one (and we conjecture exactly one) mode around the black hole background, whose contribution 
to the determinant diverges in the limit $\omega \rightarrow 1$. The mode in question arises from the quantization of the `centre of mass' motion of the black hole, and accounts for the contribution of a series of descendants (of the primaries that make up the black hole). Accounting for this divergence gives rise to the Revolving Black Hole solutions we have already introduced above. 
We demonstrate in  Appendix \ref{descendants}, that the singular part of the contribution of these modes to $\ln Z$ is  given by 
\begin{equation}\label{divcontint}
	-\ln (1-e^{-\beta(1-\omega)}) \approx 
	-\ln \left( \beta(1-\omega)\right) 
\end{equation} 
The relatively mild singularity in \eqref{divcontint} is subleading compared to the divergence of the gas (see \eqref{zfull}), and so will play no role in the new `Grey Galaxy'  solutions that we construct later in this paper. The subleading nature of this divergence is the canonical ensemble's version of the micro-canonical observation that revolving black hole saddles are entropically subleading compared to Grey Galaxy saddles.
\footnote{We emphasize that the two different divergences we have described above are physically rather different. \eqref{divcontint} is associated with the infinite occupation of a single mode and is reminiscent of Bose condensation. \eqref{zfull}, on the other hand,  is a consequence of the finite, but equal, occupation of an infinite number of modes, and is conceptually similar to a high-temperature divergence.} 
However, the divergence discussed in this subsection permits the construction of a new subleading  `revolving black hole' solution, that we study in Appendix \ref{rbs}.  

As we discuss in section \ref{susy},  new solutions appear to have interesting implications for the spectrum of supersymmetric black holes.

\subsubsection{Divergent contributions from families of modes}

We now turn to a discussion of the second kind of divergence, namely the divergence arising from an infinite family of modes, each of whose contribution remains finite as $\omega \rightarrow 1$. In subsection 
\ref{tce} we have already studied this question around the thermal AdS 
saddle. As argued in that section, the summation over modes with large values of $l$ (see \eqref{mutparp}) gives rise to the divergent contribution 
\eqref{zfull} to the partition function. We emphasize that the divergence in \eqref{zfull} has its origin in modes 
with large values of $l$. As is intuitively clear from considerations of the centrifugal force (and as we have already explained in the introduction and explain in much more detail in  Appendix \ref{fnll}), these modes live at large values of $r$, so their divergent contribution to the determinant around the Kerr-AdS saddle is identical to \eqref{mutparp}, the divergent contribution around the pure $AdS$ saddle.  
\footnote{More formally, as we have explained above, the black hole determinant is a product over terms, one associated with each quasi-normal mode. In the limit $l \to \infty$, quasi-normal mode frequencies of the counterparts of those in \eqref{sumang} become $\Delta +j+2n + {\cal O}(e^{-bl})$, where $b$ is a positive constant whose value depends on the 
	size of the black hole in the centre (the parameter $a$ in subsection 
	\ref{omegone}). The small correction to the normal mode frequency 
	includes both real and imaginary pieces. As the quasinormal mode frequencies tend to the normal mode frequencies, their contribution 
	to the DHS determinant (\cite{Denef:2009kn, Denef:2009yy}) reduces 
	to their contribution to the global $AdS$ determinant.} 

It follows from \eqref{sej} that the gas contribution to the  angular momentum and energy is given by  
\begin{equation}\label{hyt}
	\Delta E\sim \Delta L
	\sim -\frac{C_\Delta(\beta)}{\beta^2(1-\omega)^2}>0\ .
\end{equation}
Note the perfect agreement with the formula \eqref{sejint} from the introduction. As we have already noted under \eqref{sejint}, when $1-\omega \propto  \sqrt{G}$, $\Delta E \sim 
\frac{1}{G}$ is comparable to the classical or saddle point value of the black hole energy \footnote{At the same values of $\omega$, the divergent contribution of zero modes to the 
	energy from  (\ref{divcontint}) 
	scales like $\frac{1}{\sqrt{G}}$, and so is subleading compared to the classical result.}. At these values of $\omega$, however, the contributions of the determinant to the entropy and $\ln Z$ are both of order $\frac{1}{\sqrt{G}}$ (see \eqref{sej} and \eqref{zfull})  and so are parametrically subleading compared to the classical black hole contributions, which are of order $\frac{1}{G}$. 

Let us summarize the situation from the viewpoint of the canonical ensemble.   Classical black holes that lie above 
the red curve in Fig. \ref{kerr-ads4} are legitimate saddle points. Classical black holes that lie below the red curve in Fig. \ref{kerr-ads4} receive infinite one-loop contributions, and so do not exist, but are replaced by a new two-parameter family of saddle points.
All of the new saddles have values of $\omega$ just less than unity, i.e. lie just above the red curve in Fig. \ref{kerr-ads4}. More precisely $1-\omega= \sqrt{G} \alpha$ where $G$ is 
the (parametrically small) Newton constant and $\alpha$ is an arbitrary number. Our new saddles are parameterized by $\alpha$ and 
where on the red curve in Fig. \ref{kerr-ads4} they lie (equivalently by $\alpha$ their temperature $\beta$).  Two saddle points with the same temperature but different values of $\alpha$ represent the same classical black hole dressed with different one-loop gas 
contributions.\footnote{As we will see below, these distinct one-loop contributions modify the metric, but in a simple and controllable way, and far away from the black hole.} Two such saddle points have the same leading order entropy, and the same value of $E-J$, but differ in their energies (and angular momenta).

\section{The stress tensor of the bulk gas} \label{tbs}

In this section, we compute the bulk stress tensor for a minimally coupled scalar propagating in 
$AdS_4$ space in the `thermal gas' phase at inverse temperature $\beta$ and angular velocity $\omega$. 

We parameterize $AdS_{4}$ space as 
\begin{equation}\label{coordinates}\begin{split} 
		&X_{-1} = \cosh \rho \cos t \\
		&X_{0}= \cosh \rho \sin t \\
		& X_1 = \sinh \rho \sin \theta \cos \phi \\
		& X_2 = \sinh \rho \sin \theta \sin \phi \\
		& X_3 = \sinh \rho \cos \theta\ . \\
	\end{split} 
\end{equation} 
Through this paper, we will trade the coordinate $\rho$ for the coordinate $r$ defined by 
\begin{equation} \label{rhor} 
	\sinh \rho= r\ .
\end{equation} 
Note that $\cosh^2 \rho= 1+r^2$ and $\tanh^2 \rho = \frac{r^2}{1+r^2}$. 
The metric in these coordinates is given by 
\begin{equation}\label{metriccord} \begin{split} 
		ds^2&= -\cosh^2 \rho dt^2 + d\rho^2 + \sinh^2 \rho d \Omega_2^2\\ 
		&= -(1+r^2) dt^2 + \frac{dr^2}{1+r^2} + r^2 d \Omega_2^2\\
		d\Omega_2^2&= d \theta^2+ \sin^2 \theta d \phi^2\ .\\
	\end{split} 
\end{equation}         

The energy is the charge generated by the killing vector $\partial_t$, while $J_z$ is the charge generated by rotations on the $X_{1}$-$X_2$ plane, i.e. by the killing vector $\partial_\phi$. 

In this section, we study a free real minimally coupled scalar field of mass $M$ (chosen so that $\Delta(\Delta-3)= M^2$) propagating in the bulk 
$AdS_4$. Our bulk system is taken to be at inverse temperature $\beta$ and angular velocity $\omega$. Thermal excitations of the scalar produce a net effective bulk stress tensor. In this section, we evaluate the expectation value 
(ensemble average) of this bulk stress tensor using two different methods: first by using the Hamiltonian method, and second by using the Euclidean method. The advantage of the first method is that 
it is physically very transparent. The advantage of the second method is that it is algebraically 
convenient. As we will see below, our two methods yield identical answers. 

\subsection{Hamiltonian method} 

\subsubsection{Mode wavefunctions}

The real massive scalar field described above  can be expanded as a linear combination of solutions to the Klein-Gordon equation as:
\begin{equation}\label{expoffield}
	\begin{split}  
		\phi(r,t,\theta,\phi)&= \sum_{n, l, m} \frac{1}{\sqrt{2 E_{nl} }} \left( e^{-i(\Delta + 2n +l)t} Y_{lm} F_{nl}(r) a_{nl m} + e^{i(\Delta + 2n +l)t} Y^*_{lm} F^*_{n,l}(r) a^*_{n, l, m} \right) \\
		& \equiv \sum_{n, l, m} \psi_{n l m} (x^\mu) a_{n, l, m}+ \psi^*_{nl m}(x^\mu)  a^*_{n, l, m}
	\end{split} 
\end{equation}  
where we have defined 
\begin{equation}\label{enl} 
	E_{nl}= \Delta + 2n +l\ .
\end{equation} 
Our spherical harmonics are normalized in the usual manner, i.e. so that they obey 
\begin{equation}\label{sphericalharm}
	\int d^2 \Omega \;  Y_{lm}(\theta, \phi) Y^*_{l'm'} (\theta, \phi)   
	= \delta_{l l'} \delta_{m m'}\ .
\end{equation}
We demand that the coefficient wave functions $\psi_{nlm}$ are orthogonal in the Klein-Gordon norm, i.e. that 
\begin{equation}\label{fullnnorm}
	2E_{nl} 	\int \sqrt{-g} (-g^{00}) \psi_{nlm}  \psi^*_{n'l'm'}(r) = \delta_{n, n'} \delta_{l, l'} \delta_{mm'} \ .
\end{equation}

Using \eqref{sphericalharm}, \eqref{fullnnorm} reduces to the requirement 
\begin{equation}\label{radfunnorm} 
	\int \sqrt{-g} (-g^{00}) F_{n l}(r) F^*_{n' l'}(r) = \delta_{n n'} \delta_{l l'} \ .
\end{equation} 
By performing a canonical quantization of this scalar field, we demonstrate in Appendix \ref{cq} that 
the classical numbers $a_{nl m}$ and $a^*_{n, l, m}$ are promoted, in the quantum theory, to operators that 
obey the standard canonical commutation relations 
\begin{equation}\label{sccr} 
	[a_{nlm},a_{n'l'm'}^{\dagger}]= \delta_{n, n'} \delta_{l, l'} \delta_{mm'} \ .
\end{equation} 
Even though we will not use this explicit expression in the rest of this section, we note for completeness that the exact form of the function $F_{n,l}(\rho)$ is known \cite{kaplan}, 
which is given by 
\begin{equation}\label{fnl} 
	\begin{split} 
		F_{nl}(r)&= N_{n, l} \frac{r^l} {(1+r^2)^{\frac{l+\Delta}{2}}} ~{}_2F_1\left( -n, \Delta+l+n, l+\frac{3}{2}, \frac{r^2}{1+r^2} \right) \\
		N_{nl} &= (-1)^n \sqrt{2(\Delta + 2n +l)} \sqrt{ \frac{ \Gamma(n+l+\frac{3}{2}) \Gamma(\Delta+ n+l)}{n! \Gamma^2(l+\frac{3}{2}) \Gamma(\Delta+n-\frac{1}{2})} }\ .\\
	\end{split} 
\end{equation} 
Note also that, in the large $l$ limit, 
\begin{equation}
	N_{nl}\simeq (-1)^n\sqrt{\frac{2  l^{\Delta +2 n-\frac{1}{2}}}{n! \Gamma \left(n+\Delta -\frac{1}{2}\right)}}\ .
\end{equation}

In this paper, we are most interested in large values of $l$. Both the spherical harmonics and the functions $F_{nl}$ simplify in this limit, as we explain in Appendix \ref{shll} and Appendix
\ref{fnll} respectively.

\subsection{Expression for the bulk stress tensor as a sum over modes}\label{bst-H}

The bulk stress tensor for our free scalar field is given by 
\begin{equation}\label{stresstensor} 
	T_{\mu\nu}= \partial_\mu \phi \partial_\nu \phi  -\frac{g_{\mu \nu}}{2} \left( (\partial \phi)^2+M^2 \phi^2 \right)   \ .
\end{equation} 
Let us study the scalar field in a thermal ensemble with inverse temperature $\beta$ and angular velocity 
$\omega$. Because we are dealing with a free system, it follows that the density matrix, $\rho$, for our system is given by 
\begin{equation}\label{denmatfree}
	\rho = \prod_{n ,l ,m} \rho_{n l m} 
\end{equation} 
where $\rho_{n l m}$ is the density matrix associated with the single particle state with the specified quantum numbers. Explicitly 
\begin{equation} \label{rhonlm} 
	\rho_{nlm} = \sum_{k} e^{-\beta k \left( (\Delta + 2n + l) -\omega m\right)  } |k \rangle \langle k |
\end{equation} 
where $k$ is the occupation number of the state with quantum numbers $n, l, m$, and the states $|k\rangle$ are of unit norm. The expectation value of 
the stress tensor only receives contributions from terms in which the same mode is both created and destroyed and so  
\begin{equation} \label{sten1} 
	\langle T_{\mu \nu} \rangle=	\frac{ { \Tr} (\rho T_{\mu\nu})}{Z} = \sum_{n, l, m}  \frac{1}{ Z_{nlm}}{\Tr}  \left( \rho_{n l m} T^{nlm} _{\mu\nu} \right) 
\end{equation} 
where $Z= \Tr \rho$ is the partition function, $Z_{nlm}=\Tr \rho_{nlm}=\frac{1}{1-e^{-\beta(\Delta +2n +l -\omega m)}}$ is the partition function over the given state and 
\begin{equation}\label{tnlm} 
	\begin{split} 
		&T^{n lm}_{\mu\nu}= t^{nlm}_{\mu\nu} a^\dagger_{n,l,m} a_{n, l, m} \\
		&t^{nlm}_{\mu\nu}=\left[\partial_\mu \psi_{n,l, m}  \partial_\nu \psi^*_{n,l, m} + \partial_\mu \psi^*_{n,l, m}  \partial_\nu \psi_{n,l, m}  - g_{\mu \nu} \left( g^{\alpha\beta} \partial_\alpha \psi_{n,l,m} \partial_\beta \psi_{n,l,m}^*  +M^2 |\psi_{n,l,m}|^2  \right) \right]\ .
	\end{split} 
\end{equation} 	
To evaluate the traces in \eqref{sten1}, we need to evaluate the sum over $k$ in \eqref{rhonlm}. As the operator in \eqref{tnlm} is just the number operator, it follows that 
\begin{equation}\label{nom} 
	T^{n lm}_{\mu\nu}  |k \rangle = k ~ t^{nlm}_{\mu\nu} | k \rangle \ .
\end{equation} 
Performing the sum over $k$, it follows that 
\begin{equation}\label{teon}
	\frac{ {\Tr } \left( \rho_{n l m} T^{n l m} _{\mu\nu} \right)}{Z_{nlm}}  = \frac{t^{nlm}_{\mu\nu}}{e^{\beta(\Delta +2n +l -\omega m)} -1} \ .
\end{equation} 
We conclude that 
\begin{equation}\label{tmunu} 
	\langle T_{\mu\nu} \rangle = \sum_{n, l, m} 
	\frac{t^{nlm}_{\mu\nu}}{e^{\beta(\Delta +2n +l -\omega m)} -1} \ .
\end{equation} 

\eqref{tmunu} has a simple physical interpretation. The expectation value of the stress tensor is simply the sum of the stress tensors of each of the individual modes of the scalar, weighted by the mode's bosonic thermal occupation number.

The change of variables $m=l-a$ turns \eqref{tmunu} into  
\begin{equation}\label{tmunucv} 
	\langle T_{\mu\nu} \rangle = \sum_{n=0}^\infty \sum_{l=0}^\infty \sum_{a=0}^{2l }  
	\frac{t^{nlm}_{\mu\nu}}{e^{\beta(\Delta +2n +a\omega +(1-\omega) l)} -1} \ .
\end{equation} 
In the limit $\omega \to 1$ of our interest, the summation in 
\eqref{tmunucv} receives its dominant contributions from $l$ of order $\frac{1}{\beta(1-\omega)}$. 
\footnote{We are working in the limit when $\omega \to 1$ at fixed $\beta$. All the approximations in this 
	section needs to be revisited if the temperature simultaneously scales to zero like a power of 
	$(1-\omega)$. In the physical context of this paper, however, the temperature of the gas equals the temperature of the black hole at its core. As we have seen in section \ref{kerr}, the temperature of such black holes is bounded from below by $\frac{1}{2 \pi}$, and so is never small.}
\footnote{As we will see below, $t_{\mu\nu}^{nlm}$ is proportional to $l^2$, which means stress tensor from higher $l$ modes is dominant, however, they are suppressed by the exponential in the denominator for generic values of $\omega$. However, when $\omega$ is close to unity, this suppression is suppressed, hence the stress tensor is peaked at values of $l$ which are as high as $(1-\omega)^{-1}$.} On the other hand, the summation receives its dominant contribution from values of the variables $n$ and $a$ that are of order unity. In the regime of interest, therefore, the upper limit in the summation
over $a$ in \eqref{tmunucv} is unimportant and can be dropped. 
\eqref{tmunucv} can thus be written as 
\begin{equation}\label{tummhh}
	\begin{split} 
		\langle T_{\mu\nu}\rangle&=\sum_{n=0}^\infty \sum_{a=0}^\infty  (T^{\rm eff}_{n a})_{\mu\nu} \\
		(T^{\rm eff}_{n a})_{\mu\nu}&= \sum_{l=0}^\infty \frac{t^{nlm}_{\mu\nu}}{e^{\beta(\Delta +2n +a\omega +(1-\omega) l)} -1} \ .
	\end{split} 
\end{equation} 

Let us estimate the stress tensor to the leading order in large $l$. It can be easily seen, that the leading order contribution in the stress tensor components will be from derivatives with respect to $t$ and $\phi$. For instance, taking the derivative of the term $\frac{e^\frac{-2l}{r^2}}{r^\Delta}$ with respect to $r$ gives rise to term which roughly goes like $\frac{l}{r^3}\frac{e^\frac{-2l}{r^2}}{r^\Delta}$. This term is peaked at $r\sim\sqrt{l}$, hence it is of the order $l^{-1/2}$. 
Similarly, the derivative with respect to $\theta$ is of the order $\sqrt{l}$. On the other hand, derivatives with respect to $t$ and $\phi$ are of the order $l$, and hence, the one which contributes the most to the stress tensor. 

Therefore, in leading order in $l$, only $T_{ij}$ where $i,j=t,\phi$ are significant. These terms are well approximated by 
\begin{equation}\label{stlight}
	t^{nla}_{tt}=-t^{nla}_{t\phi}= 	t^{nla}_{\phi \phi}=  2 
	|\psi_{n,l,l-a}|^2 l^2\equiv t^{nla}\ .
\end{equation}
\footnote{ This term comes entirely from the $\partial_\mu \phi \partial_\nu \phi$ part of the stress tensor.
	Note that the terms in $g^{\mu \nu} \partial_\mu \phi \partial_\nu \phi$ cancel each other at leading order:  this is simply because $k_\mu k^\mu =0$. Restated, the cancellation happens because the gas is purely right-moving (chiral).} 
It follows that, to leading order 
\begin{equation}\label{tummhhn}
	\begin{split} 
		\langle T\rangle&\equiv\langle T_{tt}\rangle=-\langle T_{t\phi}\rangle
		=\langle T_{\phi\phi}\rangle
		=\sum_{n=0}^\infty \sum_{a=0}^\infty  t^{n a} \\
		t^{n a}&= \sum_{l=0}^\infty \frac{2l^2|\psi_{n,l,l-a}|^2 }{e^{\beta(\Delta +2n +a\omega +(1-\omega) l)} -1} \ .
	\end{split} 
\end{equation} 
All other components of the stress tensor vanish to the leading order.

\subsubsection{Matching the total energy with thermodynamics} 

Before proceeding to evaluate the stress tensor in more detail, we pause to check that the integral of our 
stress tensor correctly captures the correct thermodynamical energy and angular momentum of our system.

The bulk energy and angular momentum are given in terms of the bulk stress tensor by the integral expressions 
\begin{equation}\label{bulkenerg} 
	E= -\int \sqrt{-g} g^{0 \nu} T_{\nu 0}, ~~~~~~~J= \int \sqrt{-g} g^{0 \nu} T_{\nu \phi}
\end{equation} 
where the integral is taken over a constant time slice. Since the background metric is diagonal, we find 
\begin{equation}\label{bulkenergsimp} 
	E= -\int \sqrt{-g} g^{0 0} T_{00}, ~~~~~~~J= \int \sqrt{-g} g^{0 0} T_{0 \phi}
\end{equation} 
Using the AdS metric \eqref{metriccord} at large $r$ and plugging in \eqref{tummhhn} we get 
\begin{equation}\label{endenslarnmm}
	E^{nal}= \sum_{nla} \int dr d\Omega    \frac{ 2 l^2 |\psi_{n, l, l-a}|^2}{{e^{\beta(\Delta +2n +a\omega +(1-\omega) l)} -1}}\ .
\end{equation} 
\footnote{As an aside we note that it is possible to work a bit more precisely. Instead of taking the large 
	$l$ limit from the beginning, we can more accurately work out the stress tensor. 
	Putting in the form of the AdS metric and working at large $r$ (i.e. using the Poincare path metric), the expression for the energy becomes
	\begin{equation}\label{endenslar}
		E= 2 \int dr d\Omega   \left( (\Delta + 2n +l)^2 + (l \delta \theta)^2 + M^2 r^2 \right)  |\psi_{n, l, l-a}|^2\ .
	\end{equation} 
	The first term in this expression comes from the term $\partial_0 \phi \partial_0\phi$. The second term in this 
	expression comes from $g^{\theta \theta} \partial_\theta \phi \partial_\theta \phi$. We note that 
	the terms in $g^{\mu \nu} \partial_\mu \phi \partial_\nu \phi$ cancel each other at leading order (this is 
	simply because $k_\mu k^\mu =0$). Also, $r$ derivatives in that expression are completely negligible at large $l$. 
	The term $(l^2 \delta \theta)^2$ is subleading, because $(\delta \theta)^2 \sim \frac{1}{l^2}$ and so can be ignored. Also, the term $M^2 r^2$ is also of order $l$ (because $r \sim \sqrt{l}$). This term is also, therefore, 
	subleading in comparison to the first term. We conclude that, at leading order
	\begin{equation}\label{endenslarn}
		E^{nal}= \sum_{nla}   \int dr d\Omega   \frac{ 2 (\Delta + 2n +l)^2|\psi_{n, l, a}|^2}{{e^{\beta(\Delta +2n +a\omega +(1-\omega) l)} -1}}
	\end{equation} 
	Making the approximation $(\Delta + 2n +l)^2 \rightarrow l^2$ gives us the answer in the main text. }

Using \eqref{fullnnorm}, together with the fact that $g^{00} \sqrt{-g}=-1$ (in the $r$ coordinate system at large $r$) we see that 
\begin{equation}
	\int \sqrt{-g}(-g^{00})dr d\Omega|\psi_{n,l,l-a}|^2\approx \frac{1}{2E_{nl}}\approx\frac{1}{2l}\ .
\end{equation}
So the energy in \eqref{endenslarnmm} evaluates to 
\begin{equation}\label{endenslarnn}
	E= \sum_{nla}  \frac{l}{{e^{\beta(\Delta +2n +a\omega +(1-\omega) l)} -1}}
\end{equation} 
i.e. precisely to the thermodynamical energy as expected. Also notice that since $T_{t\phi}=-T_{00}$, it can be easily seen from \eqref{bulkenergsimp}, that $E=J$ as we should expect from the chiral nature of the gas.

\subsubsection{Performing the summations} 

Reassured by the check of the previous subsubsection, we now proceed with our evaluation of the detailed local form of the stress tensor. 

Using the expressions presented in Appendices \ref{shll} and \ref{fnll}, we see that the mod squared wave function to the leading order in $l$ is:
\begin{equation}\begin{split}
		|\psi_{nla}|^2&\approx \frac{1}{2l}\frac{|\chi_{ln}|^2}{r^2}\frac{|\psi_a^{HO}(\delta\theta)|^2}{2\pi }\ .\\
	\end{split}	
\end{equation}
We have used the fact that in $\psi_{nla}=\frac{1}{\sqrt
	{2 E_{nl}}}F_{nl}(r)Y_{la}(\theta\phi)$, $E_{nl}$ is approximately $l$. We have also used \eqref{rhowfn} to express $F_{nl}$ in terms of the 
radial wave function for the three-dimensional harmonic oscillator,  $\chi_{ln}(r)$, defined in 
Appendix \ref{fnll}, and have used \eqref{ylm} to approximate the spherical harmonics. Finally, we have divided our answer by $2 \pi$ to account for the normalization coming from the integral over 
$\phi$. \footnote{We have been careful to use normalized wave functions for the dependence on 
	$r$ and $\theta$. Similarly, we should use the normalized constant wave function in the $\phi$ direction, i.e. $\frac{1}{\sqrt{2\pi}}$.}

It follows that 
\begin{equation}\begin{split}
		t^{nla}&=2l^2 |\psi_{nla}|^2\\
		&= \frac{l}{2\pi r^2}|\chi_{ln}(r)|^2 |\psi^a_{HO}(\delta\theta)|^2\ .
	\end{split}
\end{equation}

Now let us compute the total stress tensor which is the sum of stress tensors in each mode, weighted by the average occupation of that mode:
\begin{equation}
	\begin{split}
		T&=\sum_{nla}\frac{t_{nla}}{e^{\beta(\Delta+2n+a\omega)+\beta(1-\omega)l}-1}\\
		&=\sum_{nla}\sum_{q=1}^\infty t_{nla} e^{-q\beta\left((\Delta+2n+a\omega)+(1-\omega)l\right)}\\
		&=\sum_{q=1}^\infty \sum_{nla}\frac{l}{2\pi r^2}|\chi_{ln}(r)|^2 |\psi^a_{HO}(\delta\theta)|^2 e^{-q\beta\left((\Delta+2n+a\omega)+(1-\omega)l\right)}\ .
	\end{split}
\end{equation}
The sums over $n$ and $a$ are now easily performed using \eqref{propsum} and \eqref{qmtext}. We find 
\begin{equation}
	\begin{split}\label{bstt}
		T&=\sum_{q=1}^\infty\sum_l \frac{l}{2\pi r^2}e^{-q\beta l(1-\omega)}\sum_{n}|\chi_{ln}(r)|^2 e^{-l(2n+\Delta-\frac{1}{2})\left(\frac{q\beta}{l}\right)} \sum_{a}|\psi^a_{HO}(\delta\theta)|^2 e^{-l\left(a+\frac{1}{2}\right)\left(\frac{q\beta}{l}\right)}\\
		&=\sum_{q=1}^\infty\sum_l \frac{l}{2\pi r^2}e^{-q\beta l(1-\omega)}K_{RHO}\left(r,r,\frac{q\beta}{l}\right)K_{HO}\left(\delta\theta,\delta\theta,\frac{q\beta}{l}\right)\\
		&=\sum_{q=1}^\infty\int dl \frac{l^{\frac{5}{2}}}{r^3 (2\pi\sinh q\beta)^\frac{3}{2} }\exp{\left[-l\left(\left(\frac{\cosh q\beta-1}{\sinh q\beta}\right)\delta\theta^2+\frac{\cosh q\beta}{r^2 \sinh q\beta}+q\beta(1-\omega)\right)\right]}\\
		&\hspace{2cm}\times I_{\Delta-3/2}\left(\frac{l}{r^2 \sinh q\beta}\right)\ . \\
	\end{split}
\end{equation} 
On the first line of \eqref{bstt} we have replaced 
$(\Delta+2n+a\omega)$ by $(\Delta+2n+a)$ (as is appropriate in the limit $\omega \to 1$).  
/on the last line, we have substituted the explicit 
expressions for $K_{RHO}$ and $K_{HO}$, and also 
replaced the summation over $l$ by an integral over $l$; again this is justified in the limit 
$\omega \to 1$.

In order to proceed we substitute the identity
\begin{equation}\label{Iident} 
	I_{n}(x)= \frac{e^{x}\left(\frac{x}{2}\right)^n }{\Gamma(n+1)}{}_1F_1(n+1/2,2n+1,-2x)
\end{equation} 
(which expresses $I_{n}(x)$ as a product of an exponentially growing term and a term that decays 
at infinity like $x^{-n}$) to obtain 
\begin{equation}\label{bsstn}
	\begin{split} 
		T&=\sum_{q=1}^\infty\int dl \frac{l^{\frac{5}{2}}}{r^3 (2\pi\sinh q\beta)^\frac{3}{2} }\exp{\left[-l\left(\left(\frac{\cosh q\beta-1}{\sinh q\beta}\right)\left(\delta\theta^2+\frac{1}{r^2}\right)+q\beta(1-\omega)\right)\right]} \\
		&\ \ \ \ \left(\frac{l}{2r^2 \sinh q\beta}\right)^{\Delta-3/2}\frac{1}{\Gamma(\Delta-1/2)}{}_1F_1\left(\Delta-1,2\Delta-2,\frac{-2l}{r^2 \sinh q\beta}\right)\ .\\
	\end{split} 
\end{equation} 

The integral over $l$ in \eqref{bsstn} is a Laplace transform of the ${}_1 F_1$ function times a power law. The 
`frequency' of this Laplace transform is 
$$ \frac{u_q}{2 r^2 \sinh q \beta}$$
where
\begin{equation}
	u_q=2 x^2\left((\cosh q\beta-1)\left(\zeta^2+\frac{1}{x^2}\right)+ \beta q\sinh q\beta \right)
\end{equation}
and we have partially moved to the scaled variables 
$$x^2=r^2(1-\omega)\ ,\ \zeta^2=\frac{\delta\theta^2}{1-\omega}\ .$$

Happily, Mathematica can evaluate this Laplace transform. We find
\begin{equation}\label{bsstnn} 
	\begin{split} 
		&=\sum_{q=1}^\infty\int dl \frac{l^{\Delta+1}}{r^{2\Delta} \pi^{3/2}(2\sinh q\beta)^\Delta }\exp{\left(-l \frac{u_q}{2 r^2 \sinh q\beta}\right)} \\
		&\frac{1}{\Gamma(\Delta-1/2)}{}_1F_1\left(\Delta-1,2\Delta-2,\frac{-2l}{r^2 \sinh q\beta}\right)\\
		&=\sum_{q=1}^\infty \left(\frac{2 r^2\sinh q\beta}{u_q}\right)^{2+\Delta}\frac{\Gamma(2+\Delta)}{\Gamma(\Delta-1/2)} {}_2 F_1\left(\Delta-1,\Delta+2,2\Delta-2,-\frac{4}{u_q}\right)\frac{1}{r^{2\Delta} \pi^{3/2}(2\sinh q\beta)^\Delta }\\
		&=\sum_{q=1}^\infty \frac{1}{(1-\omega)^2 2^\Delta x^{2\Delta}}\frac{\sinh^2 q\beta}{\pi^{3/2}\left((\cosh q\beta-1)\left(\zeta^2+\frac{1}{x^2}\right)+ \beta q\sinh q\beta \right)^{\Delta+2}}\\
		&\frac{\Gamma(2+\Delta)}{\Gamma(\Delta-1/2)} {}_2 F_1\left(\Delta-1,\Delta+2,2\Delta-2,-\frac{4}{u_q}\right)\\
		&= \frac{4 x^4}{\pi^{3/2}(1-\omega)^2 }\frac{\Gamma(2+\Delta)}{\Gamma(\Delta-1/2)} \sum_{q=1}^\infty \frac{\sinh^2 q\beta}{u_q^{\Delta+2}} {}_2 F_1\left(\Delta-1,\Delta+2,2\Delta-2,-\frac{4}{u_q}\right)\ .
	\end{split}    
\end{equation}
In going from the first to the second line of \eqref{bsstnn} we have evaluated the Laplace transform. In going from the second to the third line in \eqref{bsstnn}, we have re-expressed our result entirely in terms of scaled variables, and have performed some convenient rearrangements. In going from the third to the fourth line of 
\eqref{bsstnn} we have further rearranged our final answer, to provide it in a form which 
is convenient for taking the small $x$ limit. 

The third and fourth lines of \eqref{bsstnn} 
are the final answer for the Hamiltonian evaluation of the stress tensor for the bulk gas in $AdS_4$. Our result is expressed in terms of 
the function 
\begin{equation}\label{functionnun}
	{}_2 F_1\left(\Delta-1,\Delta+2,2\Delta-2,-\frac{4}{u_q}\right)\ .
\end{equation} 
At small $u_q$, this function decays like $\sim u_q^{\Delta -1}$. At 
large $u_q$, the function tends to unity. 

It follows from the last line of \eqref{bsstnn} that the bulk stress tensor tends to zero like 
$x^4$ in the small $x$ limit; also that the stress
tensor goes to zero like $\frac{1}{\zeta^{2 \Delta +4}}$ in the large $\zeta$ limit. At large $x$, on the other hand,  the hypergeometric function
goes to a constant, and the third line of \eqref{bsstnn} tells us that the stress tensor 
scales like $\frac{1}{x^{2 \Delta}}$, as expected for any normalizable configuration. 

These results summarize that the bulk stress tensor is well localized around small $\zeta$ and away from small $x$, and is normalizable at infinity.

\subsection{Euclidean computation} \label{eucb}

In this subsection, we rederive the thermal expectation value of the bulk stress tensor as mentioned in \eqref{bsstnn}, this time from the  Euclidean path integral point of view. From this point of view, we would like to compute the one-point function of the stress tensor \eqref{stresstensor} around the background of the 
thermal AdS saddle. We obtain this expectation value by first evaluating the two-point function 
of the bulk scalar at separated points in the one-loop approximation about the appropriate Euclidean saddle point. We then take the derivatives that appear in \eqref{stresstensor} and finally take the coincident limit. The coincident limit includes 
a temperature-independent divergent piece. After 
subtracting away this term we find a well-defined answer (see below
\eqref{stono} for a more careful discussion of this point).

The procedure described above is easier to carry out than it may first appear. The two-point function in thermal AdS is given by a simple sum over images of zero temperature two-point functions
(see \cite{Alday:2020eua} for instance). Successive images  are separated by integral multiples 
of $(\beta,-i\beta \omega)$ in $(\tau^E,\phi)$ space, where $\tau^E$ is the Euclidean time. 
(The $i$ in the last expression is a consequence of the fact that the rotation operator is $e^{iJ_z}$ whereas the density matrix has $e^{\beta\omega J_z\phi}$.)
It follows that the two-point function in our Euclidean ensemble is given by 
\begin{equation}\label{images}	\langle\Phi(r_1,\theta_1,\tau^E_1,\phi_1)\Phi(r_2,\theta_2,\tau^E_2,\phi_2)\rangle_\beta=\sum_{q=-\infty}^{\infty} \langle \Phi(r_1,\theta_1,\tau^E_1,\phi_1)\Phi(r_2,\theta_2,\tau^E_2+q\beta,\phi_2-iq\beta\omega))\rangle_{0}\ .
\end{equation}

The zero temperature two-point functions that appear on the RHS of \eqref{images}  are functions of the ``chordal distance'' $u$ between their arguments in embedding space.  Explicitly in equations (see equation 4.6 in \cite{Alday:2020eua}).  
\begin{equation} \label{proplarge}
	\langle\Phi(x_1)\Phi(x_2)\rangle_0=C_\Delta \frac{1}{u^\Delta}{}_2F_1 (\Delta,\Delta-1,2\Delta-2,-\frac{4}{u})
\end{equation}
where 
\begin{equation}
	u=-2+ 2\left(\sqrt{(1+r_1^2)(1+r_2^2)}\cosh(\tau^E_1-\tau^E_2)-r_1 r_2 \left(\cos\theta_1 \cos\theta_2+ \sin\theta_1 \sin\theta_2  \cos(\phi_1-\phi_2)\right)\right)
\end{equation}
and 
\begin{equation}
	C_\Delta=\frac{\Gamma(\Delta)}{2\pi^{3/2}\Gamma(\Delta-1/2)}\ .
\end{equation}

In the method of images, we need to take $\tau_2-\tau_1=q\beta$, $\phi_2-\phi_1=-i q\beta\omega$, $\rho_1=\rho_2, \theta_1=\theta_2$ for the $q^{\rm th}$ image. Therefore, we have
\begin{equation}\label{chord}
	u_q=-2+ 2(-r^2\sin^2\theta \cosh(q\beta\omega)-r^2 \cos^2\theta  +(1+r^2) \cosh(q\beta))\ .
\end{equation}

Before proceeding with the computation we pause to  explain, from the Euclidean point of view, 
why the stress tensor is non-negligible at large $r$ only when $\omega \to 1$ and for values of $\theta$ very close to the equator. 

At large $r$ and generic values of $\theta$ the chordal distance $u_q$ between images is very large, and the propagator \eqref{proplarge} (and consequently its contribution to the bulk stress tensor) is very small. If we stick to large values of $r$, there is only one way to make the chordal distances small: that is to make the distances between images approximately lightlike. The chordal distance due to the temporal separation between neighboring images is spacelike and of magnitude $r \sqrt{\cosh \beta}$ (recall we are in Euclidean space). On the other hand, the distance due to the (imaginary) angular separation is timelike and of magnitude $r \sqrt{\cosh (\beta \omega)} \sin \theta $. The total chordal distance is small only if these two distances become equal in magnitude: this happens when $\omega$ is near unity and $\theta$ is also near $\frac{\pi}{2}$. We see this in equations as follows. When $\omega\to 1$, $|\tau^E_1-\tau^E_1|=|\phi_1-\phi_2|$. In that case, in the large $r$ limit of \eqref{chord}, the first and last terms compete with each other. Also if $\cos\theta$ (roughly of the order $r^{-1/2}$) is very small, the chordal distance can be made order unity. Therefore we see that even at large values of $r$, the chordal distance is of the order unity when $\omega\to 1$ and $\theta=\frac{\pi}{2}-\delta \theta$ with small $\delta\theta$. Written explicitly, in the $\omega\to 1$ limit and in terms of rescaled coordinates 
\begin{equation}\label{xr}
	\begin{split}
		x=\sqrt{1-\omega}\;r,\quad\quad \zeta=\frac{\delta\theta}{\sqrt{1-\omega}}\ ,
	\end{split}
\end{equation}
the chordal distance takes the following form:

\begin{equation}
	u_q=2x^2\left((\cosh q\beta-1)\left(\frac{1}{x^2}+\zeta^2\right)+q\beta\sinh q\beta\right)\ .
\end{equation}

Now the stress tensor is a function of derivatives of the fields as written in \eqref{stresstensor}. Therefore, to compute the thermal stress tensor, we need two derivatives of the two functions for all the images and then we need to sum over all the images except at $q=0$.  This is because, at $q=0$, the two insertion points coincide and hence give an infinite contribution. However, we find precisely the same divergence at zero temperature. Since we are interested in the magnitude of the stress tensor above the zero temperature value we simply discard this divergent term. This simple manoeuvre 
renormalizes our bulk stress tensor. 

We now proceed to compute the derivatives we need. Let us suppose that the two-point function takes the following form:
\begin{equation}
	\langle\Phi(x)\Phi(x_q)\rangle\equiv F(u_q)
\end{equation}
where $x_q$ is related to $x$ by shifts in $\tau$ and $\phi$ mentioned above. 
Taking derivatives with respect to the coordinates $\mu_1,\mu_2,$  we obtain
\begin{equation}
	K^q_{\mu_1\mu_2}\equiv\partial_{\mu_1}\partial_{\mu_2}\langle\Phi(x_1)\Phi(x_2)\rangle_m=F''(u)\frac{\partial u}{\partial\mu_1}\left. \frac{\partial u}{\partial \mu_2}\right|_{x_1=x,x_2=x_q}+ F'(u)\left. \frac{\partial^2 u}{\partial \mu_1\partial \mu_2}\right|_{x_1=x,x_2=x_q}\ .
\end{equation}

Then we rescale the coordinates as $r=\frac{x}{\sqrt{1-\omega}}$ and $\delta\theta=\sqrt{1-\omega}\zeta$ and take the $\omega\rightarrow 1$ limit. We can see that the dominant derivatives (of the order $\frac{1}{(1-\omega)^2}$) occur only from taking $\tau^E\tau^E$, $\phi\phi$, and $\tau^E\phi$ derivatives. The $r$ and $\theta$ derivatives are subdominant in the $\omega\to 1$ limit.

We list $\tau^E\tau^E$, $\phi\phi$ and $\tau^E\phi$ components defined as $K^q\equiv K^q_{\tau^E\tau^E}=i K^q_{\tau^E\phi}=-K^q_{\phi\phi}$:  
\begin{equation}\begin{split}
		-K^q&=\frac{\Gamma(\Delta)}{2\pi^{3/2}\Gamma(\Delta-1/2)}\frac{4 \Delta  x^4 u^{-\Delta -2}\sinh ^2\beta  m}{(\Delta -2) (1-\omega)^2} \left[-2 (\Delta -3) \, _2F_1\left(\Delta ,\Delta,2 (\Delta -1),-\frac{4}{u_q}\right)\right.\\
		&\hspace{0.5cm}\left.
		-\left(\Delta^2-3\Delta+4+\frac{8}{u}\right) \, _2F_1\left(\Delta ,\Delta +1;2 (\Delta -1);-\frac{4}{u}\right) \right]\\
		&= \frac{\sinh^2q\beta}{(1-\omega)^2}\frac{\Gamma(\Delta+2)}{2 \pi^{\frac{3}{2}}\Gamma(\Delta-1/2)}\frac{1}{2^\Delta x^{2\Delta}}\frac{{}_2F_1\left(\Delta-1,\Delta+2,2\Delta-2,\frac{-4}{u_q}\right)}{\left((\frac{1}{x^2}+\zeta^2)(\cosh q\beta-1)+ q\beta\sinh q\beta\right)^{\Delta+2}}
	\end{split}
\end{equation}
where in going from the first line to the second, we have used the following identity relating Hypergeometric functions:
\begin{equation}\begin{split}
		&  -2 (\Delta -3) \, _2F_1\left(\Delta ,\Delta,2 (\Delta -1),-\frac{4}{u}\right)+\left(\Delta^2-3\Delta+4+\frac{8}{u}\right) \, _2F_1\left(\Delta ,\Delta +1,2 (\Delta -1),-\frac{4}{u}\right)\\
		&=\, _2F_1\left(\Delta -1,\Delta +2;2 \Delta -2;-\frac{4}{u}\right)(\Delta -2) (\Delta +1)\ .
	\end{split}    
\end{equation}

Let us try to understand the scaling of these derivatives intuitively. As we explained above, in the $\omega\to 1$ limit, the chordal distance becomes lightlike, and hence the chordal distance is of order one even at large values of $r$. When we take the derivatives with respect to $\tau$ and $\phi$, we are going away from the lightlike ray and hence the chordal distance becomes large and hence $\tau$ and $\phi$ derivatives are large at large values of $r$ and small values of $\zeta$ (more precisely when $r\sim \frac{1}{\sqrt{1-\omega}}$ and $\delta\theta\sim \sqrt{1-\omega}$).

Now that we have the appropriate derivatives of the two functions, we can simply evaluate the thermal stress tensor by summing over all the images. The leading order components of the stress tensor are $T_{tt}=T_{\phi\phi}=-T_{t\phi}\equiv T$\footnote{Remember that we computed derivatives of the two-point function with respect to Euclidean time above which is $\tau_E=it$, however, stress tensor has derivatives with respect to the Lorentzian time}: 
\begin{equation}\begin{split} \label{stoo}
		T&=-2\sum_{q=1}^\infty K^q\\
		&=\sum_{q=1}^\infty \frac{\sinh^2q\beta}{(1-\omega)^2}\frac{\Gamma(\Delta+2)}{\pi^{\frac{3}{2}}\Gamma(\Delta-1/2)}\frac{1}{2^\Delta x^{2\Delta}}\frac{{}_2F_1\left(\Delta-1,\Delta+2,2\Delta-2,\frac{-4}{u_q}\right)}{\left((\frac{1}{x^2}+\zeta^2)(\cosh q\beta-1)+ q\beta\sinh q\beta\right)^{\Delta+2}}\\
		& \equiv \frac{f(x, \zeta)}{(1-\omega)^2}
	\end{split}
\end{equation}
which is exactly the answer that we obtained in \eqref{bsstnn}. 
(The factor of $2$ on the first line of \eqref{stoo} results from the fact that 
we have to sum over both positive as well as negative values of $q$.)

\subsection{Total energy from the integral of the stress tensor} 

In the previous two subsections, we computed the bulk stress tensor of the gas using two separate methods; both methods gave us the same final answer 
\eqref{stoo}. In this subsection, we perform a further consistency check of this answer. We compute the total energy by integrating the stress tensor over the bulk using \eqref{bulkenergsimp}, and match the result with the thermodynamic expectation mentioned in \eqref{endenslarnn}. We find perfect agreement. 

To proceed with this check, we first expand the exact thermodynamical answer \eqref{endenslarnn} in a power series expansion in $e^{-\beta \Delta}$. We accomplish this by Taylor expanding the denominator in \eqref{endenslarnn}. We then convert the sum over $l$ to an integral 
(as we have explained above, this is appropriate at small $1-\omega)$. At this stage the sum over $n$ and $a$ are geometric sums, and so very easily evaluated. We find:
\begin{equation}\label{toten}
	\begin{split}
		E&=\sum_{q=1}^\infty\sum_{na}\int dl~ l e^{-q\beta(\Delta+2n+a\omega+(1-\omega)l)}\\
		&=\sum_q \frac{1}{\beta^2 q^2(1-\omega)^2}\frac{e^{-q\beta\Delta}}{(1-e^{-2q\beta})(1-e^{-q\beta})}\\
		&=\frac{1}{(1-\omega)^2}\sum_q \frac{e^{q\beta(3/2-\Delta)}}{4q^2\beta^2\sinh q\beta \sinh\frac{q\beta}{2}}\ .
	\end{split}
\end{equation}

We will now reproduce \eqref{toten} by integrating our bulk stress tensor \eqref{stoo}. According to  \eqref{bulkenergsimp}
\begin{equation}\label{ene2}
	\begin{split}
		E=&\int \sqrt{-g} (-g^{00}) T_{00}=2\pi \int dr d\theta \sin\theta T_{00}\\
		\approx& 2\pi \int dx d\zeta T_{00}(x,\zeta)\\
		=&\sum_q \frac{2\pi\sinh^2q\beta \Gamma(\Delta+2)}{2^\Delta(1-\omega)^2 \pi^{3/2}\Gamma(\Delta-1/2)}\sum_{i=0}^\infty \frac{(-2)^i}{\Gamma(i+1)}\frac{\Gamma(\Delta-1+i)}{\Gamma(\Delta-1)} \frac{\Gamma(\Delta+2+i)}{\Gamma(\Delta+2)}\frac{\Gamma(2\Delta-2)}{\Gamma(2\Delta-2+i)} \int dx x^4\\
		&\int d\zeta \left(\frac{1}{(\cosh q\beta-1)(1+x^2\zeta^2)+q\beta \sinh q\beta}\right)^{2+\Delta+i} \\
		=&\sum_q \frac{2\pi\sinh^2q\beta \Gamma(\Delta+2)}{2^\Delta(1-\omega)^2 \pi^{3/2}\Gamma(\Delta-1/2)}\sum_{i=0}^\infty \frac{(-2)^i}{\Gamma(i+1)}\frac{\Gamma(\Delta-1+i)}{\Gamma(\Delta-1)} \frac{\Gamma(\Delta+2+i)}{\Gamma(\Delta+2)}\frac{\Gamma(2\Delta-2)}{\Gamma(2\Delta-2+i)}\\
		&\frac{\sqrt{\pi}\Gamma(\Delta+i+3/2)}{\Gamma(\Delta+i+2)}\frac{1}{\sqrt{\cosh q\beta-1}}\int dx \frac{x^3}{\left((\cosh q\beta-1)+q x^2 \sinh q\beta\right)^{\Delta+i+3/2}}\ .
	\end{split}
\end{equation}
In going from the second line to the third line, we have used the series expansion of the Hypergeometric function. In going from the third line to the fourth, we have performed the elementary integration over $\zeta$. Simplifying 
the expression presented in the final line of \eqref{ene2} gives 
\begin{equation}\begin{split} \label{ene3}
		E&= \sum_q\sum_{i} \frac{2\sinh^2q\beta}{2^\Delta(1-\omega)^2}\frac{4^{-i-1} \left(4 (\Delta +i)^2-1\right) \Gamma (2 (i+\Delta -1))}{\Gamma (i+1) \Gamma (i+2 \Delta -2)}  \\
		&\frac{(-2)^i}{\sqrt{\cosh q\beta-1}}\int dx \frac{x^3}{\left((\cosh q\beta-1)+q x^2 \sinh q\beta\right)^{\Delta+i+3/2}} \\
		&=\sum_q\frac{\sinh^2q\beta \tanh ^2\left(\frac{q\beta}{2}\right)}{2^\Delta(1-\omega)^2 q^2\beta^2}\frac{1}{(\cosh q\beta-1)^{\Delta+2}}\sum_i \frac{\Gamma (2 (i+\Delta -1))}{\Gamma (i+1) \Gamma (i+2 \Delta -2)}  \frac{1}{(-2(\cosh q\beta-1))^i}\\    
		&=\sum_q\frac{\sinh^2q\beta \tanh ^2\left(\frac{q\beta}{2}\right)}{2^{2\Delta+2}(1-\omega)^2 q^2\beta^2}\frac{1}{\left(\sinh \frac{q\beta}{2}\right)^{2\Delta+4}} ~{}_2F_1\left(\Delta-1,\Delta-1/2,2\Delta-2,-\frac{1}{\sinh^2\left(\frac{m\beta}{2}\right)}\right)\\
		&=\sum_q\frac{\sinh^2q\beta \tanh ^2\left(\frac{q\beta}{2}\right)}{2^{2\Delta+2}(1-\omega)^2 q^2\beta^2}\frac{1}{\left(\sinh \frac{q\beta}{2}\right)^{2\Delta+4}} \frac{2^{-3+2\Delta}e^{q\beta(3/2-\Delta)}}{\cosh \frac{q\beta}{2}\sinh^{2-2\Delta}\frac{q\beta}{2}}\\
		&=\frac{1}{(1-\omega)^2}\sum_q \frac{e^{q\beta(3/2-\Delta)}}{4 q^2\beta^2\sinh q\beta \sinh\frac{q\beta}{2}}\ .
	\end{split}
\end{equation}

In going from the first to the second line, we have performed the elementary integral over $x$.  In going from the second to the third line we have performed the sum over $i$ in Mathematica. This sum turns out to be a Hypergeometric function. In going from the third line to the fourth line we have used the following identity:
\begin{equation}
	{}_2F_1(a,a+1/2,2a,z)=\frac{1}{\sqrt{1-z}}\left(\frac{1}{2}+\frac{\sqrt{1-z}}{2}\right)^{1-2a}\ .
\end{equation}
The transition from the second last to the last line 
of \eqref{ene3} involves elementary algebraic simplifications. 

We see that the last line of \eqref{ene3} is in perfect agreement with the last line of \eqref{toten}, 
completing our verification. 

\subsection{Summary of this section}\label{sumsubsect} 

In this subsubsection, we summarize the results of this section that we will use in the rest of this paper. 
\begin{itemize} 
	\item To leading order in $1-\omega$, the bulk stress tensor of the gas takes the form 
	\begin{equation}\label{tik} 
		T_{tt}=-T_{t\phi}= T_{\phi\phi} =T=  \frac{f(x, \zeta)}{(1-\omega)^2} 
	\end{equation} 
	where $x$ and $\zeta$ are the scaled radial and angular variables defined in \eqref{xr} 
	and $f(x, \zeta)$ was obtained in \eqref{stoo}.
	\item In the small $x$ limit $u_q\simeq 2(\cosh q\beta-1)$,
	\begin{equation} \label{smallxih} 
		f(x)={\mathcal C}^{\beta, \Delta}_0 \;x^4 
	\end{equation} 
	where ${\mathcal C}^{\beta,\Delta}_0$ is
	\begin{equation}
		\begin{split}
			{\mathcal C}^{\beta, \Delta}_0 = \frac{\Gamma(\Delta+2)}{\pi^{\frac{3}{2}}\Gamma(\Delta-1/2)}\frac{1}{2^\Delta}\sum_{q=1}^\infty \sinh^2q\beta\; {}_2F_1\left(\Delta-1,\Delta+2,2\Delta-2,-\frac{2}{\cosh q\beta-1}\right)\ .
		\end{split}
	\end{equation}
	
	Note, in particular, that 
	$f(x)$ is independent of $\zeta$ in this limit. 
	\item In the large $x$ limit $u_q\simeq 2x^2\left((\cosh q\beta-1)\zeta^2+q\beta\sinh q\beta\right)$,
	\begin{equation}\label{largexih}
		f(x,\zeta) = \frac{{\mathcal C}_\infty^{\beta, \Delta}(\zeta)}{x^{2\Delta} }
	\end{equation} 
	where the function ${\mathcal C}_\infty^{\beta, \Delta}(\zeta)$ is,
	\begin{equation}
		\begin{split}
			{\mathcal C}_\infty^{\beta, \Delta}(\zeta) = \frac{4}{\pi^{3/2} }\frac{\Gamma(\Delta+2)}{2^{\Delta+2}\Gamma(\Delta-1/2)} \sum_{q=1}^\infty \frac{\sinh^2 q\beta}{\left((\cosh q\beta-1)\zeta^2+q\beta\sinh q\beta\right)^{\Delta+2}}\ .
		\end{split}
	\end{equation}
\end{itemize}

\section{Back-reaction on the metric}\label{bst}

In the previous section, we computed the bulk stress tensor due to the gas that fills thermal $AdS$. In this section, we will compute the backreaction of this stress tensor on the background AdS (or black hole) metric.

The stress tensor defined in the previous subsection is of order unity at values of $r$ that are of order unity. 
This is clear both intuitively (values of $r$ of order unity receive contributions only from $l \leq r^2 \Delta$, see Appendix \ref{fnll})
as well as from the third of \eqref{stresstensor}. A stress tensor of order unity induces a response in the metric of order $G$, i.e. a response that vanishes in the classical limit $G \to 0$. In other words, we should expect the solution sourced by this stress tensor to be classically indistinguishable from global 
$AdS$ (or the Kerr-AdS black hole, if we are studying the gas about this solution) at coordinates $r$ of order unity. On the other hand, 
we will see that the response to the bulk stress tensor `builds up' at larger values of $r$. In this section, we formulate and solve the equations that govern this behavior. 

\subsection{The equations in scaled coordinates} 

The backreaction of the bulk stress tensor becomes classically significant only at values of $r$ that 
are large enough to enclose a finite fraction of the total energy. It follows from \eqref{largexih} that this happens at $x$ of order unity, or $r$ of order $\frac{1}{\sqrt{1-\omega}}$. At these distances the bulk stress tensor is also highly localized in $\theta$ around $\frac{\pi}{2}$;  it will thus turn out that the backreaction to the metric is only significant over $\delta \theta$ of order $\sqrt{1-\omega}$, i.e. for coordinates $\zeta$ 
that is of order unity.  In order to compute the backreaction of the stress tensor on the $AdS$ metric, it is thus useful to work with the scaled coordinates \eqref{scaledcoord}. It is also useful to scale $t$ and $\phi$ in the same way as $\theta$. As the stress tensor is right-moving at the speed of light, it is particularly useful to define the left and right-moving linear combinations of these scaled coordinates 
\begin{equation}\label{newscale} 
	\sigma^+= \frac{t+\phi}{\sqrt{1-\omega}}, ~~~~\sigma^-= \frac{t-\phi}{\sqrt{1-\omega}}\ .
\end{equation} 
In these coordinates the background $AdS$ metric becomes 
\begin{equation}\label{metriccord2} \begin{split} 
		ds^2&= -(1+r^2) dt^2 + \frac{dr^2}{1+r^2} + r^2(d\theta^2 + \sin^2 \theta d \phi^2 )\\
		& \approx 		\frac{dr^2}{r^2} + r^2(d\theta^2 +  d \phi^2  -dt^2 )\\
		& = \frac{dx^2}{x^2} + x^2(d\zeta^2 - d\sigma^+ d \sigma^- )\ .\\
	\end{split} 
\end{equation}  
(In going from the first to the second line we have used the fact that $r$ is large and that $\theta$ is highly localized near $\frac{\pi}{2}$). 
The bulk stress tensor in these new coordinates is given by 
\begin{equation}\label{tsip} 
	T^{++}=T_{--} = (1-\omega) T= \frac{f(x, \zeta)}{(1-\omega)}\ .
\end{equation} 	
The Einstein equation takes the form 
\begin{equation}\label{eeqn0} 
	R_{\mu \nu }- \frac{g_{\mu \nu}R}{2}= 3 g_{\mu \nu}+ \frac{8 \pi G}{1-\omega} f(x,\zeta)   \delta_{\mu {\sigma}^-} \delta_{\nu { \sigma}^-}\ .
\end{equation} 	
As we have explained above, in the solutions of interest, $(1-\omega)^2 \sim G$. It follows that the quantity 
\begin{equation}\label{kappadef} 
	\kappa = \frac{8 \pi G}{1-\omega}
\end{equation} 
is parametrically small (it is of order $\sqrt{G}$). The Einstein equation can be rewritten as 
\begin{equation}\label{eeqn} 
	R_{\mu \nu }- \frac{g_{\mu \nu}R}{2}= 3 g_{\mu \nu}+ \kappa f(x,\zeta)   \delta_{\mu { \sigma}^-} \delta_{\nu {\sigma}^-}
\end{equation}
and we see that the correction to the background $AdS$ (sourced by the gas bulk stress tensor) is parametrically 
small. It follows that this correction can be accurately computed in the linearized approximation. 

In the rest of this paper, we work in Graham Fefferman gauge, i.e. we demand that $g_{xx}=\frac{1}{x^2}$  and $g_{xi}=0$ 
for $i=\zeta, \sigma^+, \sigma^-$. Always working to linearized order in $\kappa$, it is not difficult to demonstrate that  the solution to \eqref{eeqn} turns out to take the form  
\begin{equation}\label{gfform} 
	\frac{dx^2}{x^2} + x^2 \left( d\zeta^2 - d\sigma^+ d \sigma^- \right)	+ \kappa x^2 A(x,\zeta) d \sigma^- d \sigma^-
\end{equation} 
where the unknown function $A(x)$ obeys the sourced minimally coupled scalar equation  
\begin{equation} \label{Aeq} 
	\frac{1}{\sqrt{-g}}\partial_\mu\left(\sqrt{-g}g^{\mu\nu}\partial_\nu A\right)=
	\frac{-2f(x,\zeta)}{\sqrt{-g}}\ .
\end{equation}
The metric that appears on the LHS of \eqref{Aeq} is the unperturbed $AdS$ metric. Completely explicitly, 
the differential equation \eqref{Aeq} can be written as 
\begin{equation} \label{Aeq2} 
	\frac{\partial^2 A}{\partial \zeta^2} +\frac{\partial}{\partial x}\left({x^4}\frac{\partial}{\partial x}A(x,\zeta)\right)
	=	-2f(x,\zeta)\ .
\end{equation}

In the rest of this section, we will proceed to solve \eqref{Aeq2}. 

\subsubsection{Solving in Fourier space} 

In order to solve \eqref{Aeq2} we Fourier transform both the unknown $A$ and the source $f$ in $\zeta$ \footnote{Note that the periodicity of $\zeta$ is $\frac{2 \pi}{\sqrt{1-\omega}}$ and so is effectively infinite in the limit under study.} 
\begin{equation}\label{Aft} 
	\begin{split} 
		A(\zeta, x)&= \int \frac{dk}{\sqrt{2 \pi}} e^{i k \zeta} {A}_k(x)\\
		f_k(x)&=\int \frac{d\zeta}{\sqrt{2\pi}}e^{-ik\zeta}f(x,\zeta)\ . \\
	\end{split} 
\end{equation} 
$A_k(x)$ obeys the equation  
\begin{equation} \label{desfk} 
	-\frac{k^2}{2}{A}_k(x)+\frac{\partial}{\partial x}\left(\frac{x^4}{2}\frac{\partial}{\partial x}A_k(x)\right)=-f_k(x).
\end{equation}
Now the solutions to the homogeneous equation 
\begin{equation} \label{desfkhomo} 
	-\frac{k^2}{2}{A}_k(x)+2x^3\frac{\partial A_k}{\partial x}+\frac{x^4}{2}\frac{\partial^2 A_k}{\partial x^2}=0
\end{equation} 
are given by 
\begin{equation}\label{homosol} 
	B^+_k e^{\frac{k}{x}}\left(1- \frac{k}{x} \right) +  	B^-_k e^{-\frac{k}{x}}\left(1+ \frac{k}{x} \right)\ .
\end{equation} 
Using standard differential equation theory (reviewed in Appendix \ref{de}) we conclude that the most general solution 
to the sourced equation \eqref{desfk} is thus given by 
\begin{equation}\label{gensoldefk} \begin{split} 
		A_k(x)=	&B^+_k e^{\frac{k}{x}}\left(1- \frac{k}{x} \right) +  	B^-_k e^{-\frac{k}{x}}\left(1+ \frac{k}{x} \right)+ \\
		&\frac{1}{ k^3}  \left( e^{\frac{k}{x}}\left(1-\frac{k}{x}\right) \int dx' e^{\frac{-k}{x'}}\left(1+ \frac{k}{x'} \right) f_k(x')   -   e^{\frac{-k}{x}}\left(1+\frac{k}{x}\right) \int dx'e^{\frac{k}{x'}}\left(1- \frac{k}{x'} \right) f_k(x')   \right)\ . \\
	\end{split} 
\end{equation} 

\eqref{gensoldefk} has two undetermined integration constants. These constants are determined from the following 
considerations. First, we require that our solution be normalizable at $x=\infty$. Second, we require 
that it be regular at $x=0$. 
The first condition is enforced by taking all the integrals in the particular solution to run from $\infty$ to $x$, and choosing the homogeneous solution so that it is purely normalizable. In other words, it follows from the condition of normalizablity that 
\begin{equation}\label{solinf} \begin{split} 
		&K_n(k) \left( e^{\frac{k}{x}}\left(1- \frac{k}{x} \right) - e^{\frac{-k}{x}}\left(1+ \frac{k}{x} \right) \right)+ 	\\
		&  \frac{1}{ k^3}  \left( e^{\frac{k}{x}}\left(1-\frac{k}{x}\right) \int_x^\infty dx' e^{\frac{-k}{x'}}\left(1+ \frac{k}{x'} \right) f_k(x')   -   e^{\frac{-k}{x}}\left(1+\frac{k}{x}\right) \int_x^\infty dx' e^{\frac{k}{x'}}\left(1- \frac{k}{x'} \right) f_k(x') \right) \\
	\end{split} 
\end{equation} 
for an as yet undetermined constant $K_n(k)$. 

Using the fact that $f(x)$ decays, at large $x$, like $\frac{a}{x^{2 \Delta}}$, the  reader can convince herself that the term in the second line of \eqref{solinf} decays like $\frac{a}{x^{2 \Delta+2}}$ (the terms of order 
$\frac{1}{x^{2\Delta -1}}$, $\frac{1}{x^{2\Delta }}$ and $\frac{1}{x^{2\Delta +1}}$ all cancel).  For $\Delta> \frac{1}{2}$ -  i.e. for non-free fields that obey the unitarity bound -  this decay is more rapid than $\frac{1}{x^3}$.  

On the other hand the term in the first line of \eqref{solinf} (the term that multiplies $K_n(k)$) decays like $\frac{1}{x^3}$  at infinity. It follows that the solution \eqref{solinf} is normalizable and that its boundary stress tensor is proportional to $K_n(k)$ (of course Fourier transformed in $\zeta$).  

The as yet undetermined constant $K_n(k)$ is fixed by the requirement of good behaviour at $x=0$. At any fixed, 
nonzero $k$ the general solution \eqref{solinf} has a piece that diverges exponentially at $x=0$; we choose $K_n(k)$ to ensure that the coefficient of this offending divergence vanishes. We now explain in more detail 
how this works.

Let us take the case that $k$ is positive and first consider the integrals in the second line of \eqref{solinf}. In these expressions, the part of both integrals that comes from the neighbourhood of  $x=0$ 
gives expressions that are power law in $x$ (the exponentials in the prefactors cancel the exponentially large or small contributions from the integrals). In the first term on the second line of \eqref{solinf}, however, the contribution to the integral from generic (not small) values of $x$ gives rise to an exponentially growing term. This term must be cancelled by the exponentially growing term in the homogeneous solution. This is achieved if we choose $$K_n(k)=- \left( \frac{1}{ k^3} \int_0^\infty dx e^{\frac{-k}{x}}\left(1+ \frac{k}{x} \right) f_k(x)  \right) $$ 
so that the solution, for $k>0$, becomes 
\begin{equation}\label{solzero} \begin{split} 
		& A_k(x)= - \left( \frac{1}{k^3} \int_0^\infty dx'  e^{-\frac{k}{x'}}\left(1+ \frac{k}{x'} \right) f_k(x')\right) 
		\left( e^{\frac{k}{x}}\left(1- \frac{k}{x} \right) - e^{-\frac{k}{x}}\left(1+ \frac{k}{x} \right) \right) +	\\
		&  \frac{1}{k^3}  \left( e^{\frac{k}{x}}\left(1-\frac{k}{x}\right) \int_x^\infty dx' e^{\frac{-k}{x'}}\left(1+ \frac{k}{x'} \right) f_k(x')  -   e^{-\frac{k}{x}}\left(1+\frac{k}{x}\right) \int_x^\infty dx' e^{\frac{k}{x'}}\left(1- \frac{k}{x'} \right) f_k(x') \right)\ . \\
	\end{split} 
\end{equation}
Equivalently 
\begin{equation}\label{solzeronew} \begin{split} 
		& A_k(x)= \left( \frac{1}{ k^3} \int_0^\infty dx'  e^{\frac{-k}{x'}}\left(1+ \frac{k}{x'} \right) f_k(x')\right) 
		\left(e^{\frac{-k}{x}}\left(1+ \frac{k}{x} \right) \right) 	\\
		&  -\frac{1}{ k^3}  \left( e^{\frac{k}{x}}\left(1-\frac{k}{x}\right) \int_0^x dx' e^{\frac{-k}{x'}}\left(1+ \frac{k}{x'} \right) f_k(x')  +  e^{\frac{-k}{x}}\left(1+\frac{k}{x}\right) \int_x^\infty dx'  e^{\frac{k}{x'}}\left(1- \frac{k}{x'} \right) f_k(x')\right)\ . \\
	\end{split} 
\end{equation}
The properties of the solution near $\infty$ are best understood from the solution written in the form 
\eqref{solzero}. On the other hand, the properties of the solution near $x=0$ are best understood in the solution 
written in the form \eqref{solzeronew}. 

The solution for $k<0$ is obtained similarly. The analogue of \eqref{solzero} is 
\begin{equation}\label{solzeroneg} \begin{split} 
		& A_k(x)= -\left( \frac{1}{k^3} \int_0^\infty  e^{\frac{k}{x'}}\left(1- \frac{k}{x'} \right) f_k(x') \right) 
		\left( e^{\frac{k}{x}}\left(1- \frac{k}{x} \right) - e^{\frac{-k}{x}}\left(1+ \frac{k}{x} \right) \right) 	\\
		& + \frac{1}{ k^3}  \left( e^{\frac{k}{x}}\left(1-\frac{k}{x}\right) \int_x^\infty  e^{\frac{-k}{x'}}\left(1+ \frac{k}{x'} \right) f_k(x') -   e^{\frac{-k}{x}}\left(1+\frac{k}{x}\right) \int_x^\infty  e^{\frac{k}{x'}}\left(1-\frac{k}{x'} \right) f_k(x') \right) \\
	\end{split} 
\end{equation}

while the analogue of \eqref{solzeronew} is 
\begin{equation}\label{solzeronewneg} \begin{split} 
		& A_k(x)=  -\left( \frac{1}{k^3} \int_0^\infty  e^{\frac{k}{x'}}\left(1- \frac{k}{x'} \right) f_k(x') \right) 
		\left(e^{\frac{k}{x}}\left(1- \frac{k}{x} \right) \right) 	\\
		& + \frac{1}{k^3}  \left( e^{\frac{k}{x}}\left(1-\frac{k}{x}\right) \int_x^\infty  e^{\frac{-k}{x'}}\left(1+ \frac{k}{x'} \right) f_k(x') + e^{\frac{-k}{x}}\left(1+\frac{k}{x}\right) \int_0^x  e^{\frac{k}{x'}}\left(1-\frac{k}{x'} \right) f_k(x')\right)\ . \\
	\end{split} 
\end{equation}

Our final solution for $A(\zeta, x)$ is obtained by 
substituting \eqref{solzero} and \eqref{solzeroneg} 
(equivalently \eqref{solzeronew} and \eqref{solzeronewneg} ) into \eqref{Aft}.

For a value of $k$ that could be of either sign, the generalization of \eqref{solzero} is 
\begin{equation}\label{solzerog} \begin{split} 
		& A_k(x)= - \left( \frac{1}{k^3} \int_0^\infty  e^{-\frac{|k|}{x'}}\left(1+ \frac{|k|}{x'} \right) f_k(x')\right) 
		\left( e^{\frac{k}{x}}\left(1- \frac{k}{x} \right) - e^{-\frac{k}{x}}\left(1+ \frac{k}{x} \right) \right) +	\\
		&  \frac{1}{k^3}  \left( e^{\frac{k}{x}}\left(1-\frac{k}{x}\right) \int_x^\infty  e^{\frac{-k}{x'}}\left(1+ \frac{k}{x'} \right) f_k(x')  -   e^{-\frac{k}{x}}\left(1+\frac{k}{x}\right) \int_x^\infty  e^{\frac{k}{x'}}\left(1- \frac{k}{x'} \right) f_k(x') \right) \\
	\end{split} 
\end{equation}
while the analogue of \eqref{solzeronew} is 
\begin{equation}\label{solzeronewg} \begin{split} 
		& A_k(x)= \left( \frac{1}{ |k|^3} \int_0^\infty  e^{\frac{-|k|}{x'}}\left(1+ \frac{|k|}{x'} \right) f_k(x')\right) 
		\left(e^{\frac{-|k|}{x}}\left(1+ \frac{|k|}{x} \right) \right) 	\\
		&  -\frac{1}{ |k|^3}  \left( e^{\frac{|k|}{x}}\left(1-\frac{|k|}{x}\right) \int_0^x  e^{\frac{-|k|}{x'}}\left(1+ \frac{|k|}{x'} \right) f_k(x')  +  e^{\frac{-|k|}{x}}\left(1+\frac{|k|}{x}\right) \int_x^\infty   e^{\frac{|k|}{x'}}\left(1- \frac{|k|}{x'} \right) f_k(x')\right). \\
	\end{split} 
\end{equation}

The expression \eqref{solzeronew} appears to blow up as $|k| \to 0$, however, this divergence is illusory. In order to see this we use the following two approximations:
\begin{equation} \label{solapp} 
	e^{\mp\frac{k}{x}}\left(1\pm \frac{k}{x}\right)f_k(x)\approx f_0(x)+k f_1(x)+\frac{k^2}{2}\left(f_2(x)-\frac{f_0(x)}{x^2}\right)+\frac{k^3}{6}\left( f_3(x)-3\frac{f_1(x)}{x^2}\pm 2\frac{f_0(x)}{x^3} \right)+\cdots.
\end{equation} \footnote{Here, $$f_n(x) = \frac{d^n}{dk^n}\left. f_k(x)\right|_{k=0}$$}

Inserting \eqref{solapp} into \eqref{solzeronew}, it is easily verified that the integrand in this expression is actually regular as $k\to 0$. Concretely 
\begin{equation}\begin{split} \label{smallxmet} 
		A_0(x)&=\frac{1}{3}\left[\int_0^\infty \frac{f_0(x')}{x'^3}+\frac{1}{x^3}\int_0^\infty f_0(x')-\int_0^x \frac{f_0(x')}{x'^3}+\frac{1}{x^3}\int_0^x f_0(x')\right.\\
		&\left.+\int_x^\infty \frac{f_0(x')}{x'^3}-\frac{1}{x^3}\int_x^\infty f_0(x')\right]\\
		&=\frac{2}{3}\left[\int_x^\infty \frac{f_0(x')}{x'^3}+\frac{1}{x^3}\int_0^x f_0(x')\right]\ .\\
	\end{split}
\end{equation}
\eqref{smallxmet} naively appears to be singular as $x \to 0$, but once 
again this singularity is illusory. Using the fact that $f_0(x)\sim x^3$ in the small $x$ limit, we see that the second term in the last line of \eqref{smallxmet} actually scales like $x$. The first term on the same line scales like a constant, and so gives the dominant small $x$ 
contribution to $A_0$
\begin{equation}\label{smallxmets}
	A_0(x) \approx \frac{2}{3} \int_x^\infty \frac{f_0(x')}{x'^3}\ .
\end{equation}

\subsection{Metric correction in position space}\label{metpos}

In the previous subsection, we computed the correction to the metric 
in Fourier space in $\zeta$. In this subsection, we inverse Fourier transform to come back to the position space.   

Our method is simple. We insert our solution for $A_k(x)$, 
\eqref{solzerog}, into the definition 
\begin{equation}\label{defft}
	A(x,\zeta) = \int_{-\infty}^\infty \frac{dk}{\sqrt{2\pi}} A_k(x) e^{ik\zeta}\ .
\end{equation}
We then substitute the second of \eqref{Aft} into the resultant expression. At this stage our expression for $A(x, z)$ contains 
three integrals; the integral over $k$ in \eqref{defft}, the integral over $x'$ in \eqref{solzerog}  and the integral over $\zeta'$ from the second of \eqref{Aft} (we use the symbol $\zeta'$ for the dummy integration variable that appears in the second of \eqref{Aft}). The integrals over $x'$ and $\zeta'$ are weighted by the complicated 
function $f(x', \zeta')$, and so are difficult to actually perform. 
On the other hand, the integral over $k$ is weighted only by the 
elementary functions that appear in \eqref{solzerog} and the second of 
\eqref{Aft}. As a consequence, the integral over $k$ is easily performed. We relegate the details of this integral to Appendix \ref{kernel}. Performing this integral gives us an expression of the 
final form

\begin{equation}\label{metsol}
	\begin{split}
		A(x,\zeta) =& \int_{-\infty}^\infty \frac{d\zeta'}{2\pi}  \int_0^\infty dx' f(x',\zeta') {\mathcal{K}}_{x,\zeta}(x',\zeta')
	\end{split}
\end{equation}
where (as we show in Appendix \ref{kernel})
\begin{equation}\label{kern1}
	\begin{split}
		{\mathcal{K}}_{x,\zeta}(x',\zeta') & =\frac{1}{2} \left(\left(\zeta -\zeta '\right)^2+\frac{1}{x^2}+\frac{1}{x'^2}\right) \log \left[\frac{\left(\zeta -\zeta '\right)^2 +\left(\frac{1}{x}+\frac{1}{x'}\right)^2}{\left(\zeta -\zeta '\right)^2 +\left(\frac{1}{x'}-\frac{1}{x}\right)^2}\right] -\frac{2}{x x'}\ .
	\end{split}
\end{equation}
\eqref{metsol} and \eqref{kern1}, together with the formula for 
$f(x', \zeta')$ presented in \eqref{stoo}, may be regarded as our final 
result for the correction of the $AdS$ metric in response to the bulk gas. 

The function ${\mathcal{K}}_{x,\zeta}(x',\zeta')$ is an effective bulk-to-bulk Greens function for the operator that appears in \eqref{Aeq2}. 
It is easy directly check that for $x'\neq x$, 
\begin{equation} \label{Aeq3} 
	\frac{\partial^2 {\mathcal{K}}_{x,\zeta}(x',\zeta')}{\partial \zeta^2} +\frac{\partial}{\partial x}\left({x^4}\frac{\partial {\mathcal{K}}_{x,\zeta}(x',\zeta')}{\partial x}\right)
	=0
\end{equation}
and, more generally that it obeys 
\begin{equation} 
	\frac{\partial^2 {\mathcal{K}}_{x,\zeta}(x',\zeta')}{\partial \zeta^2} +\frac{\partial}{\partial x}\left({x^4}\frac{\partial {\mathcal{K}}_{x,\zeta}(x',\zeta')}{\partial x}\right)
	=-4\pi\delta(x-x')\delta(\zeta-\zeta')
\end{equation}
\footnote{One can check this equation using the following property in 2d $(\zeta,y)$ plane, we know, $\nabla_{\rm flat}^{'2}\log\left[(\zeta-\zeta')^2+(y-y')^2\right]=4\pi\delta(\zeta-\zeta')\delta(y-y')$. Then use the coordinate transform, $y=\frac{1}{x}$.}
as expected.

\subsection{The large $x$ limit}

In this subsection, we compute the correction to the metric 
in the large $x$ limit, and use our result to obtain the boundary 
stress tensor.

\subsubsection{Metric correction in the large $x$ limit} 

In large $x$ limit the Kernel ${\mathcal{K}}_{x,\zeta}(x',\zeta')$ has the following form,
\begin{equation}\label{kernlrge}
	\begin{split}
		{\mathcal{K}}_{x,\zeta}(x',\zeta') =& \frac{8}{3x^3}\frac{x'}{\left( x'^2(\zeta-\zeta')^2+1\right)^2}+\mathcal{O}\left(\frac{1}{x^4}\right)
	\end{split}
\end{equation}

Putting \eqref{kernlrge} in \eqref{metsol} at large $x$ we get,
\begin{equation} \label{largexx}
	\begin{split}
		A(x,\zeta) &= \frac{8}{3x^3} \int_{-\infty}^\infty \frac{d\zeta'}{2\pi}  \int_0^\infty    \frac{dx'\,x'\, f(x',\zeta')}{\left( x'^2(\zeta-\zeta')^2+1\right)^2}
	\end{split}
\end{equation}
As an algebraic check of \eqref{largexx} we compute its Fourier
transform in $\zeta$ to find
\begin{equation}\label{largesoln} \begin{split}
		A_k(x) 	&= \frac{1}{x^3} \times  \left( \frac{2}{3} \int_0^\infty  e^{\frac{-k}{x'}}\left(1+ \frac{k}{x'} \right) f_k(x')\right)~~~~~~~~~~~~~~~k>0 \\
		&=\frac{1}{x^3} \times  \left( \frac{2}{3} \int_0^\infty  e^{\frac{k}{x'}}\left(1- \frac{k}{x'} \right) f_k(x') \right)~~~~~~~~~~~~~~~k<0\ .
	\end{split}
\end{equation} 
This result exactly matches the large $x$ limit of \eqref{solzero}.

\subsubsection{Boundary stress tensor} \label{bdrstr}

Moving back to unscaled coordinates, we find that the normalizable tail of the metric at infinity is given by 
\begin{equation}\label{expeq}
	\frac{(dt -d\phi)^2}{1-\omega} \times  \frac{\frac{8}{3}\int _0^{\infty }\frac{dx'}{2 \pi }\int \frac{\text{d$\zeta $}' x' f\left(x',\zeta '\right)}{\left(x'^2 (\zeta -\zeta' )^2+1\right)^2} }{r\sqrt{1-\omega}}\ .
\end{equation} 

The boundary stress tensor is then given by multiplying the coefficient of $\frac{1}{r}$ by $\frac{3}{16\pi G}$ \cite{deHaro:2000vlm}. Therefore, for the boundary stress tensor we get
\begin{equation}\label{stint} 
	T_{t t}=\frac{\kappa}{2\pi G (1-\omega)^\frac{3}{2}}\int_0^\infty \frac{dx'}{2\pi}\int d\zeta'\frac{x' f(x',\zeta')}{(x'^2(\zeta-\zeta')^2+1)^2}\ .
\end{equation}
From \eqref{bulkenerg} we see that the total boundary energy is given by 
\begin{equation}\label{totbe} 
	\int d \phi d \theta \sin^2 \theta 	T_{t t}\ .
\end{equation} 
However, the stress tensor is peaked about $\zeta=0$ and has an `order one' width in $\zeta$ (more details below). It follows that the stress tensor is highly peaked around $\theta=\frac{\pi}{2}$, and so $\sin^2\theta$ 
in \eqref{totbe} can just be set to unity, and \eqref{totbe} can be rewritten by changing integration variables (from $\theta$ to $\zeta$) as 
\begin{equation}\label{totben} 
	E= \sqrt{1-\omega} \int d \phi d \zeta	T_{t t}\ .
\end{equation}
Inserting \eqref{stint} into \eqref{totben} we find 
\begin{equation}\label{totbenn} 
	E=  \int d \phi d \zeta	\frac{4}{(1-\omega)^2}\int_0^\infty \frac{dx'}{2\pi}\int d\zeta'\frac{x' f(x',\zeta')}{(x'^2(\zeta-\zeta')^2+1)^2}\ .
\end{equation}
The integral over $\zeta$ is easily performed, and we find
\begin{equation}\label{totbennn} 
	E=  	\frac{1}{(1-\omega)^2} \int d \phi dx' 
	d\zeta' f(x',\zeta')
\end{equation} 
in perfect agreement with the bulk expression for the energy of the gas. 

The expression \eqref{stint} can also be used to estimate how peaked the stress tensor is about $\theta= \frac{\pi}{2}$. It follows from \eqref{stint} that at large values of $\zeta$, 
\begin{equation}\label{stint2} 
	T_{t t}=\frac{4}{(1-\omega)^\frac{5}{2}} \times \frac{1}{\zeta^4} \int_0^\infty \frac{dx'}{2\pi}\int d\zeta'\frac{f(x',\zeta')}{x'^3}\ .
\end{equation}
It follows that the fall off of $T_{tt}$ with $\delta \theta$ is of the form 
$$ T_{tt} \sim \frac{1}{\sqrt{1-\omega} (\delta \theta)^4} \ .$$
It follows that the energy contained at values of $\theta$ s.t. $|\frac{\pi}{2}-\theta|< (\delta \theta)_0$ 
is of order
$$ \frac{1}{\sqrt{1-\omega} (\delta \theta)_0^3}\ .$$
Provided that $(\delta \theta)_0$ is held fixed as $\omega \to 1$, it follows that this energy tail is subleading by a factor $(1-\omega)^\frac{3}{2}$ relative to the total energy. It follows, in other words, that 
the gas contribution to the boundary stress tensor is $\delta$ function localized about the equator 
$\theta = \frac{\pi}{2}$.

\subsection{The small $x$ limit}

It is easy to verify that when $x$ is small 

\begin{equation}\label{kernsm1}
	\begin{split}
		\mathcal{K}_{x,\zeta}(x',\zeta')  \approx  \frac{8 x}{3 \left(\zeta^2 x^2+1\right)^2 } \frac{1}{{x'}^3}\ .
	\end{split}
\end{equation}
\footnote{ The term $\frac{1}{\left(\zeta^2 x^2+1\right)^2 }$ in  \eqref{kernsm1} is well approximated by unity when $\zeta$ is either small or of order unity, but not when $\zeta$ is of order $1/x$, a fact that will become important below. }

It follows that the metric at small $x$ (and arbitrary $\zeta$) is given by 
\begin{equation}
	\begin{split} \label{smallx}
		A(x,\zeta)=&\int_{-\infty}^\infty \frac{d\zeta'}{2\pi}  \int_0^\infty dx' f(x',\zeta') \mathcal{K}_{x,\zeta}(x',\zeta')\\
		=& \frac{8 x}{3 \left(\zeta^2 x^2+1\right)^2 } \int_{-\infty}^\infty \frac{d\zeta'}{2\pi}  \int_0^\infty \frac{dx'}{{x'}^3} f(x',\zeta')\\
		=& \frac{c_0 \,x}{\left(\zeta^2 x^2+1\right)^2 }
	\end{split}
\end{equation}
where
\begin{equation} \label{cssmallx}
	\begin{split}
		c_0= \frac{8}{3\sqrt{2\pi}} \int_0^\infty \frac{dx'}{{x'}^3} f_0(x')\ .
	\end{split}
\end{equation}
Note this $A(x, \zeta)$ is zero at $x=0$ and increases linearly with $x$ at the leading order.

As an algebraic check of \eqref{smallx}, we use this formula to compute the Fourier transform (at zero frequency) of $A(x, \zeta)$ at small $x$. We find 
\begin{equation}\label{A0lim}
	\begin{split}
		A_0(x)=&\int_{-\infty}^\infty \frac{d\zeta}{\sqrt{2\pi}}A(x,\zeta)\\
		=&\int_{-\infty}^\infty \frac{d\zeta}{\sqrt{2\pi}} \frac{c_0 \,x}{\left(\zeta^2 x^2+1\right)^2 }\\
		=&\frac{\sqrt{\pi}}{2\sqrt{2}} c_0\\
		=& \frac{2}{3}   \int_0^\infty \frac{dx'}{{x'}^3 }  f_0(x')
	\end{split}
\end{equation}
in perfect agreement with the \eqref{smallxmets}, the small $x$ limit of \eqref{smallxmet} 

\subsection{Large $\zeta$ limit}

Let us see the behaviour of the metric at large $\zeta$. In this limit, the kernel takes the following form:
\begin{equation}\label{lzeta}
	K_{x,\zeta}(x',\zeta')\approx \frac{8}{3 \zeta^4 x^3 x'^3}\ .
\end{equation}

Integrating this with the source, we get the metric correction at large $\zeta$ as:
\begin{equation}\label{mlzeta}
	A(x,\zeta)= \frac{8}{3  x^3\zeta^4} \int_0^\infty \frac{dx'}{x'^3}f_0(x') = \frac{\sqrt{2\pi} c_0 }{x^3\zeta^4}.
\end{equation}
Note, in particular, that $A(x, \zeta)$ decays rapidly for large $\zeta$. It follows that the correction to the $AdS$ metric is everywhere sharply localized around the equator of the $S^2$. The constant $c_0$ is defined in \eqref{cssmallx}.

\subsection{Summary of our final answer for the bulk metric}

It follows from the discussion of the previous subsubsections that our final answer for the bulk metric of the Grey Galaxy solutions is given as follows. 
\begin{itemize} 
	\item For $r \ll \frac{1}{\sqrt{1-\omega}}$
	the metric is given by a Kerr-AdS metric \eqref{bsol} with $\omega=1$, i.e. by a Kerr-AdS black hole with the paremeter $m$ given by \eqref{eqsr}. The free parameter $a$ of these solutions forms one of the parameters of the Grey Galaxy saddles. 
	\item At $r \gg 1$ our solution is given by the pure $AdS$ metric (the second of \eqref{metriccord2})  plus two small correction terms $\delta_1 ds^2$ and $\delta_2 ds^2$. 
	$\delta_1 ds^2$ is simply the tail of the black hole solution i.e.
	\begin{equation}\label{deltadssq}
		\delta_1  ds^2 = \frac{2 m}{r} \frac{\left( dt -a \, \sin ^2\theta \, d\phi \right)^2}{(1-a^2 \cos^2 \theta)^{5/2}}
	\end{equation} 
	(see \eqref{bhtail}). When this tail is written in terms of the coordinates $x$, $\zeta$, $\sigma^\pm$, this term in the  
	metric is of order $(1-\omega)^\frac{3}{2}$.
	\item The second of these corrections is the 
	metric response to the gas, and is given by 
	\begin{equation} \begin{split} \label{finsol} 
			&\delta_2 ds^2 = \kappa x^2 A(\zeta, x) d \sigma^- d \sigma^- \\
		\end{split} 
	\end{equation} 
	where $A(\zeta, x)$ is given in \eqref{metsol}. 
	The coordinates used in \eqref{finsol} are defined in terms of $r, \theta, t, \phi$
	by \eqref{xr} and \eqref{newscale}. Note that $\delta_2 ds^2$ is proportional to  $\kappa$ defined in \eqref{kappadef}, and so is of order $1-\omega$. Note that $\delta_2 ds^2$ is parametrically larger than $\delta_1 ds^2$ (by a factor of $\frac{1}{\sqrt{1-\omega}}$) when measured pointwise. However, while  
	$\delta_1 ds^2$ is nonzero over a range of order unity in $\theta$, 
	$\delta_2 ds^2$ is nonzero over a $\theta$ range of order $\sqrt{1-\omega}$. As a consequence, these two metric contributions correspond to stress tensors that carry total boundary energies of the same order.
	While the stress tensor from $\delta_1 ds^2$ is smoothly spread over the boundary sphere, the stress tensor from $\delta_2 ds^2$ is $\delta$ function localized on the boundary sphere.
\end{itemize}

\section{Comparison with other solutions} \label{chesler} 

\subsection{Comparison with numerical evolutions}

Consider a Kerr-AdS black hole with parameters
\cite{Chesler:2021ehz}
\begin{equation}
	m=0.2375,\quad a=0.2177\ .
\end{equation}
The reader can verify that this black hole lies in the shaded blue region of Fig. \ref{extvsunstab} (it sits on the black point in that diagram)  and so is superradiant unstable. In an interesting paper
\cite{Chesler:2021ehz}, Chelser numerically perturbed this black hole and followed its subsequent evolution. He constructed two different numerical solutions, starting with distinct initial conditions. On both occasions he found that the solution initially evolved rapidly in time, settling down into ever slower evolution at later times until the simulation was halted. Chesler observed that his two initial conditions did not evolve to the same final configuration, but that their two entropies (at the time of stopping the simulation) were very similar. In particular, the horizon entropy, in each simulation, had increased by about $11.2 \%$. 

In this section, we will examine the implications of these findings for our proposal for the endpoint of the superradiant instability of Kerr-AdS black holes. 
\begin{itemize}
	\item Let us first analyze our proposed end point of this instability quantitatively. Using \eqref{outerhor} we find that the original Kerr-AdS black hole had an outer horizon radius given by 
	\begin{equation}
		r_+ = 0.2650\ .
	\end{equation}
	Then using \eqref{chargeshh}, it follows that the scaled charges and the chemical potentials of this black hole are given by
	\begin{equation}
		\begin{split}
			\epsilon = 0.26172,\quad j = 0.05698, \quad s = 0.12345,\quad \omega = 1.9810.
		\end{split}
	\end{equation}
	As $\omega > 1$, the black hole is the superradiant unstable.
	Now using \eqref{chargessr}, it is a simple matter to compute the parameters of the black hole that sits at the centre of the Grey Galaxy solution with the same charges: these are given by  
	\begin{equation}
		\begin{split}
			\omega = 1,\quad a = 0.1276, \quad \epsilon = 0.2346,\quad s = 0.1462, \quad j = 0.0299.
		\end{split}
	\end{equation}
	Note that the increase in entropy of the Grey Galaxy solution over the initial black hole is by $18.4\%$. It follows that our Grey Galaxy black hole {\it could} be the eventual end point of Chesler's evolution, as it has a higher entropy than the configuration reached at the time of 
	stopping the simulation. 
	\item The fact that Chesler's two different initial 
	conditions did not approach the same configuration is also easily explained within our framework. Recall that the Grey Galaxy solution represents a thermodynamical ensemble of configurations rather than a particular field configuration. While we have argued that the `coarse-grained metric' (i.e. metric in scaled coordinates) and, particularly, the boundary stress tensor has small fluctuations, our solution still represents an ensemble, and
	we should not expect two different initial conditions to reach precisely the same field configuration at any given finite time. 
	\item When Chesler stopped his simulation, his configurations were continuing to evolve slowly. The rate of evolution was slower at the end than at the beginning of the simulation. This fact matches the observation (see the introduction) 
	that the time scale for superradiant decays of large $l$ modes diverges exponentially with $l$; 
	as a consequence, we should expect the time scale for settling into the final Grey Galaxy saddle 
	to be exponentially large. 
\end{itemize} 

In summary, while the intriguing detailed results 
of \cite{Chesler:2021ehz} certainly do not 
concretely verify our conjecture, they do not 
contradict it either. In fact, the observations of 
\cite{Chesler:2021ehz} appear to us to be in 
broad general agreement with one quantitative 
aspect (the entropy increase) and a couple of 
qualitative aspects (ensembles and slow evolution) 
of the picture spelt out in this paper. It would certainly be interesting to have further concrete 
evidence - either for or against our conjecture for Grey Galaxies as endpoints of the superradiant instability of Kerr-AdS black holes. 

\subsection{Comparison with Black Resonators} 

We also compare our solutions with 
the black resonators in AdS$_4$ \cite{Dias:2015rxy,Cardoso:2013pza}. Black resonators are 
solutions whose charges are below the $\omega=1$ (red) line of Fig. \ref{kerr-ads4}. 
The black resonators by themselves have extra instabilities but have larger entropies than 
the unstable Kerr-AdS black holes at the same charges if they exist in the blue region of Fig. \ref{kerr-ads4} \cite{Dias:2015rxy}. If our Grey Galaxies are the final states at given
charges $E,J$ below the $\omega=1$ line of Fig. \ref{kerr-ads4},
perturbating black resonators should also evolve to our solutions. So the Grey Galaxies 
should have larger entropies than all known black resonator solutions at the same charges. 
To test this, we compare the entropies of the black resonators reported in 
\cite{Dias:2015rxy,Cardoso:2013pza} with those of our solutions at the same charges.

We consider the case in which $E$ and $J$ are small, meaning that they 
are given by $G^{-1}$ times a parametrically small number independent of $G$. (Still, $E,J$ 
are at $G^{-1}$ order so we can rely on the two-derivative gravity.) In this case,
\cite{Cardoso:2013pza} found an analytic expression for the entropy of small black resonators for 
the `scalar mode' in the metric perturbation with $l=m$, based on earlier works on 
the `non-interacting mix' picture of hairy black holes \cite{Basu:2010uz,Bhattacharyya:2010yg,Dias:2011tj}. 
The expression is given by
\begin{equation}\label{small-resonator}
	s_{\rm br}=4\left[\epsilon-(1+m^{-1})j\right]^2\ .
\end{equation}

We expressed their formula with our rescaled quantities defined in 
(\ref{scaledcharges}), and $m$ is the angular momentum parameter of the perturbation mode. 
The subscript `br' stands for the black resonator. Since the numerical resonator 
solutions of \cite{Dias:2015rxy} satisfy 
(\ref{small-resonator}) very well (see their Fig. 2), say for $\epsilon\lesssim 0.11$ 
and $j\lesssim 0.008$, we will show here that the entropies 
of small Grey Galaxies are always larger than (\ref{small-resonator}) for arbitrary values of 
$m$. 

To see this, recall that the entropy 
of our Grey Galaxy is given by that of the core black hole at $\omega=1$. 
The charges carried by the core black hole and the total system have the same $\epsilon-j$. 
So $\epsilon-j$ can be parametrized by the rotation parameter $a$ of the core 
black hole at $\omega=1$, which is obtained from (\ref{chargessr}) by
\begin{equation}
	\epsilon-j=\frac{a^{\frac{1}{2}}}{2(1-a)}\ .
\end{equation}
To stay in the regime $\epsilon,j\ll 1$ in which the approximation (\ref{small-resonator}) 
is valid, we take $a\ll 1$, thus obtaining
\begin{equation}\label{e-j:small}
	\epsilon-j\approx \frac{a^{\frac{1}{2}}}{2}\ .
\end{equation}
The entropy of the core black hole, or that of the Grey Galaxy, is given 
from (\ref{chargessr}) by
\begin{equation}\label{entropy:small}
	s_{\rm gg}=\frac{a}{1-a}\approx a\ .
\end{equation}
Now consider the entropy (\ref{small-resonator}) of small black resonators at the same 
$\epsilon,j$. One finds
\begin{equation}
	s_{\rm br}<4[\epsilon-j]^2\approx a\approx s_{\rm gg}\ .
\end{equation}
The first inequality is only asymptotically saturated for $m=\infty$, and we plugged in 
(\ref{e-j:small}) and (\ref{entropy:small}) at the second/third steps, respectively. 
We have thus shown that the small Grey Galaxy solutions have larger entropies 
than small black resonators of \cite{Dias:2015rxy,Cardoso:2013pza} 
at arbitrary $m$.

As we shall briefly comment in section \ref{discussion}, the key ideas of our Grey Galaxies are 
quite simple and should generalize to other AdS$_D$ with $D>4$. In particular, in AdS$_5$, 
there are two angular momenta $J_1,J_2$ and two angular velocities $\omega_1,\omega_2$. 
For instance, for an over-rotating black hole satisfying $\omega_1,\omega_2>1$, we naturally
conjecture that 
its endpoint is given by the Grey Galaxy which contains the core black hole at $\omega_1=\omega_2=1$. 
(See section \ref{discussion} for more discussions on the shape of the full solution.) 
Generalizing the discussions of section \ref{me}, we consider the constant 
$E-J_1-J_2$ plane in the 3-dimensional space of $(J_1,J_2,E)$ passing through 
the point of our Grey Galaxy charges. The core black hole charge is where this plane meets 
the $\omega_1=\omega_2=1$ curve. This way, we can identify the core black hole parameters and 
compute the Grey Galaxy entropy by the Bekenstein-Hawking entropy of the core.

\begin{figure}[t]
	\centering
	\includegraphics[width=0.7\textwidth]{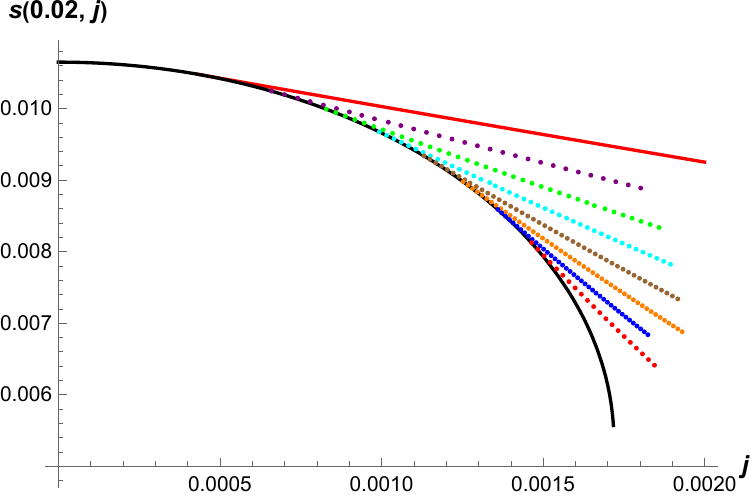}
	\caption{Entropies of Grey Galaxies (red curve), black resonators  
		of \cite{Ishii:2018oms} (various dots) and Kerr-AdS$_5$ black holes (black curve) 
		at energy $\epsilon=0.02$. (The black curve ends at the extremal black hole point 
		on the lower right side.)}\label{AdS5-res-compare}
\end{figure}

In AdS$_5$, many black resonator solutions were constructed at $J_1=J_2$ \cite{Ishii:2018oms}, 
by employing an $U(1)\times SU(2)_R\subset SO(4)$ invariant ansatz and numerically solving 
the ordinary differential equations of the radial variable. 
For instance, Fig. 7 of \cite{Ishii:2018oms} shows the entropies of various families of 
black resonators at fixed energy $\epsilon=0.02$ and different values of the angular momentum 
$j$. (See Appendix \ref{Kerr-AdS5} for our charge convention.)
By following the procedure explained in the previous paragraph, one can check that the 
AdS$_5$ Grey Galaxies have larger entropies than all the black resonators shown in Fig. 7 of 
\cite{Ishii:2018oms} at same charges. 
Fig. \ref{AdS5-res-compare} shows the entropies of the Kerr-AdS$_5$ black holes (black curve), 
various black resonators (coloured dots, for different families with $n=0,1,\cdots,6$) 
and our Grey Galaxies (red curve). They all carry energy $\epsilon=0.02$. 
The dots are taken from \cite{Ishii:2018oms},\footnote{We thank the authors of 
	\cite{Ishii:2018oms} for providing the numerical data to us.} 
while the two curves are plotted as follows. 
To plot the black curve, one first solves the equation $\epsilon(m,a)=0.02$ 
in (\ref{AdS5-bh-gibbons}) for $m(a)$, and then plug this in (\ref{m-rplus}) to obtain 
$r_+^2(a)$ by solving the cubic polonimial equation for $r_+^2$. Then plugging in 
$m(a)$ and $r_+(a)$ into $j$, $s$ in (\ref{AdS5-bh-gibbons}), one obtains 
$(j(a),s(a))$ at $\epsilon=0.02$. The black curve of Fig. \ref{AdS5-res-compare} is obtained 
by a parametric plot of the last $(j(a),s(a))$. The red curve is obtained
by plugging in (\ref{a-core}) into $s$ of (\ref{omega1}) to obtain the expression 
for $s(\epsilon-j)$, and then plotting $s(0.02-j)$. The red curve is plotted 
in the region in which the Kerr-AdS$_5$ black hole is unstable, i.e. $\omega>1$. 
We see that, for all the black resonators in this figure, our Grey Galaxies 
at the same charges have larger entropies.

\subsection{Further Collapse of the gas into a black hole?} \footnote{We thank A. Zhiboedov and Z. Komargodski for raising the question that led to this subsection. }

The Grey Galaxy configuration is one in which 
we have removed some energy (equal to angular momentum) from the rotating black hole and put it in the chiral gas (this process increases the entropy of the central black hole). The reader may wonder if it is possible to further increase the entropy of our solution by collapsing the gas
into a black hole that revolves rapidly around $AdS$. This does not work, for a reason we now explain.

To very good accuracy, the gas that makes up the GG solution has $E=J$. On the other hand a black hole revolving around the big central black hole has $E>J$. If the moving black hole is small, this difference is also small, but it must be accounted for. For this reason the gas cannot, by itself, collapse into a black hole. 

The only way that such a black hole can be formed is if the required energy is somehow extracted out of the central black hole. However such a process always turns out to lower the entropy of the central black hole by an amount that is larger than the entropy created in the newly formed small black hole. For this reason such a process is entropically 
disfavourable, and so will not happen. In the rest of this subsection we explain this point in equations.

Let us suppose that the central black hole has energy $E_1$ and angular momentum $J_1$. Let us suppose that the smaller black hole has rest energy and rest angular momentum  
$E_2$ and $J_2$, and that it has been sent zooming around $AdS$ by acting on the corresponding `black hole creation' boundary operator with $\partial^n_z$. It then follows that the total 
energy, angular momentum, and entropy of  our system is 
\begin{equation}\label{totenangent}\begin{split} 
		&E= E_1+E_2+n\\
		&J= J_1+J_2+n\\
		&S=S_{BH}(E_1, J_1) + S_{BH}(E_2, J_2)\\
	\end{split}
\end{equation} 
Let us now compare this setup with  a  Grey 
Galaxy type configuration with energy and angular momentum  $n$ put in the gas. 
The configuration \eqref{totenangent} is entropically dominant only if 
\begin{equation}\label{entnnn}
	S_{BH}(E_1, J_1) + S_{BH}(E_2, J_2)> 
	S_{BH}(E_1+E_2, J_1+J_2)
\end{equation}
\begin{figure}[h!]
	\centering
	\includegraphics[width=0.9\textwidth]{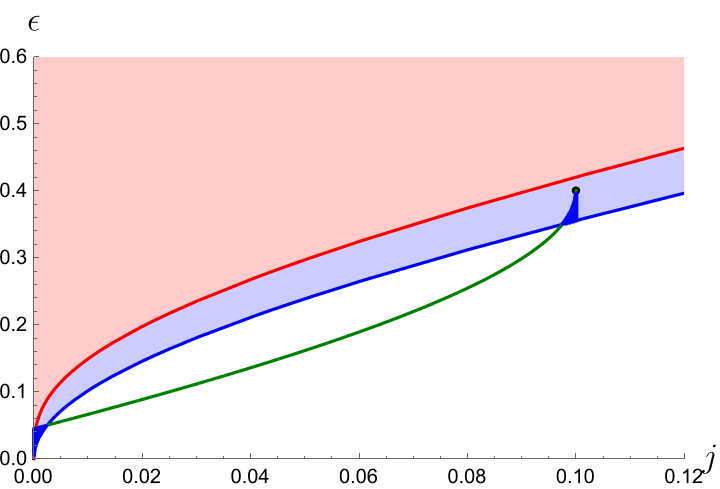}
	\caption{The black dot here depicts the Black hole with charges $E_1+E_2$ and $J_1+J_2$ defined in \eqref{totenangent}. Its position is chosen at random.  The middle blue curve is the extremal bound for the  black hole with charges $E_1$ and $J_1$ (all legal solutions lie above this curve). The green curve is the extremal line for the black hole with charges $E_2, J_2$, but plotted in the $E_1, J_1$ plane. All legal solutions lie below this curve. Therefore, the allowed values of $E_1,J_1$ lie in the shaded blue regions. We have used mathematica to create a (roughly) $40\times 120$ grid in each of these two allowed regions, and have explicitly checked, at each lattice point, that the entropy of the black hole with charges $E_1+E_2$ and $J_1+J_2$ is always greater than the sum of the entropies of the black holes with charges $E_1,J_1$ and $E_2,J_2$. We have repeated this computation for several different choices for the location of the black dot within the shaded blue region.}
	\label{binary}
\end{figure}
Note that the condition \eqref{entnnn} is now one for stationary black holes (the zooming around has left the equation). We believe
\eqref{entnnn} is never satisfied for values of $(E_1+E_2, J_1+J_2)$ that lie 
under the super-radiant curve (the red curve in Fig. \ref{kerr-ads4}). 
\footnote{In particular, if the point $(E_1+E_2, J_1+J_2)$ lies below the extremal curve (blue curve in Fig. \ref{kerr-ads4}), then it is  is impossible to decompose 
	the charges into a sum of two nonzero terms.}
Analytically, it is easy to check this is the case if $E_2$ and $J_2$ are small. At a few randomly selected values of  of $E_1+E_2$ and $J_1+J_2$, we have also numerically checked (see fig. \ref{binary}) that \eqref{entnnn} is never obeyed 
for any choice of the decomposition of charges. 

In conclusion, the chiral gas cannot gain entropy by collapsing into a black hole.

\section{Comments on supersymmetric black holes} \label{susy} 

Analogues of the solutions presented in this paper may well turn out to play a role in resolving a puzzle about the spectrum of supersymmetric black holes in $AdS_D$ spaces for $D\geq 4$. We remind the reader of the puzzle in the familiar context of ${\cal N}=4$ Yang-Mills theory. Recall that the bosonic part of the ${\cal N}=4$ superconformal algebra is $SO(4,2)\times SO(6)$. The Cartan subgroup of this algebra is spanned by the energy, the $U(1)$ rotations in $SU(2)_L$ and $SU(2)_R$, and the three Cartans, $H_1, H_2, H_3$ of $SO(6)$. To make our point in a simple manner, it is useful to specialize to a subsector of states of the theory: states with $H_1=H_2=H_3=H$, and states whose angular momentum lies entirely in $SU(2)_L$ 
(with $z$ angular momentum $J_z$). The Cartan charges in this sector are the energy, $H$
and $J_z$. The BPS bound is 
$$E \geq 2J_z+ 3 H\ .$$
One might, naively, have expected there exist supersymmetric black holes (saturating the BPS bound) for every value of $J_z$ and $H$. In fact, such black holes - the so-called Gutowski Reall black holes 
\cite{Gutowski:2004ez} - are actually known only on a curve in $H, J_z$ space. Remarkably enough, black holes everywhere on this curve have $\omega=2$ ($\omega$ is the chemical potential dual to $J_z$) and $\mu=3$ ($\mu$ is the chemical potential dual to $H$).

What is the field theory interpretation of the fact that Gutowski Reall black holes exist only on a curve in $J-H$ space? This observation would naively appear to suggest that the dual field theory hosts order $e^{N^2}$ states only at these particular charges and angular momenta. Such a result seems hard to understand from field theory. As has been suggested and studied in 
\cite{Bhattacharyya:2010yg,Markeviciute:2018yal,Markeviciute:2018cqs}, it seems 
likelier that there exist new, previously overlooked,  black hole solutions away from this curve.

In this context, it is interesting that, at least,  the revolving black hole solutions (see Appendix \ref{rbs}) have a clear supersymmetric analogue. As we have explained above, the classical entropy of Gutowski Reall black holes captures the contribution of supersymmetric primaries. 
We can act on these primaries with the two supersymmetric derivatives which, respectively, have $(J_z^L, J_z^R)=(\frac{1}{2}, \pm \frac{1}{2})$. If we act on the primary with $n$ copies of each of these derivatives, we increase $J_z^L$ by $n$ leaving $J_z^R$ unchanged. This mechanism would appear to produce a two-parameter set of new black holes - the first parameter parameterizes the Gutowski Reall black hole, and the second one characterizes its revolution. 
If the charge relation for the black hole of \cite{Gutowski:2004ez} is $J=f(H)$, 
the generalized 2-parameter solutions satisfy $J\geq f(H)$ because they are the conformal 
descendants of the former. Thus revolving black hole saddles appear to fill out at least one-half of the $H-J$ plane! 
While they may or may not be the dominant solutions, their entropy (which is easy to compute in the large $N$ limit, mimicking the work out of this  paper) gives a lower bound on the entropy of supersymmetric black holes as a function of $H$ and $J_z$, at least over the over-rotating 
half side of the $H-J$ plane.

It is also possible that supersymmetric analogues of the Grey Galaxy saddles described in this paper exist. Indeed the analysis of section 4 of the very recent paper \cite{Choi:2023znd} suggests this
may be the case.  \cite{Choi:2023znd} studied the possibilities of perturbing  
Gutowski-Reall black holes by small BPS scalar hairs, finding that such perturbed 
configurations exist for large enough mode angular momentum. At least when the core BPS black hole 
is small and the gas of BPS gravitons is dilute, it seems likely that the noninteracting mix picture 
of \cite{Basu:2010uz,Bhattacharyya:2010yg} will apply. 
We leave a detailed investigation of this fascinating possibility to future work.

\section{Discussion and future directions} \label{discussion}

\begin{figure}[t]
	\centering
	\includegraphics[width=0.7\textwidth]{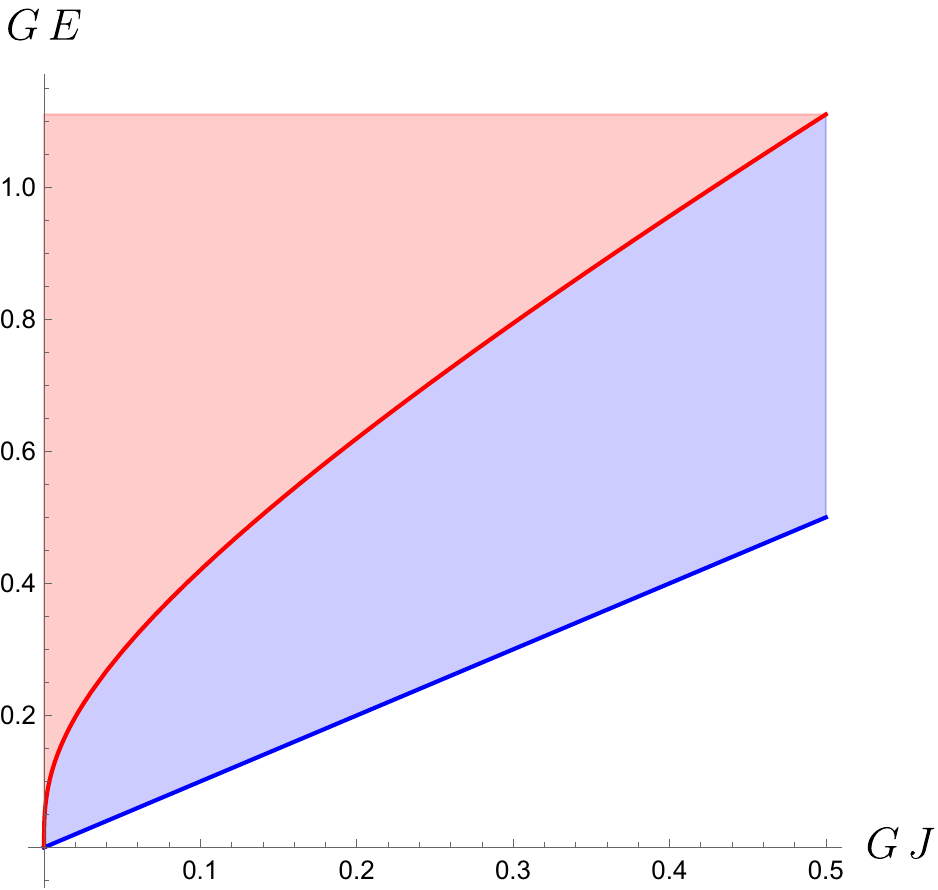}
	\caption{Phase diagram of solutions. Above the red line, the boundary field theory has a smooth tensor spread over $S^2$. Below the red line, the stress tensor has an extra delta function localized contribution on top of the smooth distribution.}
	\label{phas}
\end{figure}

In this paper, we have presented a conjecture for the microcanonical  `Phase Diagram' (entropy as a function of energy and angular momentum) of a large $N$ $CFT_3$ with a two-derivative gravity dual. 
Our phase diagram is depicted in Fig. \ref{phas}. Above the red curve
(in the pink-shaded region) the dual bulk solution is a Kerr-AdS black hole. Between the red and blue curves (in the blue-shaded region) 
the dual bulk solution is the Grey Galaxy constructed in this paper. The red curve denotes a (microcanonical) phase transition for the system. The nature of the boundary stress tensor dual to these solutions 
constitutes an order parameter for the phase transition. Just above the red line, the stress tensor is smoothly distributed over $S^2$. As one crosses the red line, along with the smooth distribution, the stress tensor includes a contribution that is highly localized around the equator of the boundary $S^2$. This localized term becomes a delta function in the large $N$ limit.  The presence of this delta function is the boundary smoking gun for the Grey-Galaxy phase. 

While we conjecture that  Grey Galaxies dominate 
the micro-canonical ensemble in the blue region of Fig. \ref{phas}, we emphasize that these solutions never dominate the canonical or grand-canonical ensembles: they are always sub-leading compared to the thermal gas. Viewed as saddle points of these ensembles, they are not even stable. Their status in this regard, we believe, is similar to that of small Schwarzschild black holes in $AdS$.

It is interesting to contrast the central proposal of this paper with the endpoints of the superradiant instabilities of {\it charged} black holes in AdS, that were extensively studied about 15 years ago. In that case, the endpoints were superfluid hairy black holes - see e.g. \cite{Gubser:2008px,Hartnoll:2008kx, Hartnoll:2008vx}. There are three main qualitative differences between these hairy black hole solutions and the `Grey Galaxies' 
presented in this paper.  First, Grey Galaxies are black holes in the background of a gas of a very large number of modes no one of which is macroscopically occupied, while the superfluid hairy black holes describe black holes immersed in a macroscopically occupied Bose condensate of a single mode. Second, while the condensate that supports hairy black holes lives in the vicinity of the black hole, the gas that supports Grey Galaxies dominantly lives far away from the black hole. Third, while hairy black holes represent a single classical configuration, Grey Galaxy saddles are an ensemble of configurations. Moreover, the ensemble in question does not exist in the purely classical theory but is stabilized by quantum mechanics. \footnote{The thermodynamics of the gas of any classical field theory suffers from the ultraviolet catastrophe. This paradox is famously resolved in quantum mechanics. In the current context, the ultraviolet catastrophe shows up in \eqref{sumover}. The classical version of \eqref{sumover} is obtained by Taylor expanding the exponential in that formula at a small value of its argument, and is given by 
	\begin{equation}\label{sumoverclass} 
		\ln Z= \sum_{n, l=0}^\infty 	\sum_{a=0}^{2l} - \ln \left(\beta(\Delta + 2n +\omega a) 
		-\beta(1-\omega) l \right), 
	\end{equation}  
	an expression that is clearly divergent. It follows that the quantum nature of the underlying bulk theory is essential to the Grey Galaxy solutions. As we have explained in the introduction, the Grey Galaxy solution is classical in a coarse grained sense, even though it is sourced by a quantum gas,  only because all fluctuations - quantum as well as statistical - are suppressed, in this solution, by the central limit theorem. In other words, the classical description applies to our solution for the same reason that the hydrodynamics of a highly quantum liquid - like strongly coupled ${\cal N}=4$ Yang-Mills - is parametrically well described by classical equations at large $N$.}

In this paper, we have focussed on the study of 
large $N$ CFTs whose dual description is Einstein gravity in the bulk. While the physics that underlies our Grey Galaxy solutions makes crucial use of the large $N$ limit, it seems to us that the two derivative nature of the bulk description did not play a qualitatively crucial role.  For concreteness let us consider ABJM theory parameterized by $N$ and $\lambda$. In this paper, we have constructed the `phase diagram' of this theory (as a function of $E$ and $J$) at large $N$ and large $\lambda$.  Let us now consider the same theory at large $N$ but a fixed finite value of $\lambda$. Since the thermodynamic charges of black holes are, presumably, continuous functions of $\frac{1}{\lambda}$, it follows that the `phase diagram' of Kerr-AdS black holes in such a theory 
will be qualitatively similar to Fig. \ref{kerr-ads4}, at least when the fixed value of $\lambda$ is large. It seems likely to us that the dominant gravitational phase below the (analogue of the) red line in Fig. \ref{kerr-ads4} will continue to be 
given by the analogues of Grey Galaxy solutions, which, we conjecture, will continue to be distinguished by the presence of a $\delta$ function contribution to the boundary stress tensor even at finite values of $\lambda$. It would be very interesting to further investigate this question. 

From the point of view of the field theory, the black hole part of the Grey Galaxy solution is a `quark-gluon plasma'; an interacting configuration of $N^2$ `gluons'. On the other hand, the gas part of the Grey Galaxy solution is composed of local singlets, and so can be thought of as the moral analogue of `glue-balls'. In field theory terms,  as $\omega$ is scaled to unity, the quark-gluon plasma manages to increase its entropy by expelling fast-rotating glue-balls. These glue-balls carry very large angular momentum, and so are sharply localized around the equator of the boundary $S^2$, accounting for the $\delta$ function contribution to the stress tensor. It would be very interesting to understand this phenomenon - in any reasonable approximation - directly in the field theory. In this context 
we note that Grey Galaxy solutions have several similarities with the field theory configurations studied in \cite{Cuomo:2022kio}. It is possible that the effective descriptions of \cite{Cuomo:2022kio} will allow us to better understand the dynamics of Grey Galaxies. 

As we have mentioned in section \ref{susy}, it is even possible that Grey Galaxy solutions have supersymmetric analogues. If this turns out to be the case, it would likely allow for a more precise 
understanding (e.g. at finite $N$) of the solutions presented in this paper.

As we have explained in subsection \ref{esri}, $\omega>1$ Kerr-AdS black holes are classically unstable, and we expect classical perturbations of such black holes to evolve into Grey Galaxy solutions. It is interesting that quantum effects also lead to a similar evolution, even in the absence of a classical perturbation. The relevant quantum effects are simply Hawking radiation. Over time scales of order unity, this effect will populate low $l$ modes, at (classical) amplitudes of order ${G}$. These amplitudes will then be amplified to order unity over time scales of order $(-\ln G)$ by the exponential classical instabilities. From this point on we expect the subsequent evolution of the system to proceed along the lines of the discussion in subsection \ref{esri}. Note that the delay in the quantum process, compared to its classical counterpart, is much smaller than the time scales associated with the complete formation of the Grey Galaxy solution.

Although the focus of this paper has been on Grey Galaxy solutions, in Appendix \ref{rbs} we have also presented a detailed construction of another 
class of new solutions, the so-called Revolving Black Hole saddles. These new solutions are intriguing because they describe large classical black holes in a coherent quantum state. These configurations are essentially descendants of classical black holes that sit at the centre of $AdS$, and so are extremely easy to construct. As their construction involves only symmetry analysis, it seems to us that the construction is precise enough to be applied to supersymmetric contexts. As we have explained in section \ref{susy}, the simple construction of Appendix 
\ref{rbs} already appears to have nontrivial implications for the spectrum of supersymmetric states in, e.g., $AdS_5 \times S^5$. It would be very interesting to further investigate this point. 

As RBHs are (marginally) entropically subdominant compared to Grey Galaxy 
solutions, we expect that RBHs will eventually decay into GGs. The net effect of such a decay process will be to transfer the rotational angular momentum 
from the RBH orbital motion into the gas. This dynamics would presumably be initiated by the Hawking radiation of gas modes (recall that the gas would like to be very heavily populated at $\omega= 1 -{\cal O}(G)$ as is appropriate for RBHs), but the qualitative nature of the subsequent dynamics is not completely clear to us. We leave further exploration of this interesting question to future work.  

An unusual feature of RBHs is that they are macroscopic objects that occupy 
highly quantum wave functions. Standard lore asserts that quantum states for macroscopic objects decohere rapidly. It would be interesting to 
work out this lore in the context of RBHs. In this context we emphasize that the charges of any particular localized revolving black hole are very different from 
those of the RBH. Perhaps decoherence effects effectively turn the 
RBH state into a density matrix that gives equal weight to every position of the 
(now localized) revolving classical black hole. It would be interesting to better understand this point. 

The bulk gas stress tensor, in the Grey Galaxy solution, is a sum of terms, one associated with each bulk field. In this paper, we have presented detailed formulae for the `gas stress tensor' (and consequent metric back-reaction) due to a bulk scalar field dual to 
a boundary operator of dimension $\Delta$. It would be useful to 
work out the corresponding expressions for bulk fields of higher spin (and, in particular, the bulk metric). Given that we already know that the 
total energy in the higher spin gas is given in terms of the energy 
of associated effective scalars by \eqref{zfin} and \eqref{zshort}, it is tempting to conjecture that a similar formula holds also for the corresponding local bulk stress tensors. We leave these issues to future work.

Through this paper, we have focussed on the special case of the superradiant instability of Kerr-$AdS_4$. As we have mentioned in the introduction, similar instabilities exist in $AdS_{D}$ for all $D >4$. \footnote{In $D=3$, on the other hand, the angular velocity of black holes never exceeds unity ($\omega$ tends to unity from below as the black hole 
	energy is lowered to extremality). As a consequence, black holes in $D=3$ do not suffer from the superradiant instability, and Grey Galaxy solutions do not exist in this dimension.}  Although the precise technical details will change from dimension to dimension, it seems clear that Grey Galaxy (and, certainly, revolving black hole) solutions will have analogues in every $D>4$. It would be interesting to explicitly construct these solutions. 

While we expect some qualitative aspects of the $D=4$ solutions to persist in all dimensions, we also expect some qualitatively new features to arise in $D>4$. Our expectation stems from the fact that the number of inequivalent angular velocities is $[\frac{D-1}{2}] >1$ for every $D>4$. For instance, $AdS_5$ contains a spatial $S^3$. Let us imagine that this $S^3$ is embedded into $C^2$, and the two complex numbers in $C^2$ are called $z_1$ and $z_2$. In this situation, we have two angular velocities, $\omega_1$ and $\omega_2$, that are respectively dual to rotations in the $z_1$ and 
$z_2$ planes. This observation suggests that we will now have three qualitatively different varieties of black holes. The first variety consists of Kerr-AdS black holes with $|\omega_1|<1$ and $|\omega_2|<1$. The second variety consists of black holes with one of the two angular velocities - let us say $\omega_1$ - parametrically near to unity, while the second angular velocity is less than and well separated from unity. We expect the Grey Galaxy solution in this situation to be closely analogous to those presented in this paper. In particular, 
we expect that the gas that makes up the galaxy will continue to be centred around a two-dimensional disk of a large radius; the disk in question will live on the plane $z_2=0$. 

The third kind of saddle is one in which $\omega_1$ and $\omega_2$ are both parametrically close 
to unity. In this case, we expect a halo spread on a large 4-dimensional spatial ball. For instance,
when $\omega_1=\omega_2$ approach $1$, the $U(2)\subset SO(4)$ 
symmetry rotating $z_1,z_2$ is unbroken. Since there are no preferred directions for the disks
in this case, we expect a graviton halo whose stress-energy tensor depends only on the radial
coordinate. It would certainly be interesting to work all this out in detail. We leave this to future work.

Another new feature in $AdS_D$ in $D>4$ is that a larger variety of black holes with angular momentum exist in this case. For instance, there exist black rings \cite{Emparan:2001wn} and black Saturns \cite{Elvang:2007rd}. It thus is possible that thermodynamically dominant solutions in $D>4$ involve these new elements in 
a nontrivial manner - perhaps together with Grey Galaxy-like saddles. We also leave investigations of this point to future work.

\acknowledgments

We would like to thank O. Aharony, J. Choi, F. Denef, S. Hellerman, I. Halder,  S. Hartnoll, G. Horowitz, T. Ishii, D. Jain, D. Jatkar, V. Krishna, K. Lee, S. Lee, 
G. Mandal,  H. Ooguri, O. Parrikar,  D. Simons-Duffin, S. Trivedi,  A. Virmani, S. Wadia, H. Yavartanoo and especially 
N. Benjamin, A. Gadde and J. Santos for very useful discussions. We would also like to thank N. Benjamin, I. Halder, S. Hartnoll, G. Horowitz, T. Ishii, Z. Komargodski,  J. Maldacena, K. Murata, J. Santos, A. Virmani, S. Wadia and A. Zhiboedov for comments on the manuscript. The work of SK and EL was supported by the NRF grant 2021R1A2C2012350. The work of CP and SM was supported by the Infosys Endowment for the study of the Quantum Structure of Spacetime. The work of SM and CP is supported by the J C Bose Fellowship JCB/2019/000052. The work of SK was supported in part by an ISF, center for excellence grant (grant number 2289/18), Simons Foundation grant 994296, and Koshland Fellowship. The work of JL was supported in part by the US Department of Energy under the award number DE-SC0011632. SK, SM and CP would also like to acknowledge their debt to the people of India for their steady support to the study of the basic sciences.

\appendix
\section{A brief review of superradiant instabilities in 
	$AdS$} \label{superradiance} 

Consider a charged rotating black hole in a flat space. Let the chemical potential and angular velocity of this black hole, respectively, be denoted by $\mu$ and $\omega$. Now consider a fluctuation about the black hole background,  of frequency $f$, angular momentum $j$ and charge $q$. 
If such a fluctuation is sent toward the black hole, one might expect that some of it would be absorbed and the rest will be reflected. However, it was discovered over 60 years ago \cite{Zel'Dovich71} that this is not always the case. When the fluctuation is such that 
\begin{equation}\label{superradcond}
	q \mu + j \omega > f,
\end{equation} 
it turns out that more of the wave comes out than was sent in. In other words the wave extracts energy, angular momentum and charge from the black hole. This remarkable phenomenon is called superradiance. 

Superradiance for $AdS$ black holes differs from the flat space version in two qualitative respects.
\begin{itemize} 
	\item In flat space one asks about the fate of a wave packet, at a particular frequency, that is incident on a black hole. The question is posed in the language of scattering (computation of a reflection amplitude) and makes sense precisely because free asymptotic states exist in flat space, and their properties are unaffected by the presence or absence of a black hole at the centre. In contrast, $AdS$ is a box. The form (and spectrum) of generic fluctuations about empty $AdS$,  and $AdS$ with a black hole, are different from each other. Consequently, a scattering-type superradiance question cannot even be posed for generic fluctuations, about a generic $AdS$ black hole solution. There are two exceptions to this rule. First, if the black hole under study is small, in 
	AdS units, then local physics is approximately that of flat space and flat space results apply. \footnote{This intuition was used by  \cite{Basu:2010uz} to compute the threshold of superradiant instabilities for small charged black holes in $AdS$ and to compute the endpoint of the superradiant instability in this case.} 
	Second, if the fluctuation mode under study has parametrically large angular momentum then 
	the centrifugal barrier ensures that it dominantly lives at a very large value of the radial parameter, and interacts very weakly with the black hole. Superradiant-type scattering computations are, thus, well-defined for such modes even for large black holes.
	\item When the superradiant phenomenon is well defined in AdS, its consequences are very different from 
	flat space. Recall that the superradiant enhancement for modes that obey \eqref{superradcond},  is a `one off' condition in flat space.  In AdS, however, this is not the case. A mode incident on the black hole is amplified. This mode hits the AdS wall and is reflected back onto the black hole, which once again amplifies it.  This process continues ad infimum giving rise to exponential instability. 
\end{itemize} 

Let us now suppose we are in a parametric regime where the superradiant phenomenon is approximately well-defined in AdS (either because the black hole is small or because the modes in question have high angular momentum). In that case, the mode that first goes unstable is the one with the largest ratio of $\frac{q}{f}$ (in the case of charged black holes) or of $\frac{j}{f}$ (in the case of rotating black holes). Let us first consider the case of a charged field of charge $q$  propagating in the background of a charged, non-rotating,  black hole. In this case, all fluctuation modes carry a charge $q$. Consequently, the highest ratio of $\frac{q}{f}$ is achieved for the fluctuation mode with the lowest energy. This mode is the `primary state' of the corresponding charged field (its frequency is the scaling dimension $\Delta$ of the dual operator). The mode in question carries zero angular momentum. As we have explained above, the language of superradiance is only appropriate for the study of such modes for small black holes. In this case, it was demonstrated in \cite{Basu:2010uz} that the black hole is unstable if $\mu q> \Delta$, and that the end point of the instability is a charged black hole sitting 
inside a cloud of the scalar (which is approximately in its primary state). As the black hole is so much smaller than the scalar cloud, these two elements interact only weakly with each other. Roughly speaking, the superradiant instability produces a Bose condensate of a single scalar mode, weakly interacting with the black hole. This picture is accurate when the black hole is small. It is corrected, order by order, in a power series in the black hole radius, and can be qualitatively inaccurate when the black hole radius exceeds the  AdS scale. 

With all this preparation, let us now consider the case of prime interest to this paper, namely the case of a rotating black hole in AdS that is not necessarily small in AdS units. According to \eqref{superradcond}, the modes that first go unstable are those with the highest 
ratio of $\frac{j}{f}$.  Now the CFT unitarity bound tells us that $\frac{j}{f}$ is bounded from above by unity. The fluctuation modes that most closely approach this bound are those with very large angular momentum and correspondingly large energy. In the large $j$ limit, the ratio of $\frac{j}{f}$ tends to unity for such modes.  \footnote{Consider an example. Under the state operator map these modes map to the operator 
	$\partial_z^{j-2} T_{zz}$. This operator has $z$ angular momentum $j$ and dimension $j+1$. In the 
	limit that $j$ is large $\frac{f}{j}$ tends to unity, and so these modes lead to a superradiant instability in the background of a black hole with $\omega>1$.} But, as we have explained above, these large angular momentum modes are precisely those for which the superradiant analysis is self-consistent (for black holes of arbitrary size).  Referring back to \eqref{superradcond}, we conclude that all rotating AdS black holes are unstable when $\omega >1$. Given any 
$\omega>1$ (no matter how close $\omega$ is to unity) we find that all modes with $j$ greater than a critical value - and therefore an infinite number of modes - are unstable. For this reason, the endpoint of the superradiant instability of rotating black holes is qualitatively different from the endpoint of the instability for charged (especially small charged) black holes. 

\section{Free particle motion in $AdS_{d+1}$} 
\label{fpa} 

In this Appendix, we review aspects of the classical and quantum description of free particles in $AdS_{d+1}$. The particles we study are dual to CFT operators with dimension $\Delta$ and `spin' $s$ (for $d>3$ the `spin' $s$ is a catchphrase for an irreducible representation of $SO(d)$).

\subsection{Quantization and group cosets} \label{qgc}

First, consider the motion of a single spinless particle in $AdS_{d+1}$. The phase space of this system is the set of all inequivalent geodesics and is $2d$ dimensional (`$d$ positions and $d$ velocities').
Satisfyingly, this $2d$ dimensional phase space is completely generated by the action of the symmetry group.  Consider, for instance,  a geodesic that sits at the centre of $AdS$. $SO(d) \times SO(2)$ transformations map this geodesic back to itself. However the remaining symmetry generators - those that live in the coset $SO(d,2)/(SO(d) \times SO(2))$ - map it to a distinct geodesic. As the coset $SO(d,2)/(SO(d) \times SO(2))$ is $2d$ dimensional
\footnote{The infinitesimal generators of this coset are precisely the $2d$ $P_\mu$ and $K_\mu$. We emphasize that it is not the case that the classical geodesic at the centre of AdS is annihilated by $K_\mu$: if this had been the case we would have too few parameters in our geodesics. The fact that $K_\mu$ does not annihilate the geodesic at the centre,  even though it quantum mechanically annihilates the primary state- is analogous to the fact that $a$ annihilates the vacuum of the Harmonic oscillator, even though it does not act as zero on the phase space point $x=p=0$. }
, it is plausible - and true - that the action of this coset action generates the full phase space. In other words, the phase space is isomorphic to 
{\it } $SO(d,2)/(SO(d) \times SO(2))$. The quantization of this phase space - with respect 
to the group invariant symplectic form inherited from particle action - produces the spin-zero, dimension $\Delta$ representation of $SO(d,2)$. From the spacetime point of view, this Hilbert Space is the space of solutions of the relativistic Klein-Gordon equation with $m^2=\Delta (\Delta -d)$. 

The generalization to particles with spin is complicated by the fact that the spin degrees of freedom of a particle are already quantum. These quantum degrees of freedom may be obtained by the quantization of the appropriate co-adjoint orbits of $SO(d)$(see e.g. \cite{Kirillov}). These orbits are generated as follows. We start with an $SO(d)$ charge `matrix' (this is an $SO(d)$ Lie algebra element - but with quantized but otherwise arbitrary coefficients, see below). Without loss of generality, we choose this algebra element to lie in the Cartan of $SO(d)$. The co-adjoint orbit is the submanifold, in the space of (adjoint valued) charges, that is generated by the action of the symmetry group 
on this chosen Cartan element. When the initial Cartan is chosen so that all quantized coefficients are nonzero, this subspace is isomorphic to 
$SO(d)/H$ ($H$ is the Cartan subgroup of $SO(d)$). The quantization of the coefficients of the Cartan generator is necessary for the consistency of the quantum problem.
\footnote{For instance, consider the special case $d=3$. In this case, the co-adjoint orbits are generated by the action of the rotation group on charge vectors and so consist of two spheres. The spin $s$ representation is obtained by taking the symplectic form to be the round magnetic field on the $S^2$ with $s$ units of magnetic flux. Flux quantization tells us that $s$ must be an integer.}. When the $SO(d)$ system is taken in isolation, the endpoint of this quantization is 
the irreducible representation of $SO(d)$ whose highest weights are determined by the quantized coefficients that define the original Cartan element.

It follows that the full phase space for the motion of spinning particles in $AdS_d$ is given by specifying both a geodesic and an $SO(d)$ charge (an $SO(d)$ coadjoint element). Let us choose our starting phase space element to be a geodesic sitting at the centre of $AdS_d$, with co-adjoint charges lying in the Cartan. Let us suppose that the co-adjoint representation under study is generic (the coefficients of all Cartan elements are nonzero). In this case, the only symmetry group elements that leave both the geodesic and the co-adjoint element unchanged are those in $SO(2) \times H$. So the phase space, in this case, is identified with the group coset $SO(d,2)/(SO(2)\times H)$. \footnote{ In the special case $d=3$, any particular unit vector on this $S^2$ is left invariant only by rotations around the vector's axis, i.e. by an $SO(2) \subset SO(3)$. Consequently, the phase space is now $SO(3,2)/(SO(2) \times SO(2))$.} The quantization of this coset produces the corresponding representation of the conformal group  $SO(d,2)$. \footnote{Even though the phase space now has a dimension that is larger than $2d$, all but $2d$ parameters on this phase space are compact, and their quantization produces the particles' spin.} From the spacetime point of view, the Hilbert Space associated with this quantization is the space of solutions of the relativistic higher spin equations in $AdS_{d+1}$.

Notice that $H'=SO(2)\times H$ is simply the Cartan subgroup of $SO(d,2)$. Consequently $SO(d,2)/(SO(2)\times H) = SO(d,2)/H'$. It follows that the quantization described above is simply the 
co-adjoint quantization of the conformal group itself.
This last statement is exact and applies without restrictions. \footnote{In contrast, the description as the quantization of $SO(d,2)/(SO(2)\times H) = SO(d,2)/H'$ applies only for co-adjoint orbits that are generic, i.e. that have nonzero entries for all quantized coefficients.}

One way of understanding the connection between particle motion and co-adjoint quantization of $SO(d,2)$ is to realize that every geodesic 
defines an $SO(d,2)$ charge vector simply because it carries specific values of each of the $(d+2)(d-1)/2$ $SO(d,2)$ Noether charges.  
The discussion of the previous paragraphs explains that the charges label geodesics (no two geodesics carry the same charge). It follows that the phase space of geodesics 
can be identified with the coadjoint phase space. 

The map between geodesics and charges can be developed as follows. To start with let us evaluate the charges of our reference point in phase space: i.e. the geodesic at the centre of $AdS_d$ - with $SO(d)$ charge vectors to be $H_i$ that lie in the Cartan of $SO(d)$. The spacetime $SO(d,2)$ Noether charges for this system lie in the Cartan of  $SO(d,2)$, and in fact are given by $(\Delta, H_i)$ ($\Delta$ is the energy, i.e. the $SO(2)$ charge).
As we have mentioned above, the most general classical solution is an $SO(d,2)$ rotation of our reference solution. The charges of this new solution are thus the $SO(d,2)$ rotation of the adjoint element $(\Delta, H_i)$. 

In the next few subsections, we present a completely explicit construction of the 
`right moving part of the phase space', and also its quantization,  the especially simple case $d=2$. We present this analysis for the following reasons: 
\begin{itemize}
	\item The right moving $AdS_3$ phase space (and its quantization) constitute a submanifold of the phase space of the motion of spinning particles in $AdS_{d+1}$ (and a sub Hilbert Space of this larger quantization). Moreover, this submanifold of phase space (and subspace of Hilbert Space) turns out to be precisely the ones that will prove to be relevant to the construction of revolving Black Hole solutions (see the next 
	Appendix).
	\item  The discussion of spinning particles is particularly simple for the special case $d=2$. As the co-adjoint orbits of $SO(2)$ are a point, the full phase space is simply the space of geodesics. Consequently, it is much simpler to directly study this simple subspace (and its quantization) than to study the full set of 
	geodesics in $AdS_{d+1}$, and then focus on the solutions of interest.
\end{itemize}

\subsection{Coordinates and symmetries in $AdS_3$}

Let us first study the action of the six-dimensional symmetry group $SO(2,2)$ on coordinates in $AdS_3$. $AdS_{3}$ can be thought of as the submanifold 
\begin{equation}\label{consttwo}
	-X_{-1}^2 - X_0^2+X_1^2+X_2^2 = -1 
\end{equation} 
of ${\mathcal R}^{2,2}$. The four embedding space coordinates can usefully be packaged into a $2 \times 2$ matrix 
\begin{equation}\label{esu}
	X= \left(
	\begin{array}{cc}
		X_{-1}-X_1 & X_2-X_0 \\
		X_0+X_2 & X_{-1}+X_1 \\
	\end{array}
	\right) \hspace{2cm}\text{where~~~~} \det X=1 
\end{equation} 
(the condition $\det X=1$ reproduces \eqref{consttwo}). Clearly the transformation  
\begin{equation}\label{actso2}
	X' = f X \bar{f}^T
\end{equation} 
(where $f$ and $\bar{f}^T$ are independent real
\footnote{The reality condition follows from the requirement that $X$ and $X'$ are both real.} $2 \times 2$ matrices, each of unit determinant) 
preserves the condition $\det X=1$. It follows that 
$SO(2,2) = SL(2, R)\times SL(2,R)$, a fact that is, of course, well known. 

In order to obtain intuition for the matrix parameterization of $AdS_3$, (\eqref{esu}) it is useful to perform a few exercises. If we parameterize global $AdS_3$ as 
\begin{equation}\label{gath}
	\begin{split} 
		&X_{-1}= \cosh \rho \cos \tau\\
		&X_0 = \cosh \rho \ \sin \tau \\
		& X_1= \sinh \rho \cos \phi \\
		& X_2= \sinh \rho \sin \phi \\
	\end{split}
\end{equation} 
then the  time translation $\tau \rightarrow \tau + T$
is generated by the $SO(2,2)$ transformation that acts as 
\begin{equation} \label{ttrans}
	X_{-1}'=\cos T X_{-1} - \sin T X_0, ~~~~X_0'= \cos T X_0 +\sin T X_{-1}, 
\end{equation} 
while the rotation $\phi \rightarrow \phi +\Theta$  is the transformation that acts as 
\begin{equation} \label{thetatrans}
	X_{1}'=\cos \Theta X_{1} - \sin \Theta X_2, ~~~~X_2'= \cos \Theta X_2 +\sin \Theta  X_{1} 
\end{equation}

In other words, the $f$ matrix 
\begin{equation}\label{rottimf}
	f_A=\left(
	\begin{array}{cc}
		\cos A & -\sin A \\
		\sin A & \cos A \\
	\end{array}
	\right)
\end{equation} 
generates time translation by $A$ accompanied by a  rotation by $A$, while the ${\bar f}$ matrix 
\begin{equation}\label{rottimbarf}
	{\bar f}_B=\left(
	\begin{array}{cc}
		\cos B & \sin B \\
		-\sin B & \cos B \\
	\end{array}
	\right)  
\end{equation} 
generates time translation by $B$ together with  rotation by $-B$. \footnote{ It follows that the choice  $A=-B=-\frac{\Theta}{2}$ 
	gives a rotation by angle $\Theta$ while the choice  $A=B=\frac{T}{2}$ gives  a time translation by time $T$. } 
Following usual conventions, we say that the matrix \eqref{rottimf} generates left moving time translations by $A$, while the matrix 
\eqref{rottimbarf} generates right moving time translations by $-B$.

Now consider a group $f$ in the neighbourhood of identity, i.e. one that takes the form 
\begin{equation}\label{finfin} 
	I + f_{\alpha, \beta, \gamma}, 
	~~~f_{\alpha, \beta, \gamma}=\alpha\sigma_3+\gamma \sigma_1 + \beta (i \sigma_2) 
\end{equation}
where $\alpha$, $\beta$ and $\gamma$ are infinitesimals. $f_{\alpha, \beta, \gamma}$ is a generator of $SL(2,R)$, and transforms in the adjoint representation  of $SL(2, R)$. In particular under a left moving time translation by $A$ this generator transforms as 
\begin{equation}\label{simtrnfA}
	f_A f_{\alpha, \beta, \gamma} f_A^{-1} 
	= f_{\alpha', \beta', \gamma'}
\end{equation} 
with
\begin{equation}\label{abcg}\begin{split}
		&\alpha'= \cos 2A ~\alpha - \sin 2A~ \gamma \\
		&\beta'= \beta \\
		&\gamma'=\alpha  \sin 2 A+\gamma  \cos 2 A\\
	\end{split} 
\end{equation} 

It follows that the generator $\beta i \sigma_2$ commutes with left moving time translations, and so 
$i\sigma_2$ should be identified with $L_0$. On the other hand, $\alpha$ and $\gamma$ transform like a two-dimensional vector under left moving time translations. We can diagonalize this action as follows. Define 
\begin{equation}\label{diagalph}
	\begin{split} 
		&\alpha = A_z + A_{\bar z}\\
		&\gamma = i \left( A_{z}- A_{\bar z} \right) 
	\end{split}
\end{equation}
then the transformation of $A_z$ and 
$A_{\bar z}$ under left moving time translations is given by 
\begin{equation}\label{abcgdiag}\begin{split}
		&A_z'= e^{-i 2 A} A_z\\
		&A_{\bar z}'= e^{i 2 A} A_{\bar z} \\
		&\beta'=\beta\\
	\end{split} 
\end{equation} 
It follows that $A_z$ and $A_{\bar z}$, respectively, parameterize infinitesimal translations and special conformal transformations. Notice that $A_z$ and $A_{\bar z}$ are complex conjugates of each other. The factors of $i$ in \eqref{diagalph} are one way of understanding the well-known equation 
$$P_z= K_z^\dagger.$$

\subsection{Geodesics in $AdS_3$}

In this subsection, we present a detailed description of the space of geodesics in the special case $d=2$, 
i.e. geodesics in $AdS_3$. In this special case, we have 4 parameter space of geodesics. 

Consider a geodesic sitting in the centre of $AdS_3$. The geodesic sits at $X_{-1}= \cos \tau$, $X_0=\sin \tau$
with $X_1=X_2=0$. The geodesic is obtained by varying over $\tau$. The $X$ matrix corresponding to this configuration is 
\begin{equation}\label{esu2}
	X= \left(
	\begin{array}{cc}
		\cos \tau & -\sin \tau \\
		\sin \tau  & \cos \tau \\
	\end{array}
	\right) 
\end{equation}
The only nonzero charges for this solution are 
\begin{equation}\label{origcharge}
	L_0= ih  \sigma_2, ~~~{{\bar L}_0}= i {\bar h} 
	\sigma_2
\end{equation}
$h$ and ${\bar h}$ are the left and rightmoving weights of the dual $d=2$ primary operator. 

We now produce a new right-moving geodesic by acting on the original geodesic by $SL(2,R)/U(1)$, i.e. by 
the action
\begin{equation}\label{matrixxp}
	X'= X \bar f^T
\end{equation} 
where $f$ lies in $SL(2,R)/U(1)$. To start with let us take 

\begin{equation}\label{fa}
	\bar f_a= \left(
	\begin{array}{cc}
		\cosh a & \sinh a \\
		\sinh a & \cosh a \\
	\end{array}
	\right)  
\end{equation} 
It is easy to verify that the coordinates of this geodesic are given by 
\begin{equation} \label{newelem}
	\begin{split}
		X'_{-1}&=\cosh a \cos \tau \\
		X'_0&= \cosh a \sin \tau\\
		X'_1&= \sinh a \sin \tau\\
		X'_2&= \sinh a \cos \tau\\
	\end{split}
\end{equation}
We see that this geodesic sits at the fixed value of $\rho$, $\rho=a$ and rotates in $\phi$ so that $\frac{d \phi}{d \tau}=1$. This rotation is purely right-moving and approaches the speed of light as $\rho \to \infty$. Note that the particle is displaced in the $X_2$ (rather than $X_1$) direction at $\tau=0$.

As the $SL(2, R)$ transformation $\bar f_a$ is purely right-moving, it leaves all left-moving charges unchanged. However right-moving charges are transformed. The right-moving $SL(2, R)$ charges for this new geodesic are 
\begin{equation}\label{cnewsol}
	Q_a= (\bar f_a^T)^{-1} Q_0 \bar f_a^T
	= E \left( \cosh 2 a (i \sigma_2)
	+\sinh 2a  ~\sigma_3 \right)
\end{equation} 
Note that $Q_a$ is a boost of $Q_0$ into the $\sigma_3$ direction. The fact that the boost is 
in the $\sigma_3$ rather than the $\sigma_1$ direction 
is related to the fact that our geodesic is displaced
in the $X_2$ (rather than the $X_1$) direction 
at $\tau=0$.

We can obtain a more general right-moving geodesic 
by choosing $f$ in \eqref{matrixxp} to be $f_A f_a$, i.e. by the construction 
\begin{equation} \label{newgeo} 
	{\tilde X}=  X\bar f_a^T \bar f_A^T
\end{equation} 
where, as above, 
\begin{equation}\label{fA}
	\bar f_A= \left(
	\begin{array}{cc}
		\cos A & -\sin A \\
		\sin A & \cos A \\
	\end{array}
	\right)  
\end{equation} 
The coordinates of this more general geodesic ${\tilde X}$ are given by
\begin{equation} \label{newelem2}
	\begin{split} 
		& {\tilde X}_{-1}=\cosh a \cos (\tau-A) \\
		& {\tilde X}_0= \cosh a \sin (\tau-A)\\
		&{\tilde X}_1= \sinh a \sin (\tau+A)\\
		&{\tilde X}_2= \sinh a \cos (\tau+A)\\
	\end{split}
\end{equation} 
This geodesic traverses the same orbit as \eqref{newelem} except that it is rotated on the circle by $\phi \rightarrow \phi +2 A$. 

The charges corresponding to these configurations are also rotated 
\begin{equation}\label{cnewsolA}
	Q_{A,a}= (\bar f_A^T)^{-1} (\bar f_a^T)^{-1} Q_0 \bar f_a^T \bar f_A^T
	= h \left( \cosh 2 a (i \sigma_2)
	+\sinh 2 a (\sin 2 A ~ \sigma_1 + \cos 2 A ~ \sigma_3) \right)
\end{equation} 
So we see $\bar f_A$ has rotated the charge - which was proportional to  $\sigma_3$ in \eqref{cnewsol} to a charge proportional to  $(\cos 2A ~\sigma_3 + \sin 2A~\sigma_1)$ in the new solution.  In other words, the charge of this new solution tracks the position of the particle at $\tau=0$. \footnote{Let us say this more clearly. From \eqref{newelem2} it is clear that the effective geodesic equation is $\tau' = \phi' - 2A$. This $2A$ offset in the initial condition is precisely reflected in the rotation of charges $Q_{A,a}$ by an angle $2A$.}.

We have, so far, described the construction and charges of the $2$ parameter set of purely right-moving geodesics. The construction (and charges) of the $2$ parameter set of purely left-moving geodesics - and indeed of the most general 4-parameter set of left and right-moving geodesics - proceeds along entirely similar lines. We do not pause to present these constructions, as only the right-moving geodesics appear in the study of Revolving Black Hole solutions in the next Appendix.

\subsection{Solutions of the  wave equation in $AdS_3$}

In this section, we study the right-moving solutions of the wave equation in $AdS_3$. In other words, we find the quantum states that result from the quantization of the two-parameter set of right-moving geodesics constructed in the previous subsection. 

We will start by this section studying the motion of scalars, and then turn to the study of spinning particles. 

\subsubsection{Scalar motion in $\rho, \tau, \phi$
	coordinates. }

Let us first study the quantization of geodesic motion in $AdS_3$ in the most straightforward manner imaginable, i.e. by simply working out the minimally coupled scalar equation in convenient coordinates, and examining the interplay with the group structure. This 
subsubsection is a review of the analysis of \cite{Maldacena:1998bw}.

The metric of $AdS_3$ is given by 
\begin{equation}\label{adsthmet}
	ds^2= -\cosh^2 \rho \, d\tau^2 + \sinh^2 \rho \,  d \phi^2  + d \rho^2 
\end{equation} 
Let us define $u= \tau + \phi$ and $v=\tau-\phi$. The paper of Maldacena and Strominger \cite{Maldacena:1998bw} informs us that the Killing vectors for the leftmoving $SL(2, R)$  are given by 
\begin{equation}\label{sl2r} 
	\begin{split} 
		&L_0 =i \partial_u \\
		&L_{-1} = e^{-iu} \left[ 
		\frac{\cosh 2 \rho}{\sinh 2 \rho } \partial_u 
		- \frac{1}{\sinh 2 \rho} \partial_v + 
		\frac{i}{2} \partial_\rho \right]  \\
		&L_{1} = e^{iu} \left[ 
		\frac{\cosh 2 \rho}{\sinh 2 \rho } \partial_u 
		- \frac{1}{\sinh 2 \rho} \partial_v -
		\frac{i}{2} \partial_\rho \right]  \\
	\end{split} 
\end{equation}
The generators above obey the commutation relations 
\begin{equation}\label{comrelL}
	\begin{split} 
		&[L_0 , L_{\pm 1} ] = \mp L_{\pm 1} \\
		&[L_1 , L_{-1} ] = 2L_0\\
	\end{split}
\end{equation} 
(we have discussed one $SL(2, R)$. The second is given by 
the interchange $u \leftrightarrow v$).

Define  the Casimir operator $L^2$ by  
\begin{equation}\label{casimir} 
	L^2= -L_0^2 + \frac{ L_1L_{-1}+L_{-1}L_{1}}{2} = L_{-1} L_1 -L_0(L_0-1)
\end{equation}

Consider a primary state (annihilated by $L_1$) with weight $L_0=h$. When acting on this state, the Casimir evaluates to $-h^2+h$. Consequently, the wave equation takes the form 
\begin{equation}\label{waveeq} 
	L_{-1} L_1 -L_0(L_0-1)= -h^2+h
\end{equation} 
A similar derivation \eqref{waveeq} also holds when we add a bar to every term on the LHS of \eqref{waveeq}
(recall that, for the operator dual to a scalar field, 
${\bar h}=h$). These two equations are consistent: using \eqref{sl2r}, and its barred version, it is easy to verify that 
\begin{equation} \label{lun}
	L_{-1} L_1 -L_0(L_0-1)= {\bar L}_{-1} {\bar L}_1 -{\bar L}_0({\bar L}_0-1)= \frac{\nabla^2}{4}
\end{equation} 
so that \eqref{waveeq} becomes 
\begin{equation}\label{waveeqn} 
	\nabla^2 \Phi= - (2h)(2h-2) \Phi
\end{equation} 
Recall that in the case of a scalar field, the famous $AdS/CFT$ formula asserts that 
$$\Delta(\Delta-d)= m^2$$
(here $d$ is the field theory dimension, and $\Delta$ is the scaling dimension of the field). 
In the case $d=2$ under study here, we find 
\begin{equation}\label{hmass}
	2h(2h-2)= m^2
\end{equation} 
and so \eqref{waveeqn} is the minimally coupled scalar equation. 
\footnote{Equivalently 
	\begin{equation} \label{scalmassrel}
		h(h-1)=\frac{m^2}{4}\quad \longrightarrow\quad h= \frac{1}{2}\left(1+\sqrt{1+m^2}\right)
\end{equation}}

The explicit form of the primary function (obtained from the condition that it is annihilated by $L_{-1}$ and $\bL_{-1}$) is,
\begin{equation}\label{wfform}
	\psi_{h, 0, 0}  = c \frac{e^{-i h (u+v)}}{(\cosh\rho)^{2h}}
\end{equation}
where $c$ is a normalization constant. 

Let us now define the descendants of the primary states as
\begin{equation}\label{primstates}
	\psi_{h, n, n'}= \left(L_{-1}\right)^n  \left(\bL_{-1}\right)^{n'}\psi_{h, 0, 0}
\end{equation}
Clearly the $L_0$ and ${\bar L}_0$ weights of these states are given by $h+n$ and ${\bar h}+n'$.  By using the explicit form of $L_{-1}$ and ${\bar L}_{-1}$ 
we find the explicit expressions 
\begin{equation} \begin{split} \label{finstates}
		&   \psi_{h, n, 0} =c_n  \left(e^{-i\,u}\tanh\rho\right)^n  \frac{e^{-i h (u+v)}}{(\cosh\rho)^{2h}}\\
		&   \psi_{h, 0, n} =\bar c_n  \left(e^{-i\,v}\tanh\rho\right)^n  \frac{e^{-i h (u+v)}}{(\cosh\rho)^{2h}}
	\end{split} 
\end{equation}
here, $c_n$ and $\bar c_n$ are normalization constants.
The formula for more general mixed descendants - those 
obtained is much more complicated (it is the 2+1 dimensional analogue of \eqref{fnl}). We will not need these expressions and do not present them here. 

As we have mentioned above, our interest in right-moving solutions of the wave equation in $AdS_3$ comes from the fact that these solutions effectively produce solutions to the wave equation in higher dimensional $AdS$ spaces. In the next few paragraphs, we illustrate this fact by direct comparison.

In \eqref{fnl} we have presented the most general solution to the radial part of the wave equation in $AdS_4$. Let us translate the wave functions \eqref{finstates} into the coordinates 
used in \eqref{fnl}. Using \eqref{gath}, the first of 
\eqref{finstates} can be rewritten in terms of the embedding space coordinates. \eqref{coordinates} and \eqref{rhor} can then be used to rewrite the embedding space coordinates in terms of the coordinates used in 
\eqref{fnl}. The net result of this exercise is the following - the wave function \eqref{finstates} can be reexpressed in terms of the coordinates of \eqref{fnl} by making the replacements 
$$\cosh \rho \rightarrow \sqrt{1+r^2}, ~~~\tanh \rho \rightarrow \frac{r\sin \theta}{\sqrt{1+r^2}}.$$
(the time and $\phi$ dependences remain unaffected between the two expressions). This replacement rule turns the second of \eqref{finstates} into 
\begin{equation}\label{psinnn} 
	\psi_{h, 0, n} =e^{-it (2h+n) +i \phi n} \frac{\sin^l \theta r^l} {(1+r^2)^{\frac{l+2h}{2}}}
\end{equation} 

Let us now compare \eqref{psinnn} with the most general solution in $AdS_4$ presented around \eqref{fnl}. This solution is proportional to  
$$e^{-i t (\Delta +l+2n)} Y_{lm}(\theta, \phi) F_{nl}(r)$$
If we choose $a=0$ then the spherical harmonic takes the simple form 
$$Y_{lm} \propto e^{il\phi} \sin^l \theta$$
If we also choose $n=0$ the hypergeometric function that appears in \eqref{fnl} reduces to unity and 
$$F_{nl}\propto \frac{r^l} {(1+r^2)^{\frac{l+\Delta}{2}}}$$
Putting everything together, we find the wave function 
\begin{equation}\label{wfnn}
	e^{-i(\Delta +l) t +il \phi}\frac{\sin^l \theta r^l} {(1+r^2)^{\frac{l+\Delta}{2}}}
\end{equation} 
We see that \eqref{psinnn} agrees exactly with \eqref{wfnn} once we make the identifications $2h=\Delta$ and $n=l$ ($n$ here is the symbol used in \eqref{finstates}, denoting the descendent level). We see that the solution to the $AdS_3$
wave equations are, indeed, also solutions to the 
$AdS_4$ wave equations, as expected.

\subsubsection{Scalar fields in embedding space} 

We will now repeat and generalize the analysis of the previous sub-subsection using the embedding space formalism. In this sub-subsection, we continue to work 
with scalars: we will generalize to the study of 
particles of spin $n$ in the next sub-subsection. 

Consider the $R^{2,2}$ space in which $AdS_{3}$ is embedded (see \eqref{consttwob}). We can coordinatize this embedding space as 
\begin{equation}\label{cordina}
	\begin{split} 
		&X_{-1}= T x_{-1}\\
		&X_0 = T x_0 \\
		& X_1= T x_1\\
		& X_2= T x_2 \\
	\end{split} 
\end{equation}
where 
\begin{equation}\label{consttwob}
	-x_{-1}^2 - x_0^2+x_1^2+x_2^2 = -1 
\end{equation} 
The $x_i$ themselves should be thought of as being determined by the three coordinates on a unit $AdS_3$.  The metric of embedding space, in these coordinates, takes the form 
\begin{equation}\label{metes}
	ds^2=-dT^2+ T^2 d\Omega_{AdS_3}^2
\end{equation} 
It follows that the Laplacian in embedding space is given by the equation 
\begin{equation}\label{les} 
	\nabla^2= -\frac{1}{T^3} \partial_T T^3  \partial_T + \frac{1}{T^2} \nabla_{AdS_3}^2
\end{equation} 

Let us now study harmonic functions (i.e. functions that obey 
$\nabla^2 f=0$) in $R^{2,2}$. Let us also demand that our functions are of homogeneity (degree) $-2h=-\Delta$ in the Cartesian embedding space coordinates.  It follows that, when acting on such functions, 
the operator 
$$\frac{1}{T^3} \partial_T T^3  \partial_T $$
gives $-m^2/T^2 $ where $m^2=\Delta(\Delta-2)$. It follows such functions when restricted to the $AdS_3$ sub-manifold, are solutions to the $AdS_3$ Klein Gordon equation with the above-quoted value of the mass. 

We have thus found a very simple way to generate solutions of the scalar wave equation in $AdS_3$. We need simply construct Harmonic functions, of a specified homogeneity, in $R^{2,2}$.

Let us now study the symmetry algebra, $SL(2,R)\times SL(2,R)$,  in embedding space language. It proves useful to define
\begin{equation}\label{defw}\begin{split} 
		&W= \frac{X_{-1}+ i X_0}{\sqrt{2}}\\
		&{\bar W}= \frac{X_{-1}- i X_0}{\sqrt{2}}\\
		&Z= \frac{X_{1}+ i X_2 }{\sqrt{2}}\\
		&{\bar Z}= \frac{X_{1}- i X_2}{\sqrt{2}}\\
	\end{split} 
\end{equation} 

Generators of the left moving $SL(2,R)$,  given in equation \eqref{sl2r},  take the following form form embedding space
\begin{equation}\label{pdses}
	\begin{split}
		&L_0=\frac{1}{2}
		\left( W \partial_W-{\bar W} 
		\partial_{{\bar W}} - Z \partial_Z + {\bar Z} 
		\partial_{{\bar Z}} \right) \\
		&L_1=Z \partial_W + {\bar W} \partial_{\bar Z}\\
		&L_{-1}=-{\bar Z} \partial_{\bar W}- W \partial_{Z}\\
	\end{split} 
\end{equation}
Similarly, the right moving $SL(2,R)$ generators are given by 
\begin{equation}\label{pdsesb}
	\begin{split}
		&{\bar L}_0=\frac{1}{2}
		\left( W \partial_W - {\bar W}
		\partial_{{\bar W}} + Z \partial_Z - {\bar Z}
		\partial_{{\bar Z}} \right)\\
		&{\bar L}_1={\bar Z} \partial_W + {\bar W} \partial_{Z}\\
		&{\bar L}_{-1}=- Z \partial_{\bar W} - W \partial_{\bar Z}\\
	\end{split}
\end{equation}

Let us now identify the primary wave function. Any expression independent of $W$, ${\bar Z}$ and ${Z}$ is clearly annihilated by both $L_1$ and ${\bar L}_1$. Consequently, 
\begin{equation}\label{funcform}
	\begin{split}
		\psi=&\frac{1}{{\bar W}^{2h}}
	\end{split}
\end{equation} 
is clearly a primary state for a scalar of weight $\Delta= 2h$. 
As \eqref{funcform} is a function of ${\bar W}$ only, it is harmonic as required. Using  $W = \cosh \rho\, e^{-i\frac{u+v}{2}}$ and $Z=\sinh \rho\, e^{i\frac{u-v}{2}}$, it is easy to check that 
\eqref{funcform} is indeed the same as \eqref{wfform}. 

By acting $n$ times with $L_{-1}$ or ${\bar L}_{-1}$ we find that 
\begin{equation}\label{funcformdesc}
	\begin{split}
		\psi_{h, n, 0}=&\frac{1}{{\bar W}^{2h}} \left( \frac{\bar Z}{\bar W} \right)^n  \\
		\psi_{h, 0, n}  =&\frac{1}{{\bar W}^{2h}} \left( \frac{Z}{\bar W} \right)^n
	\end{split}
\end{equation} 
These expressions are also easily checked against 
\eqref{finstates}. \footnote{Note that \eqref{funcformdesc} 
	are clearly homogeneous of degree $-2h$. Moreover, these functions depend only on either
	$Z$ or ${\bar Z}$ (never both) and ${\bar W}$, 
	and so consequently are harmonic.}

Notice that the expressions in \eqref{funcformdesc}
are everywhere regular in $AdS_3$, despite the fact that they have powers of ${\bar W}$ in the denominator. This is because ${\bar W}$ unlike $Z$ and ${\bar Z}$
never vanish on the $AdS_3$ sub-manifold. This observation also explains why our expressions never have $Z$ or ${\bar Z}$ in the denominator. The reason that we have ${\bar W}$, rather than $W$ in the denominator, is that we are looking for positive energy solutions (the wave functions that multiply $a$ rather than $a^\dagger$ in canonical quantization). 

\subsubsection{Higher Spin Solutions in Embedding Space}

Let us now attempt to generalize the analysis of the previous subsubsection to fields of higher spin. 
In particular, we are interested in square integrable solutions of the Laplace equation in embedding space \eqref{metes}
\begin{equation}\label{adslap}
	\nabla_{AdS_3}^2 S^{\mu_1\cdots\mu_s} = (-2h) (-2h+2) S^{\mu_1\cdots\mu_s}
\end{equation} 

A spin-$s$ primary solution also satisfies the following equations
\begin{equation}\label{spinsEOM}
	\begin{split}
		&\mathcal{L}_{L_0}\, S^{\mu_1\cdots\mu_s} = h\, S^{\mu_1\cdots\mu_s},\quad\quad \mathcal{L}_{\bL_0}\, S^{\mu_1\cdots\mu_s} = {\bar h}\, S^{\mu_1\cdots\mu_s},\\
		&\mathcal{L}_{L_1}\, S^{\mu_1\cdots\mu_s} = 0,\quad\quad\quad\quad\quad ~\mathcal{L}_{\bL_1}\, S^{\mu_1\cdots\mu_s} = 0
	\end{split}
\end{equation}
where $\mathcal{L}$ denotes the Lie derivative. 
\footnote{Here Lie derivative $\mathcal{L}$ of tensor $T^{\mu_1\cdots\mu_a}_{\quad\quad~\nu_1\cdots\nu_b}$ with respect to vector $X^\mu$ is defined by,
	\begin{equation}\label{scEOM}
		\begin{split}
			(\mathcal{L}_X T)^{\mu_1\cdots\mu_a}_{\quad\quad~\nu_1\cdots\nu_b} =&\; X.\nabla T^{\mu_1\cdots\mu_a}_{\quad\quad~\nu_1\cdots\nu_b} \\
			&- (\nabla_c X^{\mu_1}) T^{c\mu_2\cdots\mu_a}_{\quad\quad~\nu_1\cdots\nu_b} - \cdots - (\nabla_c X^{\mu_a}) T^{\mu_1\cdots\mu_{a-1} c}_{\quad\quad~\nu_1\cdots\nu_b} \\
			&+ (\nabla_{\nu_1}X^c) T^{\mu_1\cdots\mu_a}_{\quad\quad~c\nu_2\cdots\nu_b} + \cdots + (\nabla_{\nu_b}X^c) T^{\mu_1\cdots\mu_a}_{\quad\quad~\nu_1\cdots\nu_{b-1}c}
		\end{split}
\end{equation}}
\footnote{It turns out that these equations have a consistent solution only when $\bar h - h = \pm s$.}

Primaries of the spin-$s$ field are given very simply in terms of scalar primaries. It is easy to check that these are given by 
\footnote{Using chain rule in Lie derivative, $L_1\,\psi_{h,0,0}=0$, and $[L_1,L_1] = [\bL_1,L_1] = 0$ one can easily convince that \eqref{funcformdesc} are indeed spin-$s$ primaries.},
\begin{equation}\label{sppri}
	\begin{split}
		S^{(+),\mu_1\mu_2\cdots\mu_s} &= \psi_{h,0,0} \bL_{1}^{\mu_1} \bL_{1}^{\mu_2} \cdots \bL_{1}^{\mu_s},\quad\quad h-{\bar h} = s\\
		S^{(-),\mu_1\mu_2\cdots\mu_s} &= \psi_{{\bar h},0,0} L_{1}^{\mu_1} L_{1}^{\mu_2} \cdots L_{1}^{\mu_s},\quad\quad h-{\bar h} = -s.
	\end{split}
\end{equation}

Maximally rotating descendants are also given very simply in terms of their scalar counterparts. The solutions have the following form,
\begin{equation}\label{spdes}
	\begin{split}
		S^{(+),\mu_1\mu_2\cdots\mu_s}_i &= \psi_{h,i,0} \bL_{1}^{\mu_1} \bL_{1}^{\mu_2} \cdots \bL_{1}^{\mu_s},\quad\quad h-{\bar h} = s,\\
		S^{(-),\mu_1\mu_2\cdots\mu_s}_i &= \psi_{{\bar h},0,i} L_{1}^{\mu_1} L_{1}^{\mu_2} \cdots L_{1}^{\mu_s},\quad\quad h-{\bar h} = -s.
	\end{split}
\end{equation}
\footnote{One can see for $(\mathcal{L}_{L_1})^i$ descendant we used the $S^{(+)}$ primary from equation \eqref{sppri} since $L_1$ does not interact with the $\bL_1$ vector indices and the problem reduces to exactly like the scalar problem. A similar argument works for the anti-holomorphic ($S^{(-)}$) solution.}

Using the $SL(2,R)$ commutator relations one can also check that $S^{(\pm)}_n$ transforms under $SL(2,R)$ generators as follows,
\begin{equation}
	\begin{split}
		\mL_{L_0} S^{(+),\mu_1\mu_2\cdots\mu_s}_n &= (h+n) S^{(+),\mu_1\mu_2\cdots\mu_s}_n\\
		\mL_{L_1} S^{(+),\mu_1\mu_2\cdots\mu_s}_n &= n(2h+n-1) S^{(+),\mu_1\mu_2\cdots\mu_s}_{n-1}\\
		\mL_{L_{-1}} S^{(+),\mu_1\mu_2\cdots\mu_s}_n &= S^{(+),\mu_1\mu_2\cdots\mu_s}_{n+1}\\
		\mL_{\bL_0} S^{(+),\mu_1\mu_2\cdots\mu_s}_n &= {\bar h}\, S^{(+),\mu_1\mu_2\cdots\mu_s}_n\\
		\mL_{\bL_1} S^{(+),\mu_1\mu_2\cdots\mu_s}_n &= 0\\
	\end{split}
\end{equation}

\begin{equation}\label{lkjm}
	\begin{split}
		\mL_{\bL_0} S^{(-),\mu_1\mu_2\cdots\mu_s}_n &= ({\bar h}+n) S^{(-),\mu_1\mu_2\cdots\mu_s}_n\\
		\mL_{\bL_1} S^{(-),\mu_1\mu_2\cdots\mu_s}_n &= n(2{\bar h}+n-1) S^{(-),\mu_1\mu_2\cdots\mu_s}_{n-1}\\
		\mL_{\bL_{-1}} S^{(-),\mu_1\mu_2\cdots\mu_s}_n &= S^{(-),\mu_1\mu_2\cdots\mu_s}_{n+1}\\
		\mL_{L_0} S^{(-),\mu_1\mu_2\cdots\mu_s}_n &= h\, S^{(-),\mu_1\mu_2\cdots\mu_s}_n\\
		\mL_{L_1} S^{(-),\mu_1\mu_2\cdots\mu_s}_n &= 0\\
	\end{split}
\end{equation}

\subsection{Probability distribution function and classical limit} \label{pdcl}

The solutions \eqref{funcformdesc} and \eqref{spdes} and not yet properly normalized. We work with the Klein Gordon Norm. Consider a constant time surface, $\tau = u+v = i\log\sqrt{\frac{W}{\bar W}} = \tau_0$.
The unit normal vector to this surface is given by,
\begin{equation}
	\hat n = -\frac{1}{\cosh\rho} (\partial_u +\partial_v) = \frac{i}{\sqrt{2}}\left(\sqrt{\frac{\bar W}{W}}\partial_{\bar W}-\sqrt{\frac{W}{\bar W}}\partial_W\right)
\end{equation}
The induced metric on this constant time surface takes the form
\begin{equation}\label{adsind}
	ds^2|_{\tau=\tau_0}= \sinh^2 \rho \,  d \phi^2  + d \rho^2 
\end{equation}
The volume form in the constant time surface is,
\begin{equation}
	\sqrt{g_{ind}}\, d\rho\, d\phi = \sinh \rho \,d\rho \, d\phi = \sqrt{2}|Z| \,d\rho \, d\phi
\end{equation}

Now consider the  tensor
\begin{equation}
	\begin{split}
		S^{(+)}_{\mu_1\mu_2\cdots\mu_s} &= \psi_{h,n,0}\bL_{1\,\mu_1} \bL_{1\,\mu_2} \cdots \bL_{1\,\mu_s}\\
		S^{(+)*}_{\mu_1\mu_2\cdots\mu_s} &= (-1)^s \psi^*_{h,n,0}\bL_{-1\,\mu_1} \bL_{-1\,\mu_2} \cdots \bL_{-1\,\mu_s}\\
	\end{split}
\end{equation}
Its Klein-Gordon norm is given by the integral, over the constant time surface of
\begin{equation} \label{probabdens}
	\begin{split}
		&d \phi d t \sqrt{g_{ind}}\,i \left(S^{(+)*}_{\beta_1\cdots\beta_s} \hat n.\nabla S^{(+){\beta_1\cdots\beta_s}} - S^{(+){\beta_1\cdots\beta_s}} \hat n.\nabla S^{(+)*}_{\beta_1\cdots\beta_s} \right)\\
		&=\frac{1}{2^{s-1}}\left((2h+n)\left|\frac{Z}{W}\right|+2s|Z\,W|\right)|\psi_{h,n,0}|^2\\
		&=\frac{1}{2^{s-1}}\left((2h+n)\tanh\rho+s\sinh2\rho\right)\frac{\tanh^{2n}\rho}{\cosh^{4h}\rho}\\
	\end{split}
\end{equation}

In the problem of physical interest in this paper (see the next Appendix) $h, s$ and $n$
are all very large (and, generically, comparably large). In this case the spatial distribution of the probability density 
\eqref{probabdens} is dominated by the expression 
$$\frac{\tanh^{2n}\rho}{\cosh^{4h}\rho}.$$
This function is peaked at 
\begin{equation}
	\rho_0 \simeq \sinh^{-1}\sqrt{\frac{n}{2h}}
\end{equation}
with a width of order 
$$ \frac{1}{2\sqrt{h}}.$$

Note that the parameter $s$ affects the spatial distribution of the probability density only at the first sub-leading order in a large parameter expansion. In particular, the peak of the probability density function lies at the same position for every value of $s$.

Similarly for the other solution
\begin{equation}
	\begin{split}
		S^{(-)}_{\mu_1\mu_2\cdots\mu_s} &= \psi_{{\bar h},0,n}L_{1\,\mu_1} L_{1\,\mu_2} \cdots L_{1\,\mu_s}\\
		S^{(-)*}_{\mu_1\mu_2\cdots\mu_s} &= (-1)^s \psi^*_{{\bar h},0,n}L_{-1\,\mu_1} L_{-1\,\mu_2} \cdots L_{-1\,\mu_s}\\
	\end{split}
\end{equation}
repeating the same calculation, we find that the probability density is given by 
\begin{equation}
	\begin{split}
		&\sqrt{g_{ind}}\,i \left(S^{(-)*}_{\beta_1\cdots\beta_s}\hat n.\nabla S^{(-){\beta_1\cdots\beta_s}} - S^{(-){\beta_1\cdots\beta_s}}\hat n.\nabla S^{(-)*}_{\beta_1\cdots\beta_s} \right)\\
		&=\frac{1}{2^{s-1}}\left((2{\bar h}+n)\left|\frac{Z}{W}\right|+2s|Z\,W|\right)|\psi_{{\bar h},0,n}|^2\\
		&=\frac{1}{2^{s-1}}\left((2\bar h+n)\tanh\rho+s\sinh2\rho\right)\frac{\tanh^{2n}\rho}{\cosh^{4\bar h}\rho}\\
	\end{split}
\end{equation}
The peak of probability density of the $S^{(-)}_n$ wave function lies at,
\begin{equation}
	\rho_0 \simeq \sinh^{-1}\sqrt{\frac{n}{2{\bar h}}}
\end{equation}

Let us summarise. For all values of the spin of the primary field, the probability distribution associated with the $n^{th}$ descendent wave function is highly peaked at the radial location \eqref{eqforhostar}. However, the wave function is completely delocalized in the $\phi$ direction (indeed the only dependence of the wave function on $\phi$ is through the phase $e^{i m \phi}$ where $m$ is the $z$ component of the angular momentum).  In other words, this wave function 
is extremely well localized around a particular classical orbit. But it is completely delocalized in the `where on the orbit one happens to be' coordinate.

\subsection{Matching the geodesic and the wave function calculation} \label{mgwf}

As we have explained in the previous subsection, in the limit that $n$ and $\bar h$ are both large, the wave function corresponding to the $n^{th}$ descendent is strongly localized around the radial location 
\begin{equation}\label{eqforhostar}
	\rho_0 = \sinh^{-1}\sqrt{\frac{n}{2 \bar h}}
\end{equation}

Localization of wave functions at large quantum numbers is usually associated with a classical limit. In the current instance, the classical limit of the primary wave function is the geodesic that sits at the centre of 
AdS, \eqref{esu2}. The classical limit of the 
$n^{th}$ descendent of this primary is a geodesic obtained by `boosting' the primary, (the boost in question lies in the right moving $SL(2,R)$). The boost in question should be chosen so that the resulting geodesic carries 
$\bL_0=\bar h+n$. Precisely such geodesics were studied around \eqref{newelem2}. Recall that 
these geodesics were obtained (see \eqref{newgeo})  by acting on the geodesic at the centre of $AdS$ by $f_A f_a$ from the right 
(see \eqref{fa} and \eqref{rottimf} for definitions). The $\bL_0$ value of the resultant geodesic is $\bar h \cosh 2 a$ \footnote{The 
	${L}_0$ value of this geodesic is unchanged at  ${h}$.}
and the resultant geodesic lives on an orbit located at 
\begin{equation}\label{posit}
	\rho=a.
\end{equation} 
We see that the charge and radial locations of descendants and 
geodesics both agree with each other if we choose $(h, {\bar h})$ of the primary to match the corresponding charges for the geodesic at the centre of $AdS$, and also make the identifications  
\begin{equation}\label{dictionary}
	\begin{split} 
		&n=0\\
		&n'=\bar h(\cosh 2a - 1) \\
	\end{split}
\end{equation}
We can invert the last of \eqref{dictionary} to solve for $a$ in terms of $n$ and $\bar h$. We find 
\begin{equation}\label{asol} 
	a = \sinh^{-1}\sqrt{\frac{n}{2\bar h}}
\end{equation}

Comparing \eqref{asol} and \eqref{eqforhostar}, we see 
that the wave function of the descendant state $\psi_{h,0, n}$ is peaked at exactly the same value of $\rho$ as the location of classical orbits of a massive particle of mass $m$ that carries the same value of $L_0$ and ${\bar L}_0$ charges. 

Note that our matching procedure above has completely determined the value of $a$. However, it has had nothing to say about the parameter $A$. Geodesics at the value of $a$ presented in \eqref{asol} - but at all values of $A$ - have the same charge and radial location as the descendent wave function. 
We can understand this fact in the following terms. As we have explained in subsection \ref{qgc}, the phase space of geodesics in 
$AdS_3$ is the coset $SO(2,2)/(SO(2)\times SO(2) )= SL(2,R)/U(1) \times SL(2,R)/U(1)$. 
The left-moving part of this co-set plays no role in the current discussion and so can be ignored: effectively our classical phase space is $SL(2,R)/U(1)$. The quantities $a$ and $A$
are parameters on this phase space. The quantum wave function corresponding to the $n^{th}$ descendent is a state that is highly localized in $a$ but completely delocalised in $A$. As  $A$ determines the angular location of the particle at $\tau=0$ (see \eqref{newelem2}), the fact that the wave function does not single out a particular value of $A$ is simply a restatement of an already noted fact (the last paragraph of the previous subsection), namely that the wave function is not peaked at a particular position on the orbit, but is uniformly smeared all over the orbit. 

The fact that no particular geodesic \eqref{newelem2} can well approximate the descendent wave function could have been anticipated from a consideration of $SL(2,R)$ charges. As we have explained in \eqref{cnewsolA}, the geodesics described at particular values of $a$ and $A$ all carry definite $SL(2,R)$ charges. Geodesics at a given value of $a$ (but different values of $A$) all carry the same value of $L_0$. This is the value we have already matched with the 
descendent wave function earlier in this sub-subsection. However, geodesics at different values of $A$ each carry a different, $A$ dependent,  values $\bL_1$ and $\bL_{-1}$ (see the second term on the RHS of \eqref{cnewsolA}). In contrast, our descendent wave function is such that  $\langle L_1\rangle=\langle L_{-1}\rangle=0$ on this state. 
It is thus impossible for our descendent state to be well localized around any one of the classical geodesics \eqref{newelem2}. In fact, the descendent state is an equal linear combination of these geodesics at all values of $A$, explaining how its average $L_1$ and $L_{-1}$ charges vanish.

We could obtain a state that is highly peaked in $\phi$ at $t=0$ (and so more closely resembles a classical geodesic) by considering a linear combination of descendants $n$. Let us suppose we want to achieve a definite value of the location, $\phi$, of our particle on its orbit,  with accuracy$\delta \phi$. This can be achieved by a state of the sort 
\begin{equation}\label{psithetanot}
	\psi_{h, \phi_0, \delta \phi}
	= \sum_{k=-M}^M |a_k| e^{i k \phi_0 } \psi_{h,0, n+k}
\end{equation} 
where $M \sim  \frac{1}{\delta \phi}$ but $M \ll n$.
Here the functions $|a_k|$ are some smooth envelope functions like a Gaussian that is maximum at $k=0$, is symmetric in $k \leftrightarrow -k$, and is very small at $k=M$. Roughly speaking we can think of $|a_k|= \chi(k/M)$ where $\chi$ is a smooth function that interpolates from zero (when its argument is minus one) to unity (when its argument is zero) to zero (when its argument is one). Saddle point type reasoning tells us that this wave function is peaked at a value of $\phi$ given by 
\begin{equation}\label{thetapeak} 
	\phi- t = \phi_0 
\end{equation} 
Note, comparing with \eqref{lkjm} we see,
\begin{equation}
	\phi_0 = 2A
\end{equation}

Using (see \eqref{lkjm})  
\begin{equation} \label{lkjhnorm}
	\begin{split}
		\bL_0{\tilde \psi}_{h, 0,n} &= (h+n)\tilde\psi_{h,0, n}\\
		\bL_1{\tilde \psi}_{h, 0, n}  &= \sqrt{n(2\bar h+n-1)} {\tilde \psi}_{h, 0, n-1}\\
		\bL_{-1} {\tilde \psi}_{h, 0, n}  &= \sqrt{n(2\bar h+n-1)} {\tilde \psi}_{h, 0, n+1}
	\end{split}
\end{equation}
it follows that, on such a state in large $h,\,n$ limit, 
\begin{equation}\label{naivest} 
	\langle  \bL_1 \rangle = 
	\left( \langle \bL_{-1} \rangle \right)^* 
	= e^{i\, 2A} \sqrt{ n(2\bar h+n)} 
\end{equation} 
Using \eqref{asol} allows us to re-express \eqref{naivest} 
as 
\begin{equation}\label{naivestn} 
	\langle  \bL_1 \rangle = 
	\left( \langle \bL_{-1} \rangle \right)^* 
	= e^{i\,2A} h \sinh (2 a) 
\end{equation} 
in precise agreement with the value of the off-diagonal charge in \eqref{cnewsolA}.

\subsection{Primary and descendant contributions in the saddle point approximation} \label{pdspa} 

As we have explained in great detail in subsection \ref{qgc}, the state space of an irreducible representation of $SO(d,2)$ may be obtained from the quantization of geodesics (plus the co-adjoint orbits of $SO(d)$ charges) in $AdS$. Similarly, the thermal partition function over this state space may be obtained by performing a world line path integral over particle trajectories in thermal AdS space. 
\footnote{More precisely, this partition function is obtained by restricting particle trajectories to those in which the worldline circle winds the spacetime circle precisely once. These are the trajectories we study in this subsection. Trajectories that wind $n$ times capture the contributions of $n$ particles in $AdS$: summing over all such windings yields the partition function over the Fock Space of particles in $AdS$. As mentioned above, our interest, in this section is in the partition function over the single-particle Hilbert Space, so we restrict our path integral to trajectories that wind the time circle exactly once.}

In the limit that $\Delta$ is very large (with spin held fixed) the path integral over worldline trajectories (that wind once around the thermal $AdS$ circle) may be computed in the saddle point approximation. The saddle point geodesic (the one with the shortest Euclidean length) sits at the centre of $AdS$ space\footnote{  The centre of $AdS$  has the following definition. $AdS_4$ space hosts a unique time-like geodesic that is everywhere tangent to the killing vector $\partial_t$ where $t$ is the global time. Translations $\partial_t$ are generated by the energy operator $E$. Through this paper, we refer to any point on this geodesic as lying at the centre of $AdS_4$. Note that the notion of a centre only makes sense once we have made a particular choice of $E$ (or global time). Of course the entropy $S(E, J_z)$ and partition function $Z[\beta, \omega]$ are also meaningful only once we have made such a choice.}, and its contribution to the path integral is $e^{-\beta \Delta}$.

We have still to perform the integral over the $SO(d)$ co-adjoint orbit, and also the integral over fluctuations of the trajectory 
over its saddle point value. The first of these produces the partition function over the $SO(d)$ representation in which the particle lies. Combining it with the saddle point value gives us the partition function overall primary states, $Z_{\rm Primaries}$. 

The one loop path integral over fluctuations of the trajectory about this saddle point captures the contributions of descendants, and so evaluates (in the special case $d=3$) to $\frac{1}{(1-e^{-\beta(1-\omega)})(1-e^{-\beta(1+\omega)})(1-e^{-\beta})}$. Putting these pieces together we obtain (for $d=3)$
\begin{equation}\label{Zsc}
	Z=  \frac{e^{-\beta \Delta} \sum_{m=-s}^{m=s} e^{\beta \omega m }}{(1-e^{-\beta(1-\omega)})(1-e^{-\beta(1+\omega)})(1-e^{-\beta})}.
\end{equation} 
(where we have used the fact that when $d=3$
and in the spin $s$ representation
$$Z_{\rm Primaries}  = e^{-\beta \Delta} \sum_{m=-s}^{m=s} e^{\beta \omega m }$$
Of course, this is the exact answer for the partition function over the given (long) $SO(3,2)$ representation module. 

We emphasize that the saddle point contribution to $Z$ captured only the contribution of one primary state. The contribution of all other primary states -- as well as the contribution over descendants came from the one-loop determinant over this saddle point. 

In the discussion above we were able to perform the integral over particle trajectories in the saddle point approximation, but were constrained to perform the integral over co-adjoint orbits exactly (in order to obtain the group character of the primaries in the numerator of \eqref{Zsc}). The situation in this regard simplifies if $s$, like $\Delta$, is also large. In this, the integral over the co-adjoint orbit can also be performed in the saddle point approximation. To see how this  
works let us, as above,  specialize to the case $d=3$. In this case, the saddle point configuration is the one in which the coadjoint 
charge vector - whose length is $s$ -points 
in the positive $z$ direction (we assume that 
$\omega$ is positive). Consequently, the saddle
point approximation to the partition function is simply
$$e^{-\beta \Delta +\beta \omega s}$$
The exact answer \eqref{Zsc} is this saddle point answer times the extra factor 
\begin{equation}\label{bobo}
	\frac{ \sum_{a=0}^{2s} e^{-\beta \omega a}}
	{(1-e^{-\beta(1-\omega)})(1-e^{-\beta(1+\omega)})(1-e^{-\beta})}
\end{equation} 
Up to exponential accuracy in $s$ 
\footnote{In the next Appendix $s$ will be of order $\frac{1}{G}$. Consequently, this approximation is good for all orders in perturbation theory.}
we can ignore the upper limit of the summation in \eqref{bobo} and approximate \eqref{bobo} as 
\begin{equation}\label{boboapp}
	\frac{ 1}
	{(1-e^{-\beta \omega})(1-e^{-\beta(1-\omega)})(1-e^{-\beta(1+\omega)})(1-e^{-\beta})}
\end{equation} 

\eqref{boboapp} is easy to understand. Recall that the phase space, in the case under study, is 8 dimensional. $6$ of the $8$ coordinates have to do with the position and velocity of the particle in $AdS_4$, while the remaining two coordinates are locations on the $S^2$ that parameterize $SO(3)$ co-adjoint orbits. Moving to Lorentzian space for a moment, we might the linearization about this saddle point configuration (in the 8-dimensional tangent space of phase space) to yield
4 normal mode oscillator degrees of freedom. The determinant of the  Euclidean path integral should, then, reduce to the the product of thermal partition functions of these four normal mode oscillators. This exactly matches the form of the final result \eqref{boboapp}, provided the four normal modes have the following charges. 
\begin{itemize} 
	\item One of the normal modes carries angular momentum $-1$ and energy zero. This mode arises from the degrees of freedom in $SO(3)$ co-adjoint space.
	\item The remaining three normal modes each carry unit energy, but carry angular momenta  
	$1$, $-1$ and $0$ respectively. These modes describe wiggles of the geodesic in $AdS_4$.
\end{itemize} 

The contribution of each of the four normal modes produces a factor of the form \eqref{moes}, yielding the four factors in the denominator in \eqref{boboapp}. 

We conclude that the exact answer for the thermal partition function \eqref{Zsc} can be 
reproduced, up to exponential accuracy in the saddle point approximation by retaining only the saddle point and determinant contributions. 
The determinant is generated by four normal modes

\section{Revolving black hole solutions} \label{rbs}

In the introduction (see subsection \ref{divzero}) we have briefly discussed the divergence of entropy leading to the construction of the sub-leading ‘revolving black hole’ solution. In this Appendix, we elaborate on the thermodynamics and the physical interpretation of these solutions.

\subsection{Zero mode contribution to the black hole determinant}\label{descendants} 

In subsection \ref{pdspa} we have explained that the world line path integral for large 
$\Delta$ and large spin particles in thermal AdS 
can be performed in the saddle point approximation. The saddle point captures the contribution of the highest spin state to 
the partition function, while the determinant around the saddle point captures the contribution of all other primaries plus descendants. 

Now Kerr-AdS black holes are also classical saddle points that sit at the centre of $AdS$ space. Of course, black holes capture the contribution of an ensemble of representations of $SO(d,2)$ rather than the contribution of 
a single representation. Given the discussion 
of subsection \ref{pdspa}, it is natural to 
expect (and we conjecture) that the classical entropy of a spinning black hole (whose spin, for simplicity, lies in a single two plane, let's say the 12 two plane) approximately computes the entropy of only the highest spin primaries. The contribution of all other primaries - as well as of descendants - will be captured by the one-loop determinant. 

We now describe, in much more detail, precisely how we expect this to work at the technical level. It was demonstrated in 
\cite{Denef:2009kn, Denef:2009yy} that the 
one loop partition function about a Euclidean black hole is given by a product of factors, one associated with each of the quasinormal modes of the black hole. This observation is the black hole analogue of the well-known fact (one we have used extensively in the main text, as well as at the end of Appendix \ref{pdspa}) that determinant around (say) thermal AdS is the product over terms, one associated with each of the definite energy solutions of the Lorentzian theory. Each of these individual terms is simply the partition function 
\begin{equation}\label{moe}
	{\rm Tr} e^{-\beta H+ \omega \beta J}
\end{equation}
over the Hilbert Space obtained by quantizing the phase space generated by the corresponding Lorentzian solutions. If the mode in question has energy $E$ and angular momentum 
$J$, its contribution to the partition function is given by 
\begin{equation}\label{moes}
	\frac{1}{1- e^{-\beta E+ \omega \beta J} } 
\end{equation}

The chief new element that arises in the case of Black Holes is that the quasi-normal modes have complex frequencies. As a consequence, their contributions to the determinant are that of a harmonic oscillator with friction \cite{Denef:2009kn, Denef:2009yy}.
If the black hole under study 
happens to possess some normal modes (modes whose frequencies happen to be real) then the contribution of these modes to the black hole determinant continues to take the usual form
\eqref{moe}, i.e. the form \eqref{moes}. While almost all quasi-normal modes 
around Kerr-AdS black holes have complex frequencies, one class of modes has real 
frequencies. These are the modes that are generated 
by the action of the symmetry group $SO(d,2)$ on the black hole solution. \footnote{This family of black holes may be constructed by performing coordinate transformations on the original solution. As the coordinate transformations are non-vanishing
	at infinity, the resultant solutions are related by large gauge transformations, and so are physically distinct.} The manifold of solutions so obtained is precisely the co-adjoint orbit (see subsection \ref{qgc}) associated with a primary, whose highest weights are the spins of the original black hole (the one that sits at the centre of AdS). The phase space thus obtained is precisely the one described in 
subsection \ref{qgc}. It follows that the quantization of this phase space gives the answer described in subsection \ref{qgc}. 

Let us specialize, for concreteness, to the special case of a rotating black hole with spin $s$ in $AdS_4$. This is precisely the case studied at the end of subsection \ref{pdspa}. As explained in that subsection the space of black hole solutions so obtained is 8 dimensional. At the nonlinear level, this space is isomorphic to the group Coset $SO(3,2)/(SO(2)\times SO(2))$ (see subsection \ref{qgc}). However the determinant around the Kerr-AdS black hole saddle does not see this full nonlinear manifold of gravitational solutions, but only accesses the tangent space around the saddle point. We studied the quantization of this 
tangent space under \eqref{boboapp}. As explained there, the tangent space consists of four normal modes, whose energies and angular momenta are completely determined by symmetry considerations, and are given as explained at the end of subsection \ref{pdspa}. We have one mode with zero energy and angular momentum $-1$, together with three modes with unit energy and angular momenta $1$, $0$ and $-1$. The formulae of \cite{Denef:2009kn, Denef:2009yy} then tells us that the contribution of these modes to the black hole determinant is given by \eqref{boboapp}.

To reiterate, symmetry considerations assure us that the quasinormal mode analysis about this $d=4$ Kerr-AdS black hole will have four normal modes with charges described above. The contribution of these modes to the determinant is \eqref{boboapp}. In the next subsection, we will now explore the 
physical consequences of these technical results. 

\subsection{Divergence from descendants}

The physical importance of the contribution 
\eqref{boboapp} to the partition function is 
the following. The factor $\frac{1}{(1-e^{-\beta(1-\omega)})}$,  which appears in 
\eqref{boboapp},  clearly diverges as $\omega \to 1$. This factor captures the contribution of the $J_z=1$ derivative, $\partial_+$, to the partition function. This derivative is Boltzmann unsuppressed at $\omega=1$. We conclude that, unlike thermal AdS space (see subsection \ref{divzero}) the black hole saddle hosts one mode whose contributions to the determinant diverge in the limit $\omega \rightarrow 1$. \footnote{ The mode in question is one of the `zero modes' for the centre of mass motion of the black hole, as described above. Note that this mode is missing in global $AdS$ for two related reasons. From the point of view of the bulk, the zero modes 
	are absent because global $AdS$ is empty, it contains nothing that can move. From the viewpoint of conformal representation theory, these modes are absent because the vacuum module is uni-dimensional (it has no $SO(3,2)$ descendants).} We conclude that the logarithm of the partition function receives the divergent contribution
\begin{equation}\label{divcont}
	\ln Z^{\rm descendants}=	-\ln (1-e^{-\beta(1-\omega)})
\end{equation} 
from this descendant mode.

As a brief aside, we note that the discussion of the previous paragraphs can immediately be turned into the following general theorem. The partition function of every CFT diverges for $\omega >1$. This result follows because - as we have explained above, the contributions of descendants to the partition function diverge for 
$\omega>1$. Since the result follows from simple representation theory considerations, it holds for any CFT. 

Using the usual thermodynamical formulae  
\begin{equation}\label{thermod}
	\begin{split} 
		&J = \frac{1}{\beta} \partial_\omega \ln Z\\
		&E= -\partial_\beta \ln Z + \frac{\omega}{\beta} \partial_\omega \ln Z\\
		& S= \ln Z - \beta \partial_\beta  \ln  Z\\
	\end{split} 
\end{equation} 
(where all $\beta$ derivatives are taken at constant $\omega$, and all $\omega$ derivatives are taken at constant $\beta$) we find that the charge and energy associated with the descendant derivatives is 
\begin{equation}\label{chargeen} 
	J=E= \frac{1}{\beta(1-\omega)}
\end{equation} 	
while the entropy, as usual, is negligible \footnote{While thermodynamics assigns this solution a nonzero entropy of order $\ln (1-\omega)$ this answer is an artifact of thermodynamics being only marginally applicable to the case under study. The actual entropy associated with descendants is clearly zero, as there is a unique 
	descendant state with energy and angular momentum $n$ units larger than that of the primary.}. 
It follows, in particular, that when $1-\omega \sim \sqrt{G}$ (as is 
the case for the Grey Galaxy solution), the charges in descendant derivatives are of order $\frac{1}{\sqrt{G}}$ 
and so are subleading compared to the energy and angular momentum of the gas and that of the classical black hole. \footnote{On the other hand, were we to scale $1-\omega \sim G$, then the descendants would carry the energy 
	of order $\frac{1}{G}$ - comparable to the classical black hole - but the energy of the gas, at these values 
	of $\omega$, would be $\sim \frac{1}{G^2}$, and so superclassical. We conclude there is no thermodynamically 
	sensible saddle in which descendants contribute comparably to the gas.} 

Despite the fact that the contribution of descendants is subleading compared to that of the gas, in the next subsection we will attempt to better understand the physical nature of the
gravitational solutions in which charge and energy are redistributed between the classical saddle and descendants (with the gas completely unoccupied) in a manner that maximises the black hole entropy. As descendants - like the gas - carry $E=J$, the classical thermodynamics of the solutions we study more in the next subsection are identical to that of the Grey Galaxy solutions. In fact, the discussion 
of section \ref{me} applies unchanged to the new saddles of this Appendix except for one change: the subleading entropy of the gas in the Grey Galaxy solutions - which we computed to be of order $1/\sqrt{G}$ in \eqref{sej}, is
actually zero in the current context. This is because a condensate of derivatives carries no entropy (there is exactly one way to build a state of any given charge). In fact, from the viewpoint of section \ref{me}, the reason that the Grey Galaxy saddle dominates over the `revolving black hole' saddle of this section, is that the former carries an extra entropy of order $\frac{1}{\sqrt{G}}$, while the revolving black hole solution of this Appendix carries zero entropy.

Given the fact that the revolving black hole saddles that we construct in this Appendix are always subleading, the reader may wonder why we bother to further study these solutions in the next subsection. We offer three reasons
\begin{itemize}
	\item It is often interesting to study subleading saddles, even if they are not dominant, particularly 
	if they are long-lived. In the current context, we do not see a clear classical mechanism for the decay 
	of the revolving black hole solutions to the Grey Galaxy solutions. We thus suspect that the revolving black hole solutions are very (possibly non-perturbatively) long-lived, even though they are not entropically dominant. 
	\item In the case of black holes in $AdS_4$, the temperature of black holes with $\omega=1$ was bounded from below (see subsection \ref{omegone}). It is possible that this will not be the case in some other situations (e.g. either rotating black holes in higher dimensions or black holes that have both charges as well as rotation). In this case, 
	it could well turn out that revolving black hole solutions actually dominate thermodynamics at very low 
	temperatures (e.g. temperatures of order $-\ln (1-\omega)$) because all gas modes have a twist gap (a gap in 
	$E-J_z$) while the descendant modes have a twist gap of zero. 
	\item Finally, as we have already mentioned in the introduction, while the revolving black holes may not be dominant solutions, they certainly are (very simply constructed) solutions, and so their entropy gives a lower bound for the true entropy of the saddle point. As we have explained in section \ref{susy}, this argument can be used to find a lower bound for the entropy of supersymmetric states in ${\cal N}=4$ Yang-Mills.
\end{itemize} 

\subsection{Physical interpretation of the Bose condensate of derivatives in terms of Revolving Black Holes}

As we have explained in the previous subsection, every point in the blue-shaded part of the  phase diagram Fig. \ref{phas} 
admits both a Grey Galaxy solution (as we have explained in detail in the main text) as well as another solution describing a black hole `in 
equilibrium with a Bose condensate of derivatives' (described in the previous subsection). In this subsection we explore the physical interpretation of this new solution, the black hole `in equilibrium with a Bose condensate of derivatives'.

In fact, the physical interpretation we are looking for was already worked out in Appendix \ref{fpa}. In that Appendix, the motion of a spinning particle in $AdS$ was analyzed. In 
subsections \ref{pdcl} and \ref{mgwf} we constructed the wave function corresponding to the $n^{th}$ descendant state of a particle of left moving dimension (i.e. $\frac{\Delta+s}{2}$) equal to $h$. In the limit that $h$ and 
$n$ are both large, we demonstrated there that 
the wave function in question was given by a wave function over geodesics of the form 
$f_A f_a X$. $f_a$ and $f_A$ are the $SL(2,R)$
elements respectively given in \eqref{fa} and 
\eqref{rottimf}, and $X$ is the geodesic sitting at the centre of $AdS$, given by 
\eqref{esu2}. The geodesics that make up the wave function have the definite value of $a$
given in \eqref{asol}, but do not have a definite value of $A$. In fact, the correct description of the state puts it in a wave function for $A$ proportional to 
\begin{equation}\label{wfphsp}
	\psi= \int d A e^{i n A} |A\rangle
\end{equation}

Although the discussion in \ref{pdcl} and 
\ref{mgwf} were presented in the context of 
$AdS_3$, the results apply without modification
to $AdS_4$. Consider those geodesics moving in the space \eqref{coordinates} that everywhere have $X_3=0$. Such geodesics are identical to geodesics in $AdS_3$. Similarly wave functions of the form \eqref{sppri} and \eqref{spdes} 
also represent solutions to the wave equation in $AdS_4$: they are simply those solutions that happen to be independent of $X_4$. It is also easy to convince oneself that \eqref{sppri} are maximal spin primaries from the viewpoint of  $AdS_4$. The solutions in \eqref{spdes} are precisely those corresponding to primaries of maximal spin acted on with derivatives which further increase this spin. These are precisely the modes that give rise to the divergence in 
\eqref{boboapp}. 

Although the discussion in Appendix \ref{fpa} was performed for spinning particles, all the actual analysis used only symmetry considerations, and so applies without modification to black holes. 

We conclude that the physical interpretation 
for the black hole `in equilibrium with a Bose condensate of derivatives' is the wave function 
\eqref{wfphsp}. In this context, the ket $|A\rangle$ denotes the metric of the Kerr-AdS black hole boosted (coordinate transformed) by 
the symmetry element $f_A$. 
\footnote{Note that $f_A$ belongs to the left moving SL(2,R) in 
	$SO(2,2)$, which, in turn, is a subgroup of the symmetry group $SO(3,2)$. }
The completely remarkable aspect of this interpretation is that the state in question is 
quantum in nature, even though it describes the motion of a highly macroscopic object which happens to be a black hole. 

Each individual classical configuration that contributes to this wave function is a black hole whose centre of mass revolves around one of the geodesics in \eqref{newelem2}.

It is very unusual for macroscopic objects to live in wave functions that are not well approximated by classical configurations. The standard reason offered for this observation is that macroscopic objects are easily entangled with orthogonal environmental degrees of freedom, which prevents interference, effectively reducing such systems to a classical ensemble over possible values.
There may well be a sense in which this happens in the current context. We leave this question for further investigation.

\section{Spherical harmonics at large $m$}\label{shll}

$\theta$ dependent part of the spherical harmonics with $\phi$ dependence of the form $e^{im\phi}$ satisfies the following equation:
\begin{equation}\label{sphharm} 
	-\frac{1}{\sin\theta}\frac{\partial}{\partial\theta}\left(\sin\theta \frac{\partial P_{lm}(\theta)}{\partial\theta}\right)+\frac{m^2 P_{lm}(\theta)}{\sin^2\theta}=l(l+1)P_{lm}(\theta)\ .
\end{equation}

To convert this to standard Schrodinger equation form, we perform the following change of variable:

\begin{equation}
	P_{lm}(\theta)=\frac{\psi_{lm}(\theta)}{\sqrt{\sin\theta}}
\end{equation}

$\psi(\theta)$ then satisfies the following equation:
\begin{equation}
	-\psi''_{lm}+\frac{\psi_{lm}(4m^2-(1+\sin^2\theta))}{4\sin^2\theta}=l(l+1)\psi_{lm}
\end{equation}

Clearly, this potential has a minima at $\theta=\frac{\pi}{2}$. At large values of $l$ and $m$ with $a=l-m$, fixed, the potential at $\theta=\frac{\pi}{2}$ becomes steeper and steeper. Therefore, we work in the coordinate $\delta\theta=\frac{\pi}{2}-\theta$. In the limit of small $\delta\theta$, we get the following equation:
\begin{equation}
	-\psi_{lm}''(\delta\theta)+\frac{\delta\theta^2(4m^2-1)}{4}\psi_{lm}
	=\left(l(l+1)-\frac{4m^2-2}{4}\right)\psi_{lm}
\end{equation}

Taking large $l$, large $m$ limit with $a=l-m$ fixed, we get 
\begin{equation}
	-\psi_{la}''(\delta\theta)+l^2\delta\theta^2\psi_{lm} =2 l\left(a+\frac{1}{2}\right) \psi_{la}
\end{equation}
This is the Schrodinger equation for a harmonic oscillator with unit mass, $\hbar=1$, 
frequency $l$ and energy $l\left(a+\frac{1}{2}\right)$ which means that $a$ is the level of the harmonic oscillator.

It follows that when $l$ is large but $a$ is of order unity
\begin{equation} \label{ylm} 
	Y_{l, l-a}(\theta, \phi)= \left({\frac{l}{\pi }}\right)^{\frac{1}{4}} \frac{1}{\sqrt{2 \pi } \sqrt{2^a a!}} e^{i(l-a) \phi} e^{-(l-a)\frac{(\delta \theta)^2}{2}} H_a( \sqrt{m} \delta \theta)=e^{i(l-a)\phi}\frac{\psi_a^{HO}(\delta\theta)}{2\pi}
\end{equation} 
\footnote{The Gaussian decay of this function may 
	directly be understood in the following terms. 
	A spherical Harmonic with total angular momentum $l$ and $z$ component of angular momentum $l-a$
	is proportional to $\sin^{l-a}\theta$ times a degree $a$ polynomial of $\cos \theta$. The exponential dependence in \eqref{ylm} comes from the factor of $\sin^{l-a}\theta$.  Setting $\theta= \frac{\pi}{2} - \delta \theta$, with $\delta \theta^2 \ll 1$ we have 
	\begin{equation}\label{sinthetat}
		\sin^{l-a}(\theta) \approx \left( 1 - \frac{(\delta \theta)^2}{2} \right)^{l-a}
		= \left( \left( 1 - \frac{(\delta \theta)^2}{2} \right)^{\frac{2}{\delta \theta^2}}\right)^{\frac{(l-a) \delta \theta^2}{2}}\approx \left( e^{-1} \right)^{\frac{(l-a) (\delta \theta)^2}{2}}
	\end{equation} 	
	in agreement with \eqref{ylm}. This exponential kills the function unless $\delta \theta$ is small. When $\delta \theta$ is small $\cos \theta 
	\approx \delta \theta$. The polynomial in $\cos \theta$, mentioned above, thus turns into a polynomial in $\delta \theta$, and in fact, turns out to be $H_a(\sqrt{m} \delta \theta)$. } 
where $m=l-a$ and $H_a(y)$ $a^{th}$ Hermite polynomial, normalized in the usual manner, i.e. normalized 
so that 
$$ \int_{-\infty}^\infty  e^{-y^2} H_a(y) H_b(y) = \delta_{ab} (2^a a! \sqrt{\pi})  $$

Here $\psi_a^{HO}(\delta\theta)$ is the unit normalized wavefunction of the harmonic oscillator at level $a$ with frequency $l$. There is a factor of $\sqrt{2\pi}$ in the denominator since the spherical Harmonic must be unit normalized when integrated with respect to $\theta$ and $\phi$.

In the main text, we need to compute the sum 
\begin{equation}\label{propsum0} 
	\sum_{a=0}^\infty |Y_{l, l-a}(\theta, \phi)|^2 e^{-\alpha (a+\frac{1}{2})}
\end{equation} 
Focussing on the contribution of large $l$ but $a$ of order unity we find  
\begin{equation}\label{propsum} 
	\sum_{a=0}^\infty |Y_{l l-a}(\theta, \phi)|^2 e^{-\alpha (a+\frac{1}{2})}
	\approx \sum_a\frac{|\psi_{la}|^2}{2\pi}e^{-l(a+\frac{1}{2})}
	=\frac{1}{2\pi}\langle \delta \theta| e^{-\alpha H} |\delta \theta \rangle 
\end{equation} 

The last expression on the RHS 
of \eqref{propsum} is simply  
matrix element of the evolution operator of the Harmonic oscillator. This quantity is well known (for instance it is easily computed via using Feynman's path integral method), yielding
\begin{equation}\label{propexp} 
	\langle \delta \theta| e^{-\alpha H} |\delta \theta \rangle 
	= \frac{\sqrt{l}}{\sqrt{2\pi\sinh l\alpha}}\exp{\left(-l\delta\theta^2\frac{\cosh l\alpha-1}{\sinh l\alpha}\right)}
\end{equation}

\section{The radial dependence of AdS wave functions at  large $l$} \label{fnll}

In this Appendix, we demonstrate that in the large $l$ limit, the radial part of the mode wavefunction obeys a Schrodinger equation obeyed by the angular momentum $L$ sector of a three-dimensional harmonic oscillator where $L= \Delta -2$. (Note that $L$ is a fictional angular momentum, and, 
in particular, should nowhere be confused with 
the real angular momentum $l$). 
We use this connection to 
compute the wave function - and a relevant sum over mod squared wave functions weighted by energy 
- in this limit.

Let the wave function for our scalar field of mass 
$M=\sqrt{\Delta(\Delta-3)}$ take the form 
\begin{equation}\label{sfwf} 
	\phi= Y_{lm} e^{-iE t} F(r)\ .
\end{equation} 
It was shown in \cite{kaplan} that the field redefinition 
\begin{equation}
	\label{rad}
	\begin{split}
		F(r)=\frac{1}{\sqrt{1+r^2}} \chi(r)
	\end{split}
\end{equation}
\footnote{At large $r$ this redefinition simplifies to 
	\begin{equation}\label{lrfchi}
		F(r)= \frac{\chi(r)}{r}
\end{equation} }
transforms the equation of motion into the following Schrodinger-type equation for $\chi$
where $\chi(r)$ satisfies the following Schrodinger equation in the variable 
\begin{equation}
	\label{schro}
	-\chi''(\xi)+\left(\frac{l(l+1)}{\sin^2\xi}+\frac{M^2+2}{\cos^2\xi}\right)\chi(\xi)=E^2 \chi(\xi)
\end{equation}
where the variable $\xi$, used in this equation is defined by $\cos\xi=\frac{1}{\sqrt{1+r^2}}$.\footnote{In this variable, boundary of $AdS$ is at $\xi=\frac{\pi}{2}$.}

The effective potential for mode with large angular momentum $l$ has a minima at $\xi\approx \frac{\pi}{2}-\frac{(M^2+2)^{1/4}}{\sqrt{l}}$ in large $l$ limit. Since this is very near to $\frac{\pi}{2}$, let us work in the variable $\xi=\frac{\pi}{2}-z$. Note that, to leading order, $z= \frac{1}{r}$. In the large $l$ limit, Schrodinger's equation takes the following form:
\begin{equation}
	\label{effec}
	-\chi''(z)+\left(l(l+1) z^2+\frac{M^2+2}{z^2}\right)\chi=(E^2-l(l+1))\chi\ .
\end{equation}

Now the effective radial Schrodinger equation of the $L^{th}$ angular momentum mode of a 3-dimensional harmonic oscillator with frequency $f$ and mass $1$ is given by 
\begin{equation}\label{radho}
	-\chi''(z)+ \left(f^2  z^2 +\frac{L(L+1)}{z^2}\right)\chi(z)=2\mathcal{E} \chi\ .
\end{equation}
We see that \eqref{effec} and \eqref{radho} are identical once we make the identifications 
\begin{equation}\label{identifications} 
	\begin{split} 
		&f^2=l(l+1) \\
		& 2 \mathcal{E}= E^2-l(l+1)  \implies 
		E=\sqrt{l(l+1)+2\mathcal{E}}\\
		&L(L+1)= M^2+2=(\Delta)(\Delta-3)+2 \implies
		(L+2)(L-1)= \Delta(\Delta -3) \implies L=\Delta -2\\
	\end{split} 
\end{equation} 
It follows from \eqref{identifications} that 
\begin{equation}\label{identificationsimp} 
	\begin{split} 
		&E=\sqrt{l(l+1)+2\mathcal{E}}\\
		&L(L+1)= (\Delta)(\Delta-3)+2 \implies
		(L+2)(L-1)= \Delta(\Delta -3) \implies L=\Delta -2\\
	\end{split} 
\end{equation}

The angular momentum $L$ sector of the 3d harmonic oscillator of frequency $f$ has eigenenergies given by 
\begin{equation}\label{ene}
	\mathcal{E}=(2n + L+\frac{3}{2})f, ~~~n=0, 1 \ldots 
\end{equation}
(the factor of $\frac{3}{2}$ is the ground state energy of the three oscillators, $L$ is the minimum number of excitations we need to excite angular momentum $L$, and $n$ is the number of `contracted excitations' that increase the energy but do not change the angular momentum). 
It follows that at large $l$
\begin{equation}\label{energy}
	\begin{split}
		E_{nl}&=\sqrt{l(l+1)+2\mathcal{E}}\\
		&=\sqrt{l(l+1)+2l(2n+\Delta-\frac{1}{2})}\\
		&\approx l+2n+\Delta
	\end{split}
\end{equation}
in agreement with our expectations from the state operator map (see section \ref{tggs}).

The radial wavefunctions corresponding to these eigenvalues are given by 
\begin{equation}\label{wfnrad}
	\begin{split} 
		& \psi^{RHO}_{nLM}= \frac{Y_{LM}}{z} \chi_{ln}(z)\\
		&\chi_{ln}(z)\equiv  N_{n,\Delta} z^{\Delta-1} e^{-{\frac{l z^2}{2}}}L_n^{\Delta-\frac{3}{2}}(l z^2) \\
		&N_{n,\Delta}=\sqrt{\sqrt{\frac{l^3}{\pi}}\frac{2^{n+\Delta}n! l^{\Delta-2}}{\left(2n+2\Delta-3\right)!!}}\\
	\end{split}
\end{equation}
where the wave functions $\chi_{nl}(z)$ solve 
the equation \eqref{radho} with eigenvalues 
$\mathcal{E}$ listed in \eqref{ene}.

Using \eqref{lrfchi} and the relation $z=\frac{1}{r}$ it follows that 
\begin{equation}\label{rhowfn}
	F_{nl}(r)=\frac{\chi_{ln}}{r}= \frac{N_{n,\Delta}}{r^{\Delta}} e^{-{\frac{l}{2 r^2}}}L_n^{\left(\Delta-\frac{3}{2}\right)}\left(\frac{l}{r^2}\right)
\end{equation}
where $F_{nl}$ can now be inserted in \eqref{sfwf} \footnote{In that equation we have ($F \rightarrow F_{nl}$) together with $(E \rightarrow E_{nl})$; where $E_{nl}$ was listed in 
	\eqref{energy}). }

In the main text (see the first line of 
\eqref{bstt}) we are interested in the quantity
\begin{equation}\label{qmtext} 
	\frac{1}{r^2} \sum_{n=0}^\infty|\chi_{ln}(r)|^2 e^{-l(2n+ \Delta -\frac{1}{2})\frac{q \beta}{l}}
	\equiv \frac{K_{RHO}(r,r,\tau)}{r^2} 
\end{equation} 
Using $L=\Delta -2$ and $f=l$ and \eqref{ene}, we 
see that 
\eqref{qmtext} can be recast as \begin{equation}\label{twopt3d}
	\begin{split}
		K_{RHO}(r,r,\tau)  =&\sum_{n=0}^\infty|\psi_{RHO}^{ln}(r)|^2 e^{-\mathcal{E}_{nl}\tau}\\
		&\tau= \frac{q \beta}{l} \\
	\end{split}
\end{equation}

Just as in the discussion under \eqref{propsum}, $K_{RHO}$ is an effective Feynman propagator. More precisely, it is simply the projection of the matrix elements of the Euclidean evolution operator  of the three-dimensional harmonic oscillator 
\footnote{Which is given by the product of three copies of \eqref{propexp}; one for each of the Cartesian coordinates.} onto the angular momentum 
$L$ sector. This projection yields 
(see eq. 3.3.6 of \cite{Grosche:1998yu})
\begin{equation}
	\begin{split}
		K_{RHO}(r,r,\tau)=&\frac{l}{r\sinh l\tau}\exp{\left(-\frac{l}{r^2}\frac{\cosh l\tau}{\sinh l\tau}\right)}I_{\Delta-3/2}\left(\frac{l}{r^2 \sinh l\tau}\right)\\
		=&\frac{l}{r\sinh l\tau}\exp{\left(-\frac{l}{r^2}\frac{\cosh l\tau}{\sinh l\tau}\right)} \\
		&\times\left(\frac{l}{2r^2 \sinh l\tau}\right)^{\Delta-\frac{3}{2}} \exp\left[\pm\frac{l}{r^2 \sinh l\tau}\right] \frac{1}{\Gamma(\Delta-\frac{1}{2})} {}_1F_1\left(\Delta-1,2\Delta-2,\mp \frac{2l}{r^2 \sinh l\tau}\right)\\
	\end{split}
\end{equation}

\section{The thermal stress tensor of a free conformal scalar on $S^2\times S^1$} \label{tstfb} 

In this Appendix, we compute the stress-energy tensor
of a free massless scalar field in $2+1$ dimensions, on a unit $S^2$,  at inverse temperature $\beta$ and angular velocity $\omega$. We perform this computation first using an algebraically simple 
Euclidean computation, and second using a conceptually clarifying (but algebraically more complicated) Hamiltonian method.

\subsection{Euclidean Computation}

In this subsection compute the one-point function of the thermal stress tensor directly in Euclidean space. 
We do this by2  first evaluating the Euclidean two-point function of $\phi$ on $S^2 \times S^1$, taking appropriate derivatives on the two $\phi$ insertions, and then taking the coincident limit so that the operator so obtained is the stress tensor. After an 
appropriate normal ordering, the answer is finite and gives the stress tensor on the sphere. 

\subsubsection{Euclidean two Point Function on $S^2\times R$}

As the first step in our computation, we compute the two-point function of free fields on $S^2 \times R$. 
As $S^2 \times R$ is Weyl equivalent to $R^3$, this computation is easily performed. We take the two-point function in $R^3$ and apply the conformal transformation that takes us to $S^2\times R$. 

The two-point function in $R^3$ is simply the following:
\begin{equation}
	\langle \Phi(x_1)\Phi(x_2)\rangle_{R^3}=\frac{1}{4\pi|x_1-x_2|}
\end{equation}
To go to $S^2\times R$, we need to apply the conformal scaling $e^\tau=r$. The scaling dimension of the free field in $3d$ is $\frac{1}{2}$, and so we obtain
\begin{equation}\begin{split}
		\langle \Phi(x_1)\Phi(x_2)\rangle_{S^2\times R}&=\frac{e^{\frac{\tau_1 +\tau_2}{2}}}{4\pi|e^{\tau_1}\hat n_1-e^{\tau_2}\hat n_2|}\\
		&=  \frac{e^{\frac{\tau_1 +\tau_2}{2}}}{4\pi\sqrt{e^{2\tau_1}+e^{2\tau_2}-2e^{\tau_1+\tau_2}(\cos\theta_1\cos\theta_2+\sin\theta_1\sin\theta_2\cos(\phi_1-\phi_2))}} \\
		&=  \frac{1}{4\pi\sqrt{2}\sqrt{\cosh(\tau_1-\tau_2)-\cos\theta_1\cos\theta_2-\sin\theta_1\sin\theta_2\cos(\phi_1-\phi_2)}}
	\end{split}
\end{equation}
where $\tau$ is the Euclidean time.
\subsubsection{Two Point Function on $S^2\times S^1$}

The two-point function on $S^2 \times S^1$ is obtained from that on $S^2 \times R$ by the method of images just as we did in subsection \ref{eucb}.
If $\beta$ is the inverse temperature and $\omega$, the angular velocity, the thermal two-point function can be written as a sum over zero temperature two-point functions(see \eqref{images}). Therefore, for the thermal two-point function, we have 
\begin{equation}\label{images2}	\langle\Phi(\theta_1,\tau^E_1,\phi_1)\Phi(\theta_2,\tau^E_2,\phi_2)\rangle_\beta=\sum_{q=-\infty}^{\infty} \langle \Phi(\theta_1,\tau^E_1,\phi_1)\Phi(\theta_2,\tau^E_2+q\beta,\phi_2-iq\beta\omega))\rangle_{0}\ .
\end{equation}

\subsubsection{The stress tensor} 

The stress tensor of a free CFT on the sphere can be obtained by varying the action with respect to the metric on the sphere. The action of the free scalar field on $S^2\times R$ with the conformal coupling is the following:

\begin{equation}\label{spact}
	\begin{split}
		S[\Phi]&=\frac{1}{2}\int d^3x \sqrt{-g} \left(-g^{\mu\nu}\partial_\mu \Phi\partial_\nu\Phi-\frac{1}{8}R\Phi^2\right) 
	\end{split}
\end{equation}
where $R$ is the Ricci scalar of $S^2\times R$ which for a sphere of radius $r$ is $R=\frac{2}{r^2}$. Taking the variation of the action above with respect to the metric, and using the fact that $G_{\mu\nu}=0$ on the sphere, we get the following formula for the stress tensor:
\begin{equation}\label{stcon}
	\begin{split}
		T_{\mu\nu}=&\partial_\mu \Phi\partial_\nu \Phi-\frac{g_{\mu\nu}}{2}\left((\partial\Phi)^2+\frac{\Phi^2}{2}\right) + \frac{1}{8}G_{\mu\nu}\Phi^2 + \frac{1}{8 r^2}(g_{\mu\nu}\nabla^2-\nabla_\mu\nabla_\nu)\Phi^2\\
		=& \partial_\mu \Phi\partial_\nu \Phi-\frac{g_{\mu\nu}}{2}\left((\partial\Phi)^2+\frac{\Phi^2}{2}\right) + \frac{1}{4 r^2}(g_{\mu\nu}\Phi\nabla^2\Phi+g_{\mu\nu}(\partial\Phi)^2-\Phi\nabla_\mu\nabla_\nu\Phi-\partial_\mu \Phi\partial_\nu \Phi)\\
	\end{split}
\end{equation}

where,
$$G_{\mu\nu} = R_{\mu\nu} - \frac{1}{2}R  g_{\mu\nu}.$$ 

\subsubsection{Evaluation of the Stress Tensor} 

To compute the stress tensor, we use the same procedure as in subsection \ref{eucb}. We can compute various derivatives $K^q_{ij}$ and then the stress tensor will be the sum of those derivatives over all the images. Just as in subsection \ref{eucb}, we can easily see that in the limit $\omega\to 1$, the contribution of the non-derivative quadratic term will be subdominant in $(1-\omega)$. For e.g., the $tt$ component of the stress tensor  takes the following form:
\begin{equation}\begin{split} \label{stono}
		\langle T_{tt}(x)\rangle_\beta&=i^2\sum_{q=-\infty, q\neq0}^\infty
		\partial_{\tau_1}\partial_{\tau_2} \langle \Phi(x_1)\Phi(x_2)\rangle_{S^2\times R}\Big|_{\tau_2-\tau_1=q\beta, \phi_2-\phi_1=-i q\beta\omega,\theta_1=\theta_2}\\\\
		&=-\sum_{q=1}^\infty\frac{ -4 \cosh q\beta  \left(\cos ^2\theta+\sin ^2\theta \cosh q\beta\omega\right)-\cosh2 q \beta+5}{32\pi\sqrt{2}\left(\beta  q (1-\omega ) \sin ^2\theta\sinh q\beta+\cos ^2\theta (\cosh q\beta  -1)\right)^{5/2}}\\
		&\approx \sum_{q=1}^\infty\frac{3}{8\pi\sqrt{2}}\frac{ \sinh ^2 q\beta}{ \left( \delta\theta^2 (\cosh q\beta  -1)+\beta  q (1-\omega )  \sinh q\beta\right)^{5/2}}
	\end{split} 
\end{equation}
\footnote{The factor of $i^2$ is due to the fact that stress tensor contains derivatives with respect to Lorentzian time $t=i\tau$.}
This matches with the form we obtained in \eqref{stresstensorone}.

The fact that the summation in \eqref{stono} runs over all values of 
$q$ except $q=0$ follows from the following considerations. If we were to honestly compute the expectation value of the stress tensor, we would first need to carefully specify its definition (i.e. the local 
subtractions we use to give this composite operator meaning). We sidestep all these issues using the following device. In \eqref{stono} we have really computed not the expectation value of the stress tensor itself, but, instead, the expectation value of the stress tensor minus its expectation value at zero temperature. All ambiguities of definition disappear in this difference. As the zero temperature expectation value of the stress tensor is given precisely by the 
$q=0$ term in the summation, the difference computed in \eqref{stono} 
involves a summation with $q=0$ removed.

Following along the same lines, we can also see that in the limit $\omega\to 1$, $T_{tt}=-T_{t\phi}=T_{\phi\phi}\equiv T$.

\subsection{Computation using Hamiltonian method} 

In this subsection, we present the (elementary) computation of the free CFT stress tensor on the sphere.  

The single particle eigenstates for free scalar fields on $S^2$ are labeled by the angular momentum quantum numbers, $(l,m)$. Explicitly, the eigenfunctions (units normalized in the Klein-Gordon norm) are given by 
\begin{equation}
	\psi_{l,m}(\theta,\phi,t)=\frac{1}{\sqrt{2 E_l}} Y_{lm}(\theta,\phi) e^{-i E_{l} t}
\end{equation}
where $E_l=l+\frac{1}{2}$. 
The field operator $\Phi$ admits the expansion 
\begin{equation}\label{expfop} 
	\Phi = \sum_{l, m} \psi_{l,m} a_{l,m} e^{-i E_{l} t} +  \psi_{l,m}^* a^\dagger_{l,m} e^{i E_{l} t}
\end{equation} 
where 
\begin{equation}\label{comrel} 
	\left[ a_{l, m}, a^\dagger_{l', m'} \right] = \delta_{ll'} \delta_{mm'}
\end{equation} 
The computation of the expectation value of the stress tensor in the thermal ensemble is straightforward (a similar computation is performed in more detail in subsection \ref{bst-H}). One finds that 
\begin{equation}\label{tmunuf} 
	\langle T_{\mu\nu} \rangle = \sum_{ l, m} 
	\frac{t^{lm}_{\mu\nu}}{e^{\beta(\frac{1}{2}+a\omega+l(1 -\omega))} -1} 
\end{equation} 
where $t^{lm}_{\mu\nu}$ is simply the classical stress tensor of the field configuration $\phi_{l,m}$. 
$t^{lm}_{\mu\nu}$ is easy to compute, but we do not need all its details. For very similar reasons 
to the discussion around \eqref{sumover}, the only modes at large $l$ and large $m$ contribute to the part of the stress tensor that is singular (as $\omega \to 1$). Let us consider the mode with $(J, J_z)=(l, l-a)$ with 
$l \gg 1$ and $a$ of order unity. Acting on such $\partial_\phi$ and $\partial_t$ are of order $l$, and are much larger than the action of $\partial_\theta$ on the same mode (which turns out to be of order $\sqrt{l}$)\footnote{Also note that we get some extra terms in the stress tensor in \eqref{stcon} because of the conformal coupling. However, these terms are like derivatives of $\Phi^2$ which means that in the mode stress tensor, they appear as derivatives of $\psi^*_{lm}\psi_{lm}$. Since the oscillating factors in $t$ and $\phi$ cancel, these derivatives are subdominant in large $l$ limit. Therefore it is only the first term \eqref{stcon} that contributes in the limit $\omega\to 1$}. Retaining only the leading order contributions (see around \eqref{stlight} 
for details of the computation in a very similar context) we find that $t_{\mu\nu}^{lm}$
boundary stress tensor is given by 
\begin{equation}\begin{split}\label{ston}
		t_{tt}^{lm}= -t_{t\phi}^{lm}=t_{\phi\phi}^{lm}&\approx  2 l^2  \frac{1}{2 \left( l + \frac{1}{2} \right) }  |Y_{lm}|^2\\
		&\approx\frac{l}{2\pi} \left(\frac{\sqrt[4]{\frac{l}{\pi }}}{\sqrt{2^a a!}}\right)^2 e^{-l\delta\theta^2}H_a^2(\sqrt{l}\delta\theta)\\
		&=\frac{l}{2\pi}|\psi_a^{HO}(\delta\theta)|^2
	\end{split}	
\end{equation}
where $\psi_a^{HO}(\delta\theta)$ is the $a^{\rm th}$ wavefunction of a Harmonic oscillator with unit mass and frequency $l$.

Inserting \eqref{ston} into \eqref{tmunuf} we find that, at leading order,

\begin{equation}\begin{split}\label{Tlargel}
		T=T_{tt}=-T_{t\phi}=T_{\phi\phi} &\approx \frac{1}{2\pi}\int dl\sum_{a=0}^\infty \frac{l|\psi_a^{HO}(\delta\theta)|^2 }{e^{\beta\left(\frac{1}{2}+a\omega\right)} ~e^{\beta(1-\omega)l}-1}\\
		&=\frac{1}{2\pi}\sum_{q=0}^\infty\int dl~l e^{-(q+1)\beta(1-\omega)l}\sum_{a} |\psi_a^{HO}(\delta\theta)|^2e^{-(q+1)\beta\left(\frac{1}{2}+a\omega\right)}
	\end{split}
\end{equation}
The sum over $a$ is simply the quantity $$\langle \delta\theta|e^{-\frac{\beta(q+1)}{l}H_{HO}}|\delta\theta\rangle$$ where $H_{HO}$ is the harmonic oscillator Hamiltonian.

This is the standard path integral in Euclidean time. We use the following result:
\begin{equation}
	\langle x|e^{-\tau H_{HO} }|x\rangle=\frac{\sqrt{l}}{\sqrt{2\pi\sinh l\tau}}e^{-\frac{l x^2}{\sinh l\tau}(\cosh l\tau-1)} 
\end{equation}
Using the above identity, and taking $\tau=\frac{\beta(q+1)}{l}$

\begin{equation}\begin{split}\label{stresstensorone} 
		T
		&=\frac{1}{2\pi}\sum_{q=0}^\infty\int dl~l e^{-(q+1)\beta(1-\omega)l}\sum_{a} |\psi_a^{HO}(\delta\theta)|^2e^{-(q+1)\beta\left(\frac{1}{2}+a\omega\right)}\\
		&=\frac{1}{2\pi}\sum_{q=0}^\infty\int dl~l^{3/2}\exp\left(-l\left((q+1)\beta(1-\omega)+\frac{\delta\theta^2 (\cosh((q+1)\beta)-1)}{\sinh((q+1)\beta)}\right)\right)\frac{1}{\sqrt{2\pi\sinh((q+1)\beta)}} \\
		&=\sum_{q=1}^\infty \frac{3}{8\pi\sqrt{2}}\frac{\sinh^2 q\beta}{\left(q\beta(1-\omega)\sinh q\beta+\delta\theta^2 (\cosh q\beta-1)\right)^{5/2}}
	\end{split}
\end{equation}

Let us compute the total energy by integrating the stress tensor over the sphere. Since the stress tensor is dominant at values of $\theta$ close to $\frac{\pi}{2}$.Therefore we integrate over the variable $\zeta$ which is order unity here.  We will have an extra factor of $\sqrt{\omega}$ from this change in variables. 

\begin{equation}\label{ttoten}
	\begin{split}
		E&= 2\pi\sqrt{1-\omega} \int_{-\infty}^\infty d\zeta T_{00}\\
		&= \frac{2\pi}{(1-\omega)^2} \sum_{q=1}^\infty \int_{-\infty}^\infty d\zeta \frac{3}{8\pi\sqrt{2}}\frac{\sinh^2 q\beta}{\left(q\beta\sinh q\beta+\zeta^2 (\cosh q\beta-1)\right)^{5/2}}\\
		&=\frac{1}{2(1-\omega)^{2}}\sum_q \frac{1}{(q\beta)^2 \sinh\frac{q\beta}{2}}
	\end{split} 
\end{equation}

The thermal expectation value of energy can also be computed in the same way as we did in \eqref{toten}. Following the same steps we get the following:

\begin{equation}\label{totb}
	\begin{split}
		E&=\sum_{q=1}^{\infty} \sum_a \int dl l e^{-q\beta(a+\frac{1}{2}+(1-\omega)l)}\\ 
		&=\sum_q \frac{1}{(q\beta)^2(1-\omega)^2}\frac{e^{-\frac{q\beta}{2}}}{1-e^{-q\beta}}\\
		&=\sum_q \frac{1}{2(q\beta)^2(1-\omega)^2}\frac{1}{\sinh\frac{q\beta}{2}}
	\end{split}  
\end{equation}
which is exactly the same as in \eqref{ttoten}.

\section{Sourced solutions from homogeneous solutions} \label{de} 
Let us pause to discuss some differential equation theory. Suppose we are given an equation 
of the form 
\begin{equation} 
	C(x) \psi^{''} + C'(x) \psi'(x) + E(x) \psi = s(x) 
\end{equation} 
Let us suppose that $\phi_1(x)$ and $\phi_2(x)$ are known to be solutions of the homogeneous equation. We want to solve the inhomogeneous equation. Then one particular solution of the differential equation is given by 
\begin{equation}\label{partsol} 
	\psi(x) = -\phi_1(x) \int \frac{\phi_2(x) s(x)}{C(x)(\phi_1(x) \phi_2'(x) -\phi_2(x) \phi_1'(x))} 
	- \phi_2(x) \int \frac{\phi_1(x) s(x)}{C(x)(\phi_2(x) \phi_1'(x) -\phi_1(x) \phi_2'(x))} 
\end{equation} 

When studying \eqref{desfk} we can choose $\phi_1= e^{\frac{k}{x}}\left(1- \frac{k}{x} \right)$ and $\phi_2 = e^{\frac{-k}{x}}\left(1+ \frac{k}{x} \right)$. 
We find 
$${\phi'_1(x) \phi_2(x) -\phi'_2(x) \phi_1(x)}= \frac{2k^3}{x^4}$$
It follows that the general solution to \eqref{desfk} is given by  \eqref{gensoldefk}

\section{Canonical quantization of a the free scalar in $AdS_4$} \label{cq}

The symplectic current $J_\mu$ for a free scalar field in $AdS_4$ (see around \eqref{expoffield}) is given by 
\begin{equation}\label{sympcurr}
	J_\mu =  -\delta \phi \wedge \partial_\mu \delta \phi
\end{equation}  
Using the equations of motion of the scalar field, it is easy to see that $\nabla^\mu J_\mu=0$. It follows that the symplectic form 
is given by  
\begin{equation}\label{sympform} 
	\Omega= \int \sqrt{g} J^0 = \int \sqrt{g} g^{0\mu} J_\mu 
\end{equation} 	
(the integral is taken over a constant time surface; the conservation of current guarantees that this integral 
is independent of our choice of time). Inserting \eqref{expoffield} into \eqref{sympform}, we find the symplectic form 
\begin{equation}\label{sympformexp} 
	\Omega = i \sum_{n l m} \delta a_{nl m} \wedge \delta a^*_{n l m} 
\end{equation}
It follows that the Poison Brackets are 
\begin{equation}\label{pbform} 
	\{a_{nlm}, a^*_{n' l' m'} \} = -i \delta_{l l'} \delta_{mm'} \delta_{n n'}
\end{equation}
Upon quantization, the canonical commutators are given by
\begin{equation}\label{canconform} 
	\left[a_{nlm}, a^\dagger_{n' l' m'} \right] =  \delta_{l l'} \delta_{mm'} \delta_{n n'}
\end{equation}

\section{Derivation of the Kernel in \eqref{kern1}} \label{kernel} 

In this Appendix, we will derive the kernel we used in \eqref{solzeronewg} which calculates the back-reaction of the matter stress tensor on the metric. Fourier transform of $A_k(x)$ in $k$ variable will give us the correct metric correction in the following form,

\begin{equation}\label{solzeronewgn} \begin{split} 
		A(x,\zeta) =& \int_{-\infty}^\infty \frac{dk}{\sqrt{2\pi}} e^{ik\zeta} \left[ \left( \frac{1}{ |k|^3} \int_0^\infty  e^{\frac{-|k|}{x'}}\left(1+ \frac{|k|}{x'} \right) f_k(x')\right) 
		\left(e^{\frac{-|k|}{x}}\left(1+ \frac{|k|}{x} \right) \right)\right. 	\\
		&\left.  -\frac{1}{ |k|^3}  \left( e^{\frac{|k|}{x}}\left(1-\frac{|k|}{x}\right) \int_0^x  e^{\frac{-|k|}{x'}}\left(1+ \frac{|k|}{x'} \right) f_k(x')  +  e^{\frac{-|k|}{x}}\left(1+\frac{|k|}{x}\right) \int_x^\infty   e^{\frac{|k|}{x'}}\left(1- \frac{|k|}{x'} \right) f_k(x')\right)\right] \\
		=& \int_{-\infty}^\infty \frac{dk}{\sqrt{2\pi}|k|^3} e^{ik\zeta} \left[  \int_0^\infty dx'  e^{-|k|\left(\frac{1}{x'}+\frac{1}{x}\right)}\left(1+ \frac{|k|}{x'} \right)  \left(1+ \frac{|k|}{x} \right)f_k(x') \right. 	\\
		&  - \int_0^x dx' e^{-|k|\left(\frac{1}{x'}-\frac{1}{x}\right)}\left(1-\frac{|k|}{x}\right)  \left(1+ \frac{|k|}{x'} \right) f_k(x') \\ 
		&\left. - \int_x^\infty dx' e^{-|k|\left(\frac{1}{x}-\frac{1}{x'}\right)}\left(1+\frac{|k|}{x}\right)  \left(1- \frac{|k|}{x'} \right) f_k(x')\right] \\
		=& \int_{-\infty}^\infty \frac{dk}{|k|^3} \int_{-\infty}^\infty \frac{d\zeta'}{2\pi} e^{ik(\zeta-\zeta')} \left[  \int_0^\infty dx'  e^{-|k|\left(\frac{1}{x'}+\frac{1}{x}\right)}\left(1+ \frac{|k|}{x'} \right)  \left(1+ \frac{|k|}{x} \right)f(x',\zeta') \right. 	\\
		&  - \int_0^x dx' e^{-|k|\left(\frac{1}{x'}-\frac{1}{x}\right)}\left(1-\frac{|k|}{x}\right)  \left(1+ \frac{|k|}{x'} \right) f(x',\zeta') \\ 
		&\left. - \int_x^\infty dx' e^{-|k|\left(\frac{1}{x}-\frac{1}{x'}\right)}\left(1+\frac{|k|}{x}\right)  \left(1- \frac{|k|}{x'} \right) f(x',\zeta')\right] \\
		=& \int_{-\infty}^\infty \frac{dk}{|k|^3} \int_{-\infty}^\infty \frac{d\zeta'}{2\pi} e^{ik(\zeta-\zeta')} \left[  \int_0^\infty dx' f(x',\zeta') e^{-|k|\left(\frac{1}{x'}+\frac{1}{x}\right)}\left(1+ \frac{|k|}{x'} \right)  \left(1+ \frac{|k|}{x} \right) \right. 	\\
		&  - \int_0^x dx' f(x',\zeta') e^{-|k|\left(\frac{1}{x'}-\frac{1}{x}\right)}\left(1-\frac{|k|}{x}\right)  \left(1+ \frac{|k|}{x'} \right) \\ 
		&\left. - \int_x^\infty dx' f(x',\zeta') e^{-|k|\left(\frac{1}{x}-\frac{1}{x'}\right)}\left(1+\frac{|k|}{x}\right)  \left(1- \frac{|k|}{x'} \right)\right] \\
	\end{split} 
\end{equation}
Going from the second line to the third line we used the inverse Fourier transform of $f_k(x)$. In other steps, we rearranged various terms in this expression. Now to perform the $k$ integral we can rearrange the integrals in the above expression and rewrite in the following suggestive form,
\begin{equation}\label{mt1}
	\begin{split}
		A(x,\zeta) =&  \int_{-\infty}^\infty \frac{d\zeta'}{2\pi}  \left[  \int_0^\infty dx' f(x',\zeta') \int_{-\infty}^\infty \frac{dk}{|k|^3} e^{-|k|\left(\frac{1}{x'}+\frac{1}{x}\right)+ik(\zeta-\zeta')}\left(1+ \frac{|k|}{x'} \right)  \left(1+ \frac{|k|}{x} \right) \right. 	\\
		&  - \int_0^x dx' f(x',\zeta') \int_{-\infty}^\infty \frac{dk}{|k|^3} e^{-|k|\left(\frac{1}{x'}-\frac{1}{x}\right)+ik(\zeta-\zeta')}\left(1-\frac{|k|}{x}\right)  \left(1+ \frac{|k|}{x'} \right) \\ 
		&\left. - \int_x^\infty dx' f(x',\zeta') \int_{-\infty}^\infty \frac{dk}{|k|^3} e^{-|k|\left(\frac{1}{x}-\frac{1}{x'}\right)+ik(\zeta-\zeta')}\left(1+\frac{|k|}{x}\right)  \left(1- \frac{|k|}{x'} \right)\right]
	\end{split}
\end{equation}

There is a total of three $k$ integrals but all the $k$ integrals have similar forms except changes in $\pm$ signs. So we generalized these $\pm$ signs by $c_1$ and $c_2$. Let us define,
\begin{equation}\label{kerndef}
	\begin{split}
		K^{c_1c_2}_{x,\zeta}(x',\zeta') &= \int_{-\infty}^\infty \frac{dk}{|k|^3} e^{-|k|\left(\frac{c_1}{x'}+\frac{c_2}{x}\right)+ik(\zeta-\zeta')}\left(1+ c_1\frac{|k|}{x'} \right)  \left(1+ c_2\frac{|k|}{x} \right)
	\end{split}
\end{equation}
Using this definition \eqref{kerndef} we can rewrite the metric in \eqref{mt1} in the following manner
\begin{equation}\label{metmid}
	\begin{split}
		A(x,\zeta) =&  \int_{-\infty}^\infty \frac{d\zeta'}{2\pi}  \left[  \int_0^\infty dx' f(x',\zeta') K^{++}_{x,\zeta}(x',\zeta') \right. 	 - \int_0^x dx' f(x',\zeta') K^{+-}_{x,\zeta}(x',\zeta') \left. - \int_x^\infty dx' f(x',\zeta') K^{-+}_{x,\zeta}(x',\zeta')\right]
	\end{split}
\end{equation}

where various choices of $c_1,\,c_2=\pm1$ will give us all the integrals. Now let's evaluate this integral exactly keeping $c_1$ and $c_2$ arbitrary,
\begin{equation}\label{kint}
	\begin{split}
		K^{c_1c_2}_{x,\zeta}(x',\zeta') =& \int_{-\infty}^\infty \frac{dk}{|k|^3} e^{-|k|\left(\frac{c_1}{x'}+\frac{c_2}{x}\right)+ik(\zeta-\zeta')}\left(1+ c_1\frac{|k|}{x'} \right)  \left(1+ c_2\frac{|k|}{x} \right)\\
		=& \int_{-\infty}^0 \frac{dk}{-k^3} e^{k\left(\frac{c_1}{x'}+\frac{c_2}{x}\right)+ik(\zeta-\zeta')}\left(1-c_1 \frac{k}{x'} \right)  \left(1- c_2\frac{k}{x} \right)\\
		&+\int_0^\infty \frac{dk}{k^3} e^{-k\left(\frac{c_1}{x'}+\frac{c_2}{x}\right)+ik(\zeta-\zeta')}\left(1+ c_1\frac{k}{x'} \right)  \left(1+ c_2\frac{k}{x} \right)\\
		=& \int_{\infty}^0 \frac{-dk}{k^3} e^{-k\left(\frac{c_1}{x'}+\frac{c_2}{x}\right)-ik(\zeta-\zeta')}\left(1+c_1\frac{k}{x'} \right)  \left(1+c_2\frac{k}{x} \right)\\
		&+\int_0^\infty \frac{dk}{k^3} e^{-k\left(\frac{c_1}{x'}+\frac{c_2}{x}\right)+ik(\zeta-\zeta')}\left(1+ c_1\frac{k}{x'} \right)  \left(1+c_2 \frac{k}{x} \right)\\
		=& 2\int_0^{\infty} \frac{dk}{k^3} e^{-k\left(\frac{c_1}{x'}+\frac{c_2}{x}\right)}\cos\left[k(\zeta-\zeta')\right]\left(1+c_1\frac{k}{x'} \right)  \left(1+c_2\frac{k}{x} \right)\\
	\end{split}
\end{equation}
In the above steps, we simplified the integral of $k$ by removing the modulus of $k$ and changing the limit of integration from $(-\infty, \infty)$ to $(0,\infty)$.
Mathematica can easily perform the indefinite $k$ integral and we get,
\begin{equation}
	\begin{split}
		2& \int \frac{dk}{k^3} e^{-k\left(\frac{c_1}{x'}+\frac{c_2}{x}\right)}\cos\left[k(\zeta-\zeta')\right]\left(1+c_1\frac{k}{x'} \right)  \left(1+c_2\frac{k}{x} \right)\\
		=& -\frac{1}{2} \left((\zeta-\zeta') ^2+\frac{1}{x^2}+\frac{1}{{x'}^2}\right) \left(\text{Ei}\left[k \left(-i (\zeta-\zeta') -\frac{c_1}{x}-\frac{c_2}{{x'}}\right)\right]+\text{Ei}\left[k \left(i (\zeta-\zeta') -\frac{c_1}{x}-\frac{c_2}{{x'}}\right)\right]\right)\\
		&-\frac{1}{2k} e^{k \left(-\frac{c_1}{x}-\frac{c_2}{{x'}}+i (\zeta-\zeta') \right)} \left(\frac{c_1}{x}+\frac{c_2}{{x'}}+i (\zeta-\zeta') +\frac{1}{k}\right)-\frac{1}{2k}e^{k \left(-\frac{c_1}{x}-\frac{c_2}{{x'}}-i (\zeta-\zeta') \right)} \left(\frac{c_1}{x}+\frac{c_2}{{x'}}-i (\zeta-\zeta') +\frac{1}{k}\right)\\
		=& \mathcal{I}^{c_1c_2}(k)\quad\quad\quad\quad{\rm (let)}
	\end{split}
\end{equation} 

Now we need to evaluate this function $\mathcal{I}^{c_1c_2}(k)$ at $k=0$ and $k\rightarrow \infty$. This function is zero at $k\rightarrow \infty$ though it is not regular at $k=0$. So we will evaluate this function at $k=\eta$ where $\eta$ is a small finite positive number that is approaching zero. In very small $\eta$ limit $\mathcal{I}^{c_1c_2}(\eta)$ has the following expansion
\begin{equation}
	\begin{split}
		\lim_{\eta\rightarrow 0^+}\mathcal{I}^{c_1c_2}(\eta) =&-\frac{1}{\eta^2}-\left((\zeta-\zeta')^2+\frac{1}{x^2}+\frac{1}{{x'}^2}\right)\log \eta +\zeta ^2+\left(\frac{1}{2}-\gamma \right) \left((\zeta-\zeta')^2+\frac{1}{x^2}+\frac{1}{{x'}^2}\right)\\
		&+\frac{c_1 c_2}{x {x'}} - \frac{1}{2}\left((\zeta-\zeta')^2+\frac{1}{x^2}+\frac{1}{{x'}^2}\right) \log \left(\left(\frac{c_1}{x}+\frac{c_2}{{x'}}\right)^2+(\zeta-\zeta')^2\right)
	\end{split}
\end{equation}
Hence, from \eqref{kint} we get,
\begin{equation}\label{kerdefi}
	\begin{split}
		K^{c_1c_2}_{x,\zeta}(x',\zeta')= & \;0-\lim_{\eta\rightarrow 0^+}\mathcal{I}^{c_1c_2}(\eta)\\
		=&\frac{1}{\eta^2}+\left((\zeta-\zeta')^2+\frac{1}{x^2}+\frac{1}{{x'}^2}\right)\log \eta -(\zeta-\zeta')^2-\left(\frac{1}{2}-\gamma \right) \left((\zeta-\zeta')^2+\frac{1}{x^2}+\frac{1}{{x'}^2}\right)\\
		&-\frac{c_1 c_2}{x {x'}} + \frac{1}{2}\left((\zeta-\zeta')^2+\frac{1}{x^2}+\frac{1}{{x'}^2}\right) \log \left(\left(\frac{1}{x}+\frac{c_1c_2}{{x'}}\right)^2+(\zeta-\zeta')^2\right)
	\end{split}
\end{equation}
In the above answer since $c_1$, $c_2$ comes always as a product, we can see that $K^{c_1c_2}_{x,\zeta}(x',\zeta')$ satisfies,
\begin{equation}\label{id1}
	K^{+-}_{x,\zeta}(x',\zeta') = K^{-+}_{x,\zeta}(x',\zeta')
\end{equation}
Now we can use the result of \eqref{kerdefi} and \eqref{id1} in \eqref{metmid} and after simplification we get,
\begin{equation}
	\begin{split}
		A(x,\zeta) =&  \int_{-\infty}^\infty \frac{d\zeta'}{2\pi}  \left[  \int_0^\infty dx' f(x',\zeta') K^{++}_{x,\zeta}(x',\zeta')  - \int_0^x dx' f(x',\zeta') K^{+-}_{x,\zeta}(x',\zeta') - \int_x^\infty dx' f(x',\zeta') K^{-+}_{x,\zeta}(x',\zeta')\right]\\
		=&  \int_{-\infty}^\infty \frac{d\zeta'}{2\pi}  \int_0^\infty dx' f(x',\zeta') \left[K^{++}_{x,\zeta}(x',\zeta')  - K^{+-}_{x,\zeta}(x',\zeta') \right]\\
		=&  \int_{-\infty}^\infty \frac{d\zeta'}{2\pi}  \int_0^\infty dx' f(x',\zeta') \mathcal{K}_{x,\zeta}(x',\zeta') \quad\quad\quad\quad\quad\quad{\rm (let).}
	\end{split}
\end{equation}
where, $\mathcal{K}_{x,\zeta}(x',\zeta')$. Finally, we get the metric in the form,
\begin{equation}
	\begin{split}
		\mathcal{K}_{x,\zeta}(x',\zeta') =& K^{++}_{x,\zeta}(x',\zeta')  - K^{+-}_{x,\zeta}(x',\zeta')\\
		=& \frac{1}{2} \left((\zeta-\zeta')^2+\frac{1}{x^2}+\frac{1}{{x'}^2}\right)\left( \log \left[(\zeta-\zeta')^2+\left(\frac{1}{x}+\frac{1}{{x'}}\right)^2\right]-\log\left[(\zeta-\zeta')^2+\left(\frac{1}{x}-\frac{1}{{x'}}\right)^2\right]\right)\\
		&-\frac{2}{x {x'}}
	\end{split}
\end{equation}
This is the kernel we used in subsection \eqref{metpos}.

\section{Quasi-Thermodynamic analysis}\label{quasi}

In this Appendix, we perform a quasi-thermodynamic analysis of the gas
studied in this paper at $\omega>1$. 

The reader may find the study of the system at $\omega>1$ surprising, given the fact that its unstable nature seems to rule out the use of thermodynamics.   Concretely, when $\omega>1$, modes with $l>\frac{\Delta}{\omega-1}$ are unstable, and are infinitely occupied in the canonical partition function.
As we have explained in the introduction, however, (see subsection 
\ref{esri}), even though modes at large $l$ are unstable, the time 
scale for their emission scales like $e^{ b l}$ at large $l$. At any 
given finite time, therefore, modes with $l>l_m$ (where $l_m$ is a suitably increasing function of time) are not occupied. For this reason, it seems to us that canonical thermodynamics with a cut-off on angular momenta ($l<l_m$) might well turn out to be a reasonable model for the 
nature of the configuration as a function of time (parameterized through $l_m$) as the black hole approaches the end point of super-radiance, in the manner described in subsection \ref{esri}). In this Appendix, we study this model. 

The specific question we ask is the following. Suppose we are located at a specific point in E, J Space in Fig. \ref{kerr-ads4-boost}. When $l_m$
is large compared to unity, then our solution will have traversed almost all of the distance, along the dotted line, to the solid black line 
in Fig. \ref{kerr-ads4-boost}. Let us denote the difference between $E$
and the energy of the intersection of the dotted lines with the black curve, 
as $E_{out}$. When $l_m$ is large, the energy $E_{out}$ lies outside the black hole. Note that $E_{out}$ is of order $\frac{1}{G}$. In this Appendix, we model the distribution of this energy between various bulk modes
by a quasi-thermodynamic ensemble - one in which we apply the rules 
of thermodynamics after simply throwing away all modes with $l>l_m$.
The quasi-thermodynamics of our system is defined by the quasi-partition function 
\begin{equation}\label{mutparpcut} 
	\ln Z= \sum_{n=0}^\infty   \sum_{l=0}^{l_m} \sum_{a=0}^{2l}   - \ln \left(1-e^{-\beta(\Delta + 2n-\alpha +\omega a )- \beta l(1-\omega) }\right)\ .
\end{equation} 
together with the requirement that the energy contained in this 
ensemble is $E_{out}$ (the last requirement will be used to find the effective value of $\omega$ in \eqref{mutparpcut}).

\subsection{The Bose Condensate Phase} 

If $l_m$ is not too large we might intuitively suspect that the energy of the ensemble \eqref{mutparpcut} can be very large  only if $\omega$ is less than but very close to $\omega_c$ the maximum allowed value of $\omega$, 
\begin{equation}\label{omegac} 
	\omega_c=1+\frac{\Delta}{l_m}
\end{equation} 
Let 
\begin{equation}\label{omegaoc}
	\omega= \omega_c-\delta \omega
\end{equation} 
If $\delta \omega$ is very small then the mode with $n=0$, $a=0$, $l=l_m$
will be very strongly occupied: our system will then consist of a Bose condensate in this mode. The occupation of all other modes can be ignored if the energy in these modes is much smaller than the energy contained in the mode with $l=l_m$. We will now work out the condition for this requirement to be self-consistently met.

Recall that the energy contained in a mode with $n=a=0$ is given by 
\begin{equation}
	E_l=\frac{l}{e^{\beta(\Delta+(1-\omega)l)}-1}  
\end{equation}
It follows that the energy contained in the mode with $l=l_m$ is given by 
\begin{equation}\label{lene}
	E_{l_m}=\frac{l_m}{\beta l_m(\omega_c-\omega)}= \frac{1}{\beta\delta\omega}.
\end{equation}
(where we have used the fact that $e^{\beta(\Delta+(1-\omega)l_m)}$
must be near to unity for macroscopic occupation of this state). 

If we assume that all the energy lies in this mode, it must be that
\begin{equation}\label{dom}
	\delta \omega = \frac{1}{\beta E_{out} }
\end{equation} 
Now the result \eqref{dom} is self-consistent only if the energy contained in all other modes is small
compared to $E_{out}$. Let us check when this is the case. The energy contained in all other modes $E_{oth}$ may be estimated by 
\begin{equation}\begin{split} \label{allother}
		E_{oth}&=\sum_{l=0}^{l_m-1} \frac{l}{e^{\beta(\Delta+(1-\omega)l)}-1}\\
		&=\sum_{b=1}^{l_m}\frac{l_m-b}{e^{\beta\left(\Delta+(\delta\omega-\frac{\Delta}{l_m})(l_m-b)\right)}-1}\\
		&=\sum_{b=1}^{l_m} \frac{l_m-b}{e^{\beta\left(\delta\omega l_m(1-\frac{b}{l_m})+\frac{b\Delta}{l_m}\right)}-1}\\
	\end{split}   
\end{equation}
(we have retained only modes with $n=a=0$ because the contribution of modes with nonzero $n$ and nonzero $a$ is much smaller).

Let us first consider the contribution of terms with $b< \alpha l_m$ to the summation in \eqref{allother}. For such terms, the exponent in the last of \eqref{allother} can be Taylor expanded. We will see self-consistently that 
\begin{equation}\label{selfon}
	\delta \omega  l_m^2 \ll 1
\end{equation} 
Using this condition the relevant range of summation in the last line of \eqref{allother} is 
approximately given by 
\begin{equation}\label{appi}
	\frac{l_m}{\beta \Delta} \sum_{b=1}^{\alpha l_m} \frac{l_m-b}{b} \approx \frac{l_m^2}{\beta \Delta}
	\left( \ln \alpha l_m -\alpha \right) \approx \frac{l_m^2 \ln l_m }{\beta \Delta} 
\end{equation}
(where we have kept only the leading term in the large $l_m$ limit). It is easy to convince oneself that the summation in $b$ from $\alpha l_m$ to 
$l_m$ is subleading in \eqref{appi}, and so can be ignored. 

We have a Bose condensate (almost all energy in one state) when 
\begin{equation}\label{rc}
	\frac{l_m^2 \ln l_m }{\beta \Delta}
	\ll E_{out} 
\end{equation} 
i.e., very approximately, when 
\begin{equation}\label{lmeout}
	l_m\ll  \frac{2  \sqrt{2\beta \Delta E_{out}}}{\ln (2\beta \Delta E_{out})} 
\end{equation} 
Plugging \eqref{lmeout} into \eqref{selfon} and 
using \eqref{dom}, we see that \eqref{selfon} is 
indeed obeyed (because  $\ln E_{out}$ is a large number). 

In summary, we conclude that our system is well approximated by a Bose condensate in the single mode $l=l_m$ when \eqref{rc} is met. 
When $l_m $ becomes so large that \eqref{rc} is no longer obeyed, the system is better thought of as a gas (with energy spread over many modes) than a Bose condensate.

\section{Kerr-AdS$_5$ black holes}\label{Kerr-AdS5}

The mass $E$, angular momentum $J=J_1+J_2$ (in the special case $J_1=J_2$), 
angular velocity $\omega\equiv\omega_1=\omega_2$ and the entropy $S$ of the Kerr-AdS$_5$ 
black holes are given in the convention of \cite{Ishii:2018oms} by 
\begin{equation}\label{AdS5-bh}
	\epsilon\equiv GE=\frac{\pi\mu(a^2+3)}{4}\ ,\ 
	j\equiv GJ=\pi \mu a\ ,\ s\equiv GS=\frac{\pi^2r_h^3}{2}\sqrt{1+\frac{2\mu a^2}{r_h^4}}
	\ ,\ \omega=\frac{2\mu a}{r_h^4+2a^2\mu}
\end{equation}
where $\mu,a$ are the two parameters of the solution, and $r_h$ is the location of 
the event horizon which is given by the largest root of 
\begin{equation}
	0=1-\frac{2\mu(1-a^2)}{r_h^2(1+r_h^2)}+\frac{2a^2\mu}{r_h^4(1+r_h^2)}\ .
\end{equation}
$G$ is the 5d Newton constant, and we have set the AdS$_5$ radius to $1$. 
Although we present the expressions in the convention of \cite{Ishii:2018oms} for 
comparisons, it is often more convenient to use the parameters 
$r_+,m$ of \cite{Gibbons:2004ai} instead of $r_h,\mu$ above, which are 
related by  $r_h^2=\frac{r_+^2+a^2}{1-a^2}$, $\mu=\frac{m}{(1-a^2)^3}$. 
In terms of these parameters, the expressions above are given by
\begin{equation}\label{AdS5-bh-gibbons}
	\epsilon=\frac{\pi m(a^2+3)}{4(1-a^2)^3}\ ,\ j=\frac{\pi ma}{(1-a^2)^3}\ ,\ 
	s=\frac{\pi^2(r_+^2+a^2)^2}{2r_+(1-a^2)^2}\ ,\ 
	\omega=\frac{a(r_+^2+1)}{r_+^2+a^2}
\end{equation}
and 
\begin{equation}\label{m-rplus}
	m=\frac{(r_+^2+a^2)^2(r_+^2+1)}{2r_+^2}\ .
\end{equation}
To have a black hole solution with a regular event horizon (non-negative Hawking temperature), 
$r_+,a$ should satisfy \cite{Gibbons:2004ai}
\begin{equation}
	2r_+^4+r_+^2-a^2\geq 0\ \rightarrow\  
	r_+^2\geq -\frac{1}{4}+\sqrt{\frac{1}{16}+\frac{a^2}{2}}\ .
\end{equation}
The extremal black holes saturate this inequality. $a$ satisfies $0\leq a<1$. In this range, 
$\epsilon$, $j$ satisfy the bound $\epsilon> j$.

The black hole at $\omega\geq 1$ suffer from superradiant instabilities. 
This condition is equivalent to 
\begin{equation}
	r_+^2\leq a\ .
\end{equation}
The black holes saturating this inequality, $r_+^2=a$ will be the core black holes 
of our AdS$_5$ Grey Galaxies satisfying $\omega=1$. These black holes 
satisfy $m=\frac{a(1+a)^3}{2}$ and
\begin{equation}\label{omega1}
	\epsilon=\frac{\pi a(a^2+3)}{8(1-a)^3}\ \ ,\ \ j=\frac{\pi a^2}{2(1-a)^3}\ \ ,\ \ 
	s=\frac{\pi^2 a^{\frac{3}{2}}}{2(1-a)^2}\ .
\end{equation}
In particular, one finds 
\begin{equation}
	\epsilon-j=\frac{\pi a(3-a)}{8(1-a)^2}
\end{equation}
for the core black holes at $\omega=1$. One can invert this relation to obtain
\begin{equation}\label{a-core}
	a=\frac{\frac{3}{2}+\delta-\sqrt{\frac{9}{4}+2\delta}}{1+\delta}
	\ \ ,\ \ \delta\equiv\frac{8(\epsilon-j)}{\pi}>0\ .
\end{equation}
Inserting it back to $s$ in (\ref{omega1}) expresses the entropy is given in terms
of the charge $\epsilon-2j$.
The $\epsilon,j$ charges of a Grey Galaxy and its core black hole have the same 
values of $\epsilon-2j$. Therefore, $s$ of (\ref{omega1}) expressed as a function 
of $\epsilon-2j$  is the entropy of the Grey Galaxy in terms of its independent charges 
$\epsilon,j$. 

\bibliography{biblio}\bibliographystyle{JHEP}

\end{document}